\documentclass[manuscript, screen=true, natbib=false, nonacm, 11pt]{acmart}
\usepackage{hyperref}
\usepackage{cite}
\usepackage{amsmath,amsfonts}
\usepackage{algorithmic}
\usepackage{graphicx}
\usepackage{textcomp}
\usepackage{xcolor}
\def\BibTeX{{\rm B\kern-.05em{\sc i\kern-.025em b}\kern-.08em
    T\kern-.1667em\lower.7ex\hbox{E}\kern-.125emX}}

\newcommand{\eg}{\textit{e.g., }}
\newcommand{\ie}{\textit{i.e., }}

\usepackage{subcaption}
\usepackage{tikz}
\newcommand*\circled[1]{\tikz[baseline=(char.base)]{
            \node[shape=circle,draw,inner sep=1pt] (char) {#1};}}

\usepackage{natbib}
\setcitestyle{acmauthoryear}
\begin{document}

\title{COOCK project Smart Port 2025 D3.2 \\``Variability in Twinning Architectures''}

\author{Randy Paredis}
\author{Hans Vangheluwe}
\author{Pamela Adelino Ramos Albertins}
\affiliation{
    \institution{University of Antwerp}
    \department{Department of Computer Science}
    \city{Antwerp}
    \country{Belgium}
}

\setcopyright{none}

\begin{abstract}
This document is a result of the COOCK project ``Smart Port 2025: improving and accelerating the operational efficiency of a harbour eco-system through the application of intelligent technologies''. The project is mainly aimed at SMEs, but also at large corporations. Together, they form the value-chain of the harbour. The digital maturity of these actors will be increased by model and data-driven digitization. The project brings together both technology users and providers/integrators. In this report, the broad spectrum of model and data-based digitization approaches is structured, under the unifying umbrella of ``Digital Twins''.
During the (currently quite ad-hoc) digitization process and in particular, the creations of Digital Twins, a variety of choices have an impact on the ultimately realised system. This document identifies three stages during which this ``variability'' appears: the Problem Space Goal Construction Stage, the (Conceptual) Architecture Design Stage and the Deployment Stage.
To illustrate the workflow, two simple use-cases are used: one of a ship moving in 1 dimension and, at a different scale and level of detail, a macroscopic model of the Port of Antwerp.
\end{abstract}

\maketitle

\section{Introduction}
\label{sec:intro}
A Digital Twins (DTs) are increasingly seen as a key enabler and backbone for digitization of complex Cyber-Physical Systems (CPSs). The ``Twinning Paradigm''  combines a System under Study (SuS -- also called an Actual Object (AO)) with a corresponding Twin Object (TO), \ie a model of that system. The Twin Object may be symbolic in nature, or data based.

The concept of ``twinning'' is not new.
Sand tables, for example, have been used since antiquity as ``twins'' of battlefields, which allowed generals to evaluate and plan troop movements. During the Apollo missions \cite{apollo13}, NASA used a mirrored training capsule on earth to mimic and, when needed, diagnose anomalies occurring in their deployed spacecraft.  While the ``twin'' is these historical examples is not ``digital'' (as it exist within physical space), they cover the same twinning concepts as modern-day DTs.

The core idea behind twinning is that the TOs are not just simulation models or data sets, but are continually kept ``in sync'' with their corresponding AO. The TO is fed the same sensor input as the AO receives. This allows for useful functionality such as ``anomaly detection'': significant deviation between AO measurements and TO predictions is an indication that the SuS is not functioning as intended.

Commonly, twins are built to satisfy a specific set of \textit{goals} pertaining to \textit{Properties of Interest} (PoIs). These goals are also referred to as the \textit{purpose} of a twin \cite{DALIBOR2022111361}. An example goal is ``anomaly detection'', with for example ``energy efficiency'' as a Property of Interest.
\cite{Qamar2012} defines a ``property'' as the
``descriptor of an artifact''. Hence, it is a specific concern of that artifact (\ie a system), which is either \textit{logical} 
Some of these properties can be \textit{computed} (or \textit{derived} from other, related artifacts), or \textit{specified} (by a user).
Goals consider the specific system PoIs that are of concern to certain stakeholders.

For many products, multiple \textit{variants} exist. Each of these variants have \textit{common} parts, aspects, components, \ldots, but \textit{varies} in others.
This is also the case for DTs. 
For instance, multiple ships of the same type might have different sails, different engines, different control software, \ldots These differences are called \textit{features}. One may see each variant as a separate \textit{configuration} of the product. The set of all variants forms a \textit{product family}. \textit{Feature trees/models} are often used to describe the elements of a product family.

\begin{figure}[htb]
    \centering
    \includegraphics[width=\textwidth]{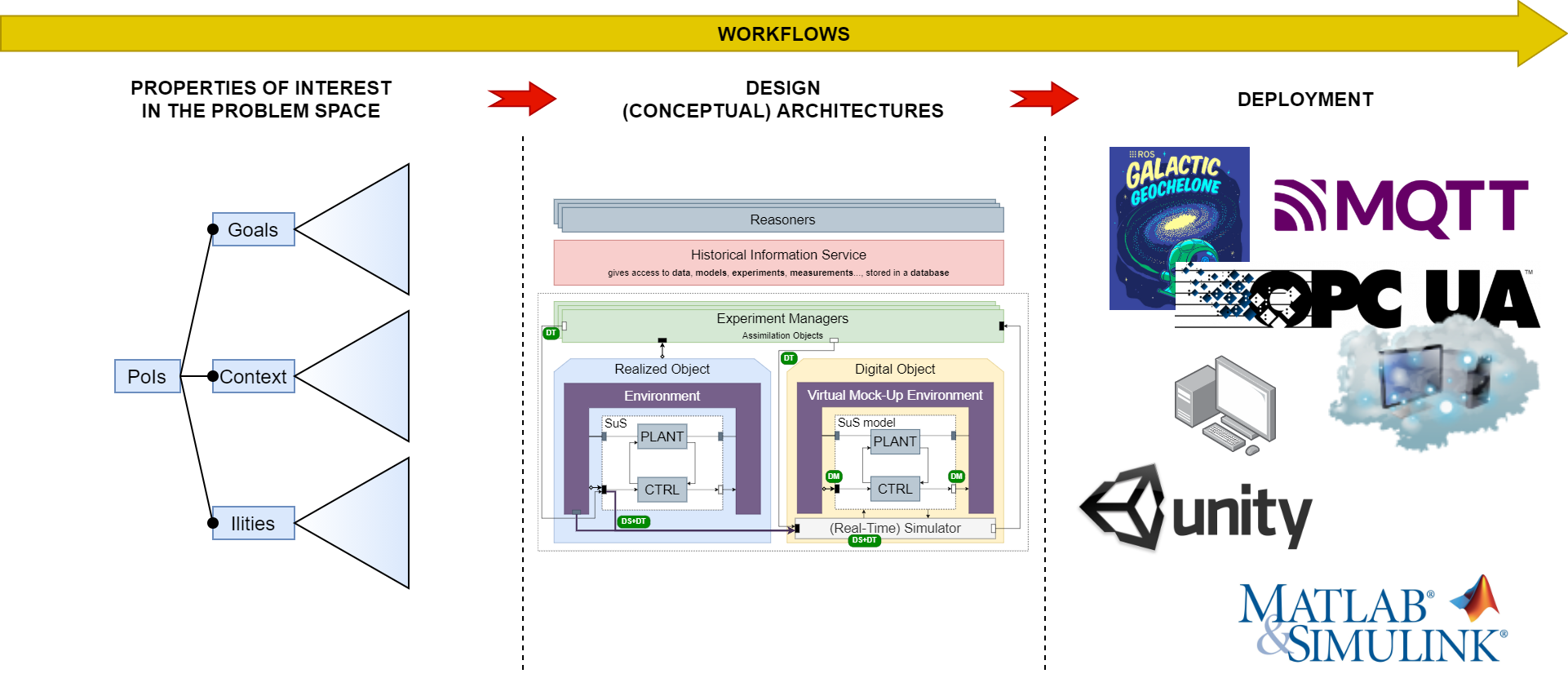}
    \caption{High-level workflow for creating (digital) twins.}
    \label{fig:timeline}
\end{figure}

Figure~\ref{fig:timeline} illustrates a high-level workflow for constructing twinning systems. 
Firstly, a set of goals (pertaining to PoIs) are identified for which the twin must be created. There can be multiple goals which the twin needs to satisfy. Section~\ref{sec:goals} goes deeper into this topic.

Secondly, based on the chosen goals/PoIs, a conceptual architecture will be chosen. Some components of this a architecture will be present/absent depending  on the choices made in the previous stage. The figure illustrates this by including a reference architecture from \cite{Paredis2022DZ} that has a set of ``toggles'' or ``presence conditions'' (\ie parts that can be enabled/disabled). This is further described in section~\ref{sec:archs}.

Finally, there is the deployment stage in which appropriate technologies are chosen/used to realise this twin. This details the communication protocols, server hardware and operating system, modelling formalism (and toolbox/simulator) etc. Section~\ref{sec:deploy} discusses this topic in more detail.

These three stages are all part of a set of workflows that allows the creation of a twin that implements the desired (and specified) behaviour.

In terms of execution, we will focus on the notion of experiments.
An experiment is an intentional set of (hierarchical) activities, carried out on a specific SuS in order to accomplish a specific (set of) goal(s).
Each experiment should have a recorded explicit description, setup and workflow, such that becomes is repeatable.

\section{Background}
\label{sec:background}
Ever since the conceptualization of Digital Twins (DTs) in \cite{grieves2017digital} and its introduction to Industry 4.0 as described in \cite{boss2020digital}, a plethora of (reference) architectures have sprouted up. Yet, most of these architectures remain conceptual, or highly abstract to actually translate into a running DT system.

\begin{figure}[htb]
    \centering
    \includegraphics[width=\textwidth]{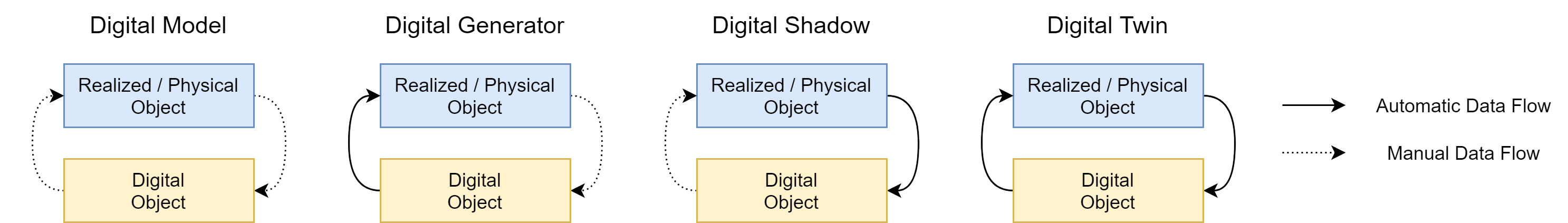}
    \caption{Generic timeline for creating (digital) twins, adapted from \cite{kritzinger2018digital} and \cite{tekinerdogan2020systems}.}
    \label{fig:kritzinger}
\end{figure}

Some architectures focus on the difference in connectivity between the AO and TO. For instance, \cite{kritzinger2018digital} introduces the notion of a Digital Model (DM -- when the AO and TO are only connected to each other via a bidirectional \textit{manual} connection), and a Digital Shadow (DS -- when the connection from the AO to the TO is automated). The paper states that a DT only appears if the bidirectional communication has no manual components. \cite{tekinerdogan2020systems} expands on this notion by introducing a Digital Generator (DG --  the dual of the DS, \ie when the connection from the TO to the AO is automated). This is illustrated in figure~\ref{fig:kritzinger}.

\begin{figure}[htb]
    \centering
    \includegraphics[width=0.5\textwidth]{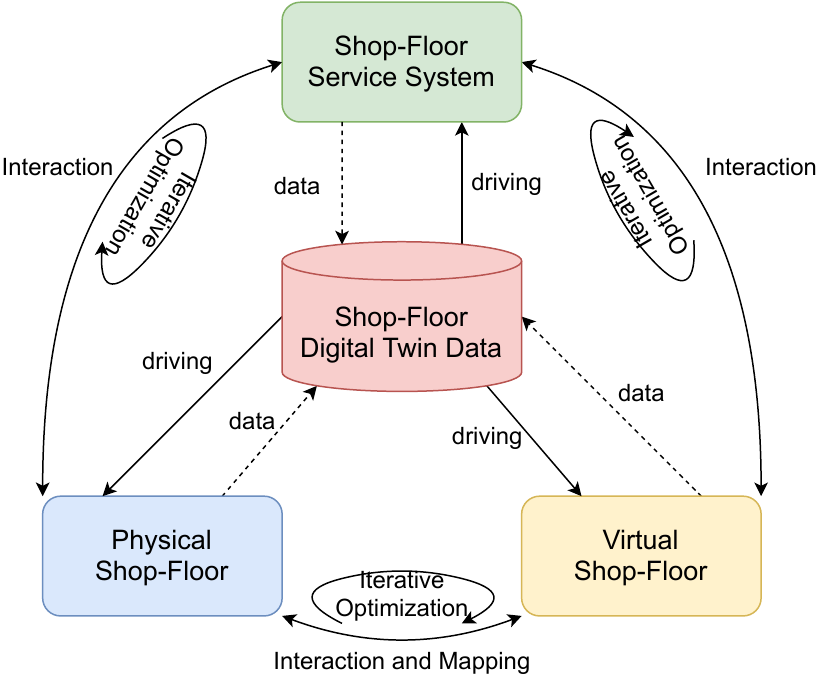}
    \caption{5D architecture for DTs, adapted from \cite{tao2017digital}.}
    \label{fig:tao}
\end{figure}

\cite{tao2017digital} leaves the communication aspect be and focuses on the individual parts and components required in a DT. As shown in figure~\ref{fig:tao}, they arrive at the so-called 5D architecture \cite{van2021models}, where five components are highlighted: (1) the ``\textit{Phyiscal Shop-Floor}'' (\ie the AO), (2) the ``\textit{Virtual Shop-Floor}'' (\ie the TO), (3) ``\textit{Shop-Floor Digital Twin Data}'' (in essence some sort of communication point or data lake), (4) ``\textit{Shop-Floor Service System} (a set of processes that use the data from the AO and the TO -- this is commonly called a set of \textit{Services}), and (5) all the communication between these components.

Note that both figures~\ref{fig:kritzinger} and \ref{fig:tao} implicitly incorporate a specific set of design and deployment choices. \cite{kritzinger2018digital} focuses on the communication channel between the AO and the TO. Yet, when considering twinning, this level of communication is the result of a specific design choice of the system, based on which goal and which corresponding PoI it tries to solve, as well as the technologies that are available.
\cite{tao2017digital} focuses on a specific domain of a shop-floor, but DTs also exist in a multitude of other domains \cite{DALIBOR2022111361}. Furthermore, the communication point/data lake is clearly already a deployment choice, subtly hinting at OPC-UA like technologies. Additionally, the indication of a ``physical'' AO and a ``virtual'' TO already shows a specific choice for the implementation of the DT.


\section{Running Examples}
\label{sec:example}
We will discuss two running examples to show the presented approach.

\subsection{Simple Port Simulation}
We present a simplified model of the Port of Antwerp-Bruges (PoAB) in which vessels
travel from the North Sea, over the Scheldt river, through canals and locks toward docks, and back. These locks and docks have a finite capacity.

\begin{figure}[htb]
    \centering
    \includegraphics[width=\textwidth]{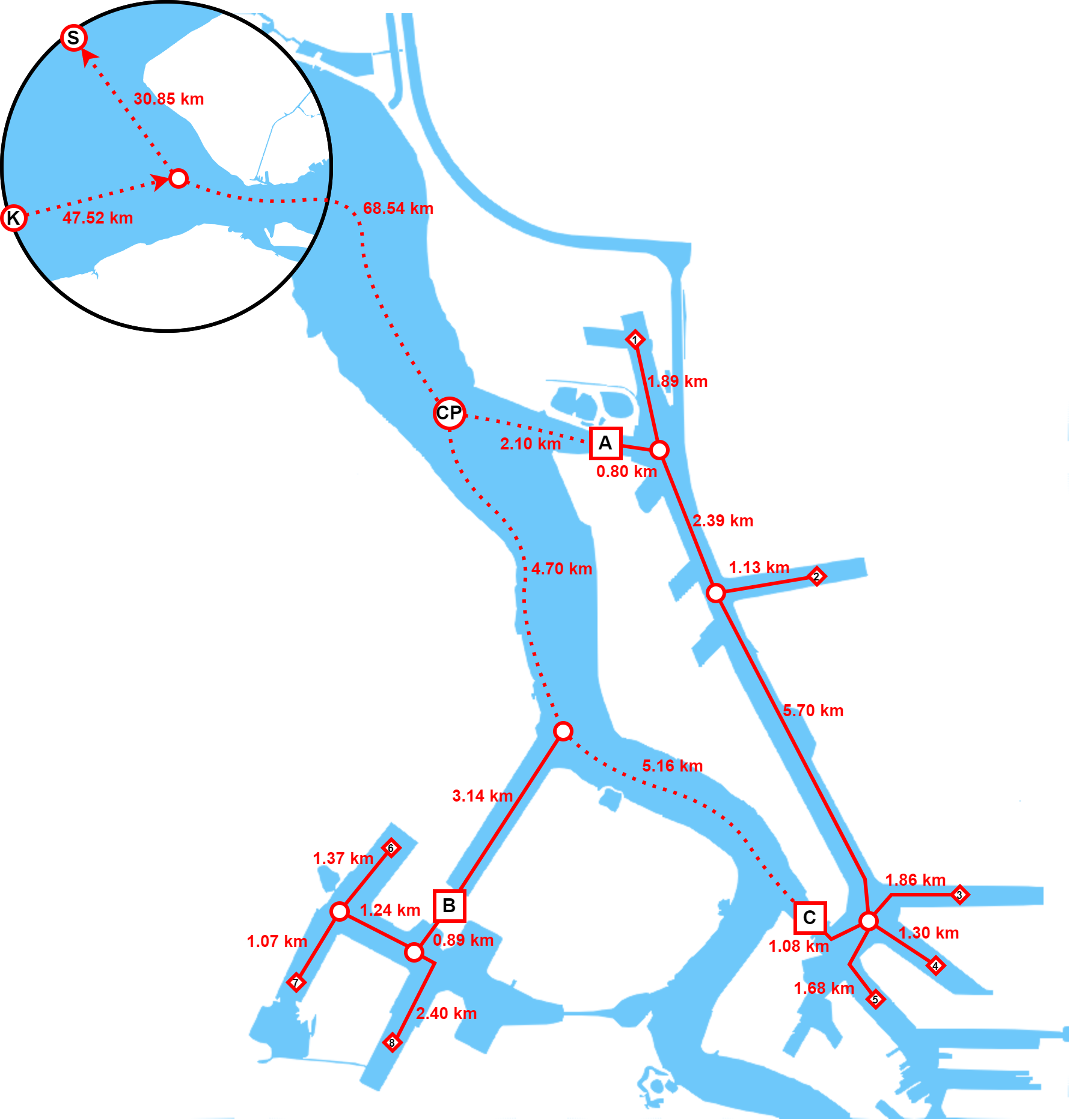}
    \caption{Simplified map of the Port of Antwerp.}
    \label{fig:PoA}
\end{figure}

Figure~\ref{fig:PoA} shows a map of the Scheldt through the city of Antwerp. For the sake of simplicity, we highlight the equivalence to road traffic.
In the figure, circles indicate junctions where multiple trajectories merge/diverge.
The three large squares are the locks and the small diamonds are the docks.
The red, dotted lines identify sections of sailing routes on the river. Generally, these are two-way connections, except for the connection from the ``loodskotter'' source (identified with ``K'') and to the sea endpoint (identified with ``S''). Canals are represented with full red lines. Next to all river sections and all canals, their distance is denoted.

Ships that arrive from ``K'' are given a target dock to sail to, where they will stay for some time, before departing towards ``S''. Each ship type is given its own unique velocity distribution, based on real-world information.
The docks can hold at most 50 vessels and the locks are area-constrained (\ie we will ignore that the top area of a ship is solid). Additionally, ships cannot overtake each other on canals.

The full specification of this PoAB example is found on \url{http://msdl.uantwerpen.be/people/hv/teaching/MoSIS/202223/assignments/DEVS}.

This system is clearly an abstraction of the real-world scenario. To make this more interesting and fit within the context of twinning, we will use another simulation of this very same system, with slightly different parameters. This second simulation will be considered the AO and the ``source of truth'' that we want the TO to conform to.

\subsection{1D Yacht}
Another example is a 1-dimensional yacht on water. This is based on an assignment given to Master-level students at the University of Antwerp, thus highlighting the simplicity of this use-case.

Let's say there exists a yacht that only sails in a single direction (i.e., turning is not incorporated in this example). Furthermore, we will assume that it is not being influenced by sea currents. We would like to construct a Digital Shadow for this yacht, such that we can visualize the differences from the expected behaviour in a simple dashboard.
We expect that the yacht follows a specific (repeating) velocity profile (which may be provided by some port authority). This can be modelled using the following Ordinary Differential Equations (ODEs):
\begin{align*}
    \begin{cases}
    F_R &= \dfrac{1}{2}\cdot\rho\cdot v^2\cdot S\cdot C_f\\
    C_f &= \dfrac{0.075}{\left(\log_{10}\left(Re\right) - 2\right)^2}\\
    Re &= v\cdot L/k
    \end{cases}
\end{align*}
Where $\rho$ is the density of sea water ($1025\ kg/m^3$, assuming the temperature is $15^\circ C$), $S$ is the submerged surface area of the vessel w.r.t. the direction of movement (approximately $261\ m^2$, based on previous experiments), $C_f$ is the friction value, $Re$ is Reynolds number, $L$ is the length of the yacht ($21.54\ m$), the dry mass $m$ equals $32,000\ kg$, and $F_R$ is the resistance force the yacht experiences when sailing at velocity $v$.

A PID controller, minimizing the error between $v$ and the velocity profile ($v_{ideal}$) provides a traction force $F_T$, such that the total force exerted on this yacht equals $F_T - F_R = m\cdot a$.

\section{Goal Feature Modeling}\label{sec:goals}
A DT is almost always built with a specific goal in mind. This is the DT's purpose \cite{DALIBOR2022111361}. There exist many different kinds of goals for DTs. While they may commonly focus on the services that must be created, the selection of certain goals will also influence specific other aspects of the architecture.

The most accepted way to model variability is by using Feature Modeling \cite{kang1990feature}. A Feature Tree (also known as a Feature Model or Feature Diagram) is a tree-like diagram that depicts all features of a product in groups of increasing levels of detail.
At each level, the Feature Tree indicates which features are mandatory and which are optional. It can also identify causality between features of the same or a different group (\ie ``if feature A is present, feature B must also be present'').
When constructing a specific product in a product family, one simply has to traverse the Feature Tree downwards, starting from the root. In each group, a set of features is selected from where new traversal is possible.
This results in a specific configuration (or feature selection) that uniquely identifies the product.

The variability of DTs has been discussed in the past \cite{DALIBOR2022111361, Paredis2021DT, van2020taxonomy, JONES202036}. While not always on the level of Feature Trees, the research can commonly easily be transformed to such a tree. In essence, in order for a DT to be built, a multitude of large choices are required. Yet, in most reference architectures, all concepts of variability pertaining the DTs purpose is bundled in the ``services'' component of many architectures.

In reality, a user conceives an experiment that includes some sort of collaboration between the AO and the TO.
This experiment's logic is described in a manner that an experiment manager (or orchestrator) can use it to viably setup, steer and tear down the full experiment.

\subsection{Goal Variability}\label{sec:ft}
\cite{paredis2024coock} and \cite{DALIBOR2022111361} identified the following twinning goals/purposes:
\begin{itemize}
    \item \textbf{Architectural Connection}: general system kind, \eg w.r.t. figure~\ref{fig:kritzinger}.
    \item \textbf{Design}: whether or not it is used for designing a SuS.
    \item \textbf{Operation}: considers the execution of the system.
    \begin{itemize}
        \item \textbf{Data Allocation}: how data is collected and stored.
        \item \textbf{Data Processing/Analysis}: which insights can be attained from the data.
        \item \textbf{Observe}: what needs to be monitored.
        \item \textbf{Modify}: what needs to be adapted.
    \end{itemize}
    \item \textbf{Visualization}: how the twin is visualized.
    \begin{itemize}
        \item \textbf{Console/Dashboard}: if a user interface is used.
        \item \textbf{Animation}: if there is 2D/3D visualization focusing on the system state.
    \end{itemize}
    \item \textbf{Maintenance}: how the system deteriorates.
    \begin{itemize}
        \item \textbf{Predictive Maintenance}: predicts when components will fail.
        \item \textbf{Fatigue Testing}: life verification of wear and tear.
        \item \textbf{Lifecycle Management}: how the system evolves.
    \end{itemize}
    \item \textbf{Quality Assurance}: verifies the quality of the system.
    \begin{itemize}
        \item \textbf{Consistency}: focuses on equivalence and differences between AO and TO.
        \item \textbf{Data Link}: are there delays allowed on the connections.
        \item \textbf{Timing}: how fast runs the system.
        \item \textbf{Ilities}: properties manifested after the system is in use.
        \item \textbf{Safety}: security, safety, possibility and what is allowed.
    \end{itemize}
    \item \textbf{Usage Context}: verifies when the system is used.
    \begin{itemize}
        \item \textbf{Sustainability}: how green/ecological is the system.
        \item \textbf{Execution}: is the system live, or is it based on historical data.
        \item \textbf{User}: who uses the system.
        \item \textbf{Education}: can the system be used to learn/teach/train.
        \item \textbf{PLM Stage}: when is the twin used.
        \item \textbf{Reuse}: can we repeat/replicate/reproduce/reuse/recondition the experiments.
    \end{itemize}
\end{itemize}

\subsection{Goals for Running Examples}
When considering the running example, there are some goals we need to select, as summarised in section~\ref{sec:ft}. Table~\ref{tab:goals} shows which goals were selected for the port example.

\begin{table}[htb]
    \centering
    \begin{tabular}{l|l}
        dimension & Properties of Interest \\ \hline
        Connection & Digital Model \\
        PLM Stage & As-Operated \\
        Operation & \begin{tabular}[c]{@{}l@{}}Data Allocation\\Monitoring\end{tabular}\\
        Visualization & Animation \\
        Timing & Real-Time \\
        Execution & Live \\
        Quality Assurance & Consistency \\
        Ilities & Reliability \\
        Safety & \begin{tabular}[c]{@{}l@{}}Legal Safety\\ Physical Laws\\ Human Safety\end{tabular}\\
        Reuse & Reproducibility\\
    \end{tabular}
    \caption{Selected Properties of interest for the port running example.}
    \label{tab:goals}
\end{table}

In terms of \textit{operation} for this \textit{Digital Model}, we would like to \textit{monitor} the current (and past) states, also enforcing \textit{data allocation}. This way we can \textit{visualize} the behaviour of the ships, which is \textit{animated} during a \textit{real-time}, \textit{live execution}.
When looking at this visualization, we can deduce the \textit{consistency} and \textit{reliability} of the constructed twin.
Furthermore we can see if (1) the velocity of the ships stay within legal bounds (\textit{legal safety}), (2) their engines can produce enough torque for the desired velocities (\textit{physical laws}), and (3) passengers do not constantly fall down due to fast changing speeds (\textit{human safety}). Note that these goals might be valid by default, if the model has been calibrated well enough.
Finally, we may want to \textit{reproduce} these experiments for a completely different port.

As for the 1D yacht example, table~\ref{tab:goals2} shows the selected goals. This is almost identical to the previous table.

\begin{table}[htb]
    \centering
    \begin{tabular}{l|l}
        dimension & Properties of Interest \\ \hline
        Connection & Digital Shadow \\
        PLM Stage & As-Operated \\
        Operation & \begin{tabular}[c]{@{}l@{}}Data Allocation\\Monitoring\end{tabular}\\
        Visualization & Plot \\
        Timing & Real-Time \\
        Execution & Live \\
        Quality Assurance & Consistency \\
        Ilities & Reliability \\
        Safety & \begin{tabular}[c]{@{}l@{}}Legal Safety\\ Physical Laws\\ Human Safety\end{tabular}\\
        Reuse & Reproducibility\\
    \end{tabular}
    \caption{Selected Properties of interest for the yacht running example.}
    \label{tab:goals2}
\end{table}

In terms of \textit{operation} for this \textit{Digital Shadow}, we would like to \textit{monitor} the current (and past) velocity/position, also enforcing \textit{data allocation}. This way we can \textit{visualize} the behaviour of the yacht, which is \textit{plotted} during a \textit{real-time}, \textit{live execution}. Other goals follow the same reasoning as mentioned before.

\section{(Conceptual) Reference Architecture}\label{sec:archs}

\begin{figure}[htb]
    \centering
    \includegraphics[width=0.6\textwidth]{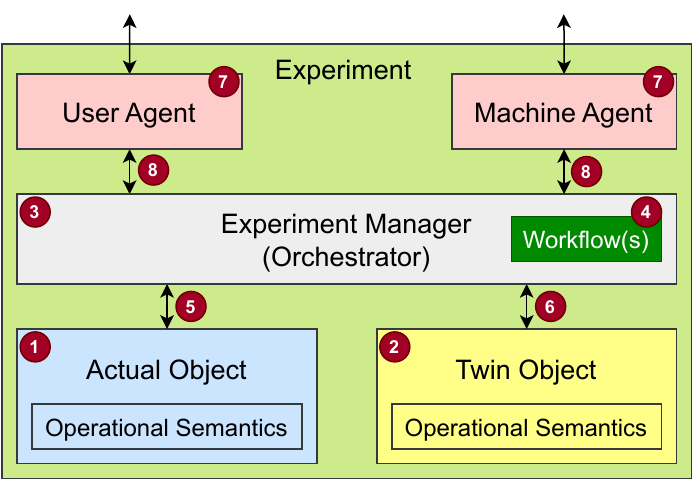}
    \caption{Annotated high-level architectural view of a twin experiment.}
    \label{fig:arch1}
\end{figure}

Figure~\ref{fig:arch1} illustrates the basic architecture to which a multitude of existing twinning experiments conform. Notice that it is high-level and abstract, allowing users to decide on a high level which components to include in their system.

To further explain the figure, each component has been annotated with a number.
The AO \circled{1} and TO \circled{2} are given a behaviour via its \textit{operational semantics}. This can be a neural network, the execution of code, real-world behaviour, etc.
It is important to denote that \circled{1} is an \textit{abstraction} of a specific \textit{view} of the actual world (and environment) in which the AO is active.
\circled{3} is a so-called ``\textit{Experiment Manger}'' (or ``\textit{Orchestrator}'') which contains a set of \textit{workflows} \circled{4} that indicate \textit{how} the experiment is to be executed. Because the experiment is created for a specific set of goals, this logic should also be contained in \circled{3}.
For instance, if we only want to have a dashboard that visualizes the current state, the collection of this state is to be done by the Experiment Manager. If instead we want some anomaly detection, the Experiment Manager needs to compute some distance metric.

\circled{5} and \circled{6} indicate the communication between the Experiment Manager and the AO or TO, respectively. Note that the downwards communication may be interpreted in a really broad manner. It may consider the intructions that need to be sent to the objects, to launch or halt their individual executions. Alternatively, it may also send data to update the objects (for a DG, DS or DT). The upwards communication can be considered the set of data that the sensors in this AO have captured.

The Orchestrator can communicate with some User Agent or Machine Agent \circled{7}, which identifies an access point for a user or another system to obtain information about this experiment. This could be a dashboard, or an API. The communication between the Agents and the Orchestrator \circled{8} can be bidirectional when \circled{7} can also steer the twin, but it might likely be one-directional when discussing a DM or a DS.

\begin{figure}[htb]
    \vspace{0.3cm}
    \begin{subfigure}[t]{0.3\textwidth}
        \centering
        \includegraphics[width=\textwidth]{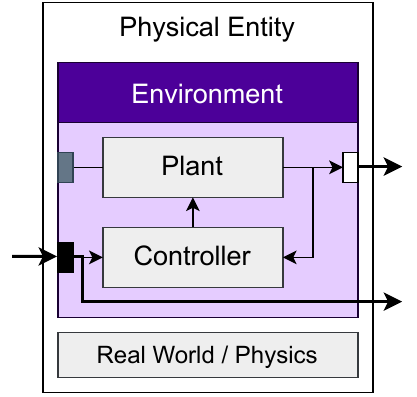}
        \caption{Architecture for a plant-controller physical entity.}
        \label{fig:ent-p}
    \end{subfigure}
    ~
    \begin{subfigure}[t]{0.3\textwidth}
        \centering
        \includegraphics[width=\textwidth]{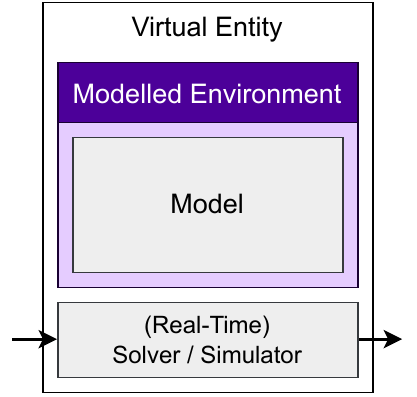}
        \caption{Architecture for a virtual entity.}
        \label{fig:ent-v}
    \end{subfigure}
    ~
    \begin{subfigure}[t]{0.3\textwidth}
        \centering
        \includegraphics[width=\textwidth]{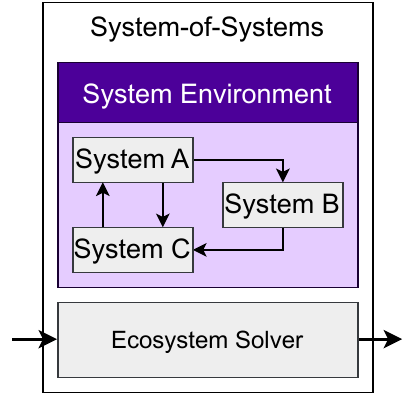}
        \caption{Architecture for a system-of-systems.}
        \label{fig:ent-ss}
    \end{subfigure}
    
    \caption{Three variability examples for AO and TO (including their operational semantics).}
    \label{fig:arch-entity}
\end{figure}

The AO and the TO themselves (including their operational semantics) can be separated into more specific architectures, based on the desired goals w.r.t. their PoIs. This is illustrated in figure~\ref{fig:arch-entity}.

Commonly in a DT, the AO is a \textit{physical entity}, which implies that it is a SuS that is active in the real, physical world. One possible architecture for such a real-world system, focusing on a plant-controller feedback loop is given in figure~\ref{fig:ent-p}. This system is active in a (subset of) the real-world environment, which implies that this should also be incorporated in the entity.
Additionally, there may be environmental influences we did not account for. For instance: a really sunny day might yield a sensor to produce invalid results.
The operational semantics of this entity equates to its behaviour in the real-world (\ie physics).

A DT has its TO in the virtual world, which is identified as a ``\textit{virtual entity}'' in figure~\ref{fig:ent-v}. Here, a model in some formalism is given operational semantics by executing a solver (or a simulator). This may potentially be a real-time simulator, or an as-fast-as-possible simulator, depending on the chosen goals.

Figure~\ref{fig:ent-ss} shows a \textit{system-of-systems} setup. For instance, collaboration of multiple machines in a factory. If this is virtual, this will likely be solved using co-simulation. If it is in the real-world, its behaviour is the union of the sub-system behaviour, including the communications.

Hence, there is no requirement that the AO is a physical entity and the TO is a virtual entity. A plethora of other options exist \cite{paredis2024coock}.
It might as well be the case that the AO can itself be a virtual entity \cite{Heithoff2023, ANGJELIU2020106282}, a \textit{biochemical entity} \cite{david2023digital, howard2021greenhouse, tekinerdogan2020systems, silber2023accelerating, lifescienceDT}, a \textit{social entity} \cite{GRAESSLER2018185, Traore2023}, an \textit{electromechanical entity} \cite{Barosan2020DevelopmentCenter, Mandolla2019, Soccer_robots}, a \textit{solid entity} \cite{DALIBOR2022111361}, a \textit{system-of-systems} \cite{Biesinger2018}, or a \textit{business process} \cite{RambowHoeschele2018CreationOA}.
Similarly, the TO also has the option to be a \textit{biochemical entity} \cite{topuzoglu2019finding}, a \textit{social entity}, an \textit{electromechanical entity} \cite{apollo13}, or a solid entity (\eg a sand table, or the Apollo training capsules).

For the AO, the incoming and outgoing arrows in figures~\ref{fig:ent-p}, ~\ref{fig:ent-v} and \ref{fig:ent-ss} relate to \circled{5} in figure~\ref{fig:arch1}.
Similarly, for the TO, the incoming and outgoing arrows represent \circled{6}.

\subsection{Common Architecture Variations}
Appendix~\ref{app:figs} shows all possible variations of the architecture shown in figure~\ref{fig:arch1}. Notice that at least an AO or a TO needs to be present and that all arrows need to be connected.

Commonly, a Machine Agent or a User Agent should be present. This only makes sense when the communication between the Machine/User Agent and the Orchestrator goes in the same direction as the communication between the Orchestrator and the AO and/or TO. For instance, figure~\ref{fig:var-15} does not make sense, as the Machine Agent influences the Orchestrator, but the latter does not use this to update the TO. For the sake of readability, we will ignore the Machine Agent and User Agent in the further description (and thus only look at the first 15 variants).

Figure~\ref{fig:var-0}, figure~\ref{fig:var-1} and figure~\ref{fig:var-2} show a simple simulation that can either be autonomous, instructed or both.
Figure~\ref{fig:var-3}, figure~\ref{fig:var-4} and figure~\ref{fig:var-5} show the same for a running SuS.

If we assume that all arrows represent data transfer (and not control information), we can identify an example of a Digital Model (as per \cite{kritzinger2018digital}) in figure~\ref{fig:var-6}.
Figure~\ref{fig:var-7} then shows a Digital Shadow and figure~\ref{fig:var-9} is indicative of a Digital Generator \citep{tekinerdogan2020systems}. Figure~\ref{fig:var-14} is the minimal setup when we consider the Digital twin from \cite{kritzinger2018digital}.
Note that for the Digital Shadow, figure~\ref{fig:var-8} makes more sense, given that the Orchestrator will also use information from the TO. The same can be said for the Digital Generator and figure~\ref{fig:var-12}.

\subsection{Choices for Running Examples}
\begin{figure}[htb]
    \centering
    \includegraphics[width=0.6\textwidth]{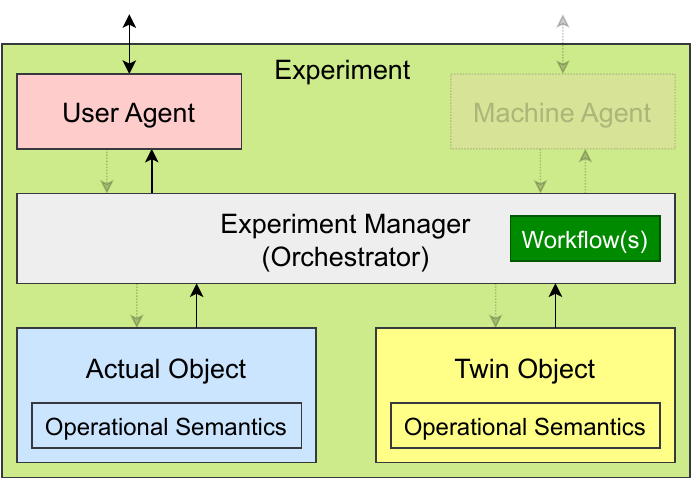}
    \caption{Architectural view for the port running example.}
    \label{fig:arch2}
\end{figure}

\begin{figure}[htb]
    \centering
    \includegraphics[width=0.6\textwidth]{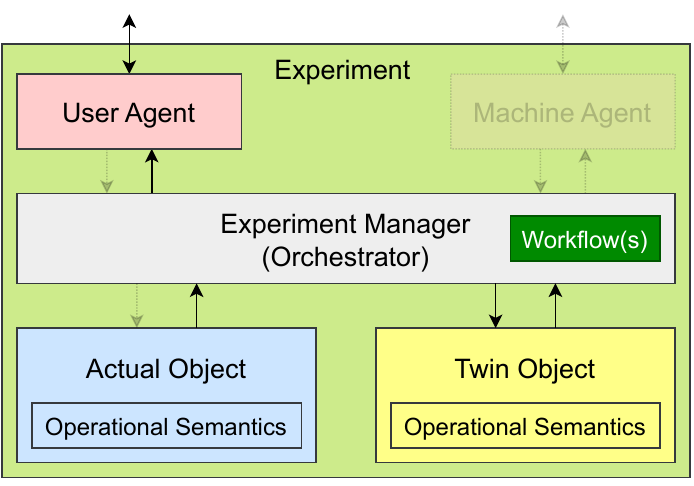}
    \caption{Architectural view for the yacht running example.}
    \label{fig:arch3}
\end{figure}

Based on the goal choices, made in table~\ref{tab:goals} and table~\ref{tab:goals2}, we can deduce that (for the use-cases), the architecture from figure~\ref{fig:arch1} is to be altered to figure~\ref{fig:arch2} and figure~\ref{fig:arch3} (respectively).

The port example is a simple Digital Model, which implies data gathering from both the AO and the TO (via the Orchestrator). Additionally, the Orchestrator may also steer (\ie setup, launch, progress, pause, halt, teardown,\ldots) the TO. The real-time animation is made possible via the User Agent.

The port use-case is a very simple example, based on real-world information.
For both the AO and the TO, figure~\ref{fig:ent-v} will be used.

The yacht example is a Digital Shadow, which implies that besides data gathering, there is also data transfer between the AO and the TO.
Whereas the AO will exist in the physical space, it is a simplification of reality, resulting in figure~\ref{fig:ent-v} for both the AO and the TO.
Inside the model, however, a plant-controller loop is constructed, to mimic the behaviour shown in figure~\ref{fig:ent-p}.

Within the COOCK project, PoAB provided access to data from their APICS system, yielding a data lake with information for all activities of tugboats in 2022. This data contains a real-world execution trace and can hence be used as AO (\ie a representation of reality) if desired.

\section{Deployment}\label{sec:deploy}
In order to deploy the ship experiment, we will use the architecture from figure~\ref{fig:arch2} as a basis to know which components to construct. Yet, there are still some technological decisions to be made.
Besides the selection of the modelling languages, tools, simulators, formalisms and exact implementation of the models, we still need to decide the Operating System (OS) on which we will deploy the components and the exact communication protocols.

Common OS choices are Windows, Apple, or Linux, but all their versions and flavours can also be considered. This choice is less important when tools are platform-independent, but this is definitely not always the case. Specific devices might have a very small OS with limited memory, disabling the possibility of memory-intensive computations (such as most geometric mathematics) that might be required to accomplish a specific goal.
When the AO and/or TO is not in the virtual space, we need to consider the real-world location where it is situated. A twin that focuses on the very specific light intensity of a sensor might be massively influenced by the amount of sunshine during the experiment.

For communication protocols, common standards for twinning are DDS (\url{https://www.rti.com/products/dds-standard}), MQTT (\url{https://mqtt.org/}) and OPC-UA (\url{https://opcfoundation.org/about/opc-technologies/opc-ua/}).
When components are present on the same device, maybe the communication might happen via shared memory. Alternatively, classic distributed computing protocols (like client-server and Peer to Peer (P2P)) might also be an option, likely implemented over HTTP, TCP/IP or UDP.
When working with robots, the industrial standard is ROS (\url{https://www.ros.org/}), which is a DDS-based technology.
In the real world it is possible to use indicator lights, displays, Morse code, flag semaphores, speech,\ldots in order to convey information.
Each of these technologies has its own advantages and downsides, making selecting ``the right one'' for your goal(s) an important question.

\subsection{Choices for Running Examples}
\begin{figure}[htb]
    \centering
    \includegraphics[width=0.7\textwidth]{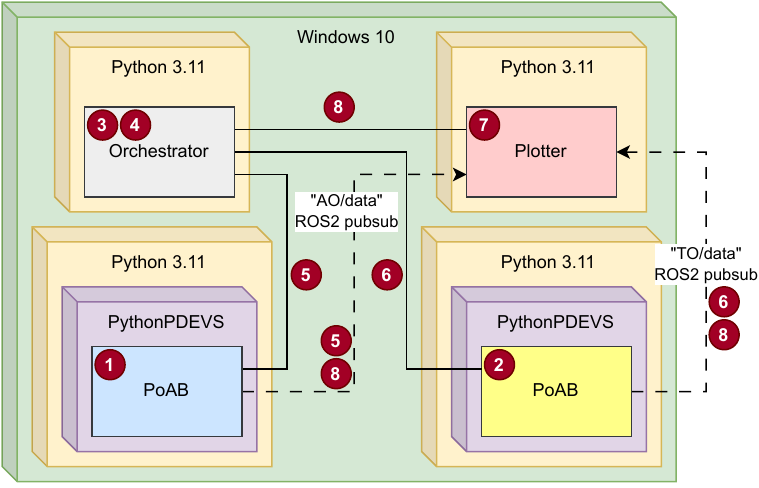}
    \caption{Deployment diagram for the port experiment.}
    \label{fig:deploy}
\end{figure}

For the port, we decided on using DEVS to model the PoAB system, such that we can implement it in PythonPDEVS \cite{PythonPDEVS}, which allows real-time execution of the simulation.
A mapping tool called \texttt{geoplotlib} is used to do the visualization of the port.
All communication happens via ROS2 (galactic geochelone).
A deployment diagram is shown in figure~\ref{fig:deploy}, following the same numbering annotations as figure~\ref{fig:arch1}.

\begin{figure}[htb]
    \centering
    \includegraphics[width=0.7\textwidth]{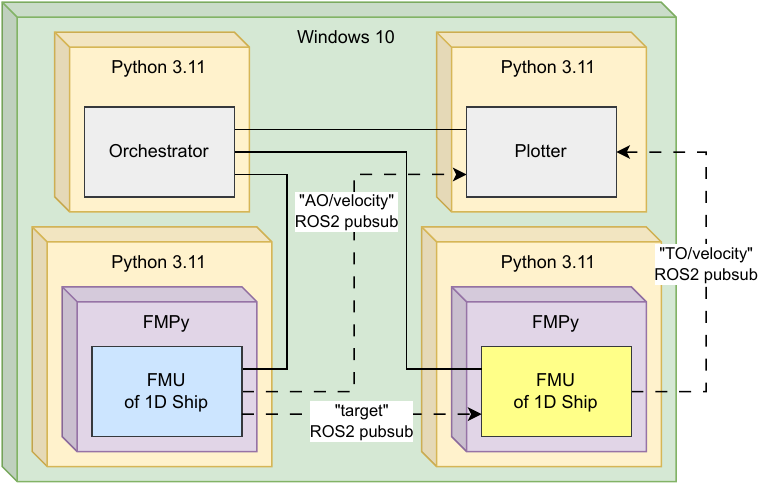}
    \caption{Deployment diagram for the yacht experiment.}
    \label{fig:deploy2}
\end{figure}

The yacht example uses ODEs that can be modelled using Modelica. We will use this to model both the AO and the TO, following a standard plant-controller pattern.
Next, these models are converted to FMUs, which can easily be co-simulated. A user-agent written in Python can visualize certain system states via the FMU interface.
As a communication protocol, we will use ROS2 (galactic geochelone) for all communications.
A deployment diagram is shown in figure~\ref{fig:deploy2}.

\section{Conclusions and Future Work}
Nowadays, industry is heavily focused on Industry 4.0 and DTs. This twinning paradigm can be applied in any domain to answer specific research questions. Hence, a vast product family of twinning experiments appears.
This paper highlights three high-level phases on which variability may appear: \textit{Problem Space Goal Construction}, \textit{Designing Architectures} and \textit{Deployment}.

We can construct a large feature tree based on existing sets of goals \cite{DALIBOR2022111361, paredis2024coock}. Based on these choices, a lot of implementation details are already made explicit.

A lot of conceptual architectures already exist. Some focus on the connectivity \cite{kritzinger2018digital}, others emphasise on the system flow \cite{eramo2021conceptualizing}, whereas others mainly focus on the internal components of the TO itself \cite{bolender2021self}.
This paper focuses on the variability that exists between the most common architectural descriptions. Next, this architecture is deployed as a running system, making even more choices on the used formalisms and technologies.

The use-case described in this paper entails a simplistic ``toy'' scenario. In the future, more complicated (and real-world) use-cases may be used to show and verify the validity of the provided architecture. Furthermore, we can take a look at the architecture within the context of twinning ecosystems.
Finally, we are planning on showing the equivalence between the architecture and the previously mentioned conceptual architectures that also have appeared in literature.

\begin{acks}
We would like to thank the Port of Antwerp-Bruges for their explanations of the internal workings of the nautical chain.
\end{acks}

\bibliography{my_papers.bib, Papers.bib}

\clearpage
\appendix
\section{Architectural Variability}\label{app:figs}
List of all architectural variations for the given image. Removed components are faded out.

\begin{figure}[htb]
	\centering
	\begin{minipage}[t]{0.3\textwidth}
		\vspace{0.6cm}
        \centering
        \includegraphics[width=\textwidth]{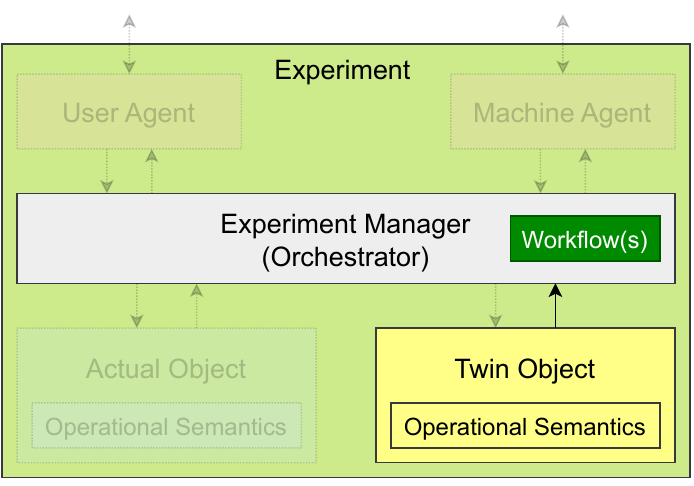}
        \caption{Reference architecture, variant 1, without Actual Object, User Agent and Machine Agent.}
        \label{fig:var-0}
    \end{minipage}\hfill
	\begin{minipage}[t]{0.3\textwidth}
		\vspace{0.6cm}
        \centering
        \includegraphics[width=\textwidth]{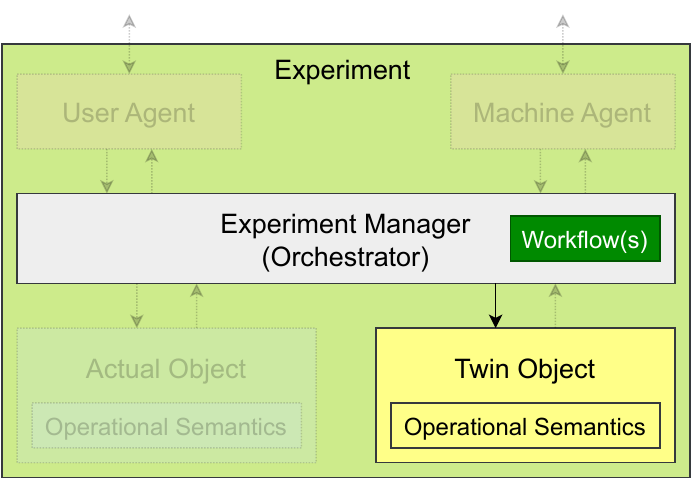}
        \caption{Reference architecture, variant 2, without Actual Object, User Agent and Machine Agent.}
        \label{fig:var-1}
    \end{minipage}\hfill
	\begin{minipage}[t]{0.3\textwidth}
		\vspace{0.6cm}
        \centering
        \includegraphics[width=\textwidth]{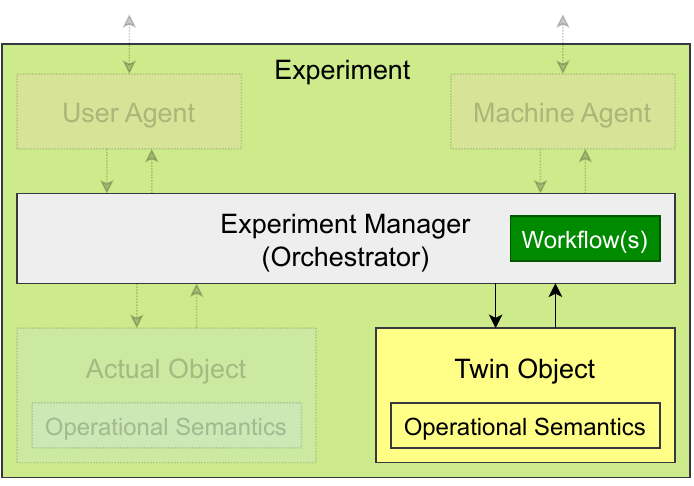}
        \caption{Reference architecture, variant 3, without Actual Object, User Agent and Machine Agent.}
        \label{fig:var-2}
    \end{minipage}\\

	\begin{minipage}[t]{0.3\textwidth}
		\vspace{0.6cm}
        \centering
        \includegraphics[width=\textwidth]{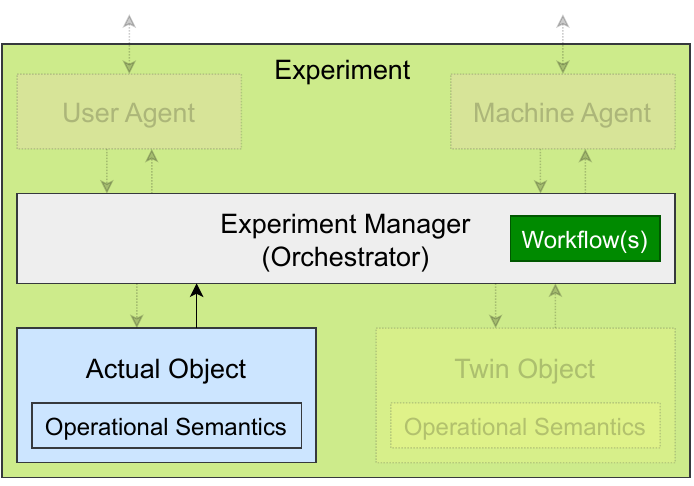}
        \caption{Reference architecture, variant 4, without Twin Object, User Agent and Machine Agent.}
        \label{fig:var-3}
    \end{minipage}\hfill
	\begin{minipage}[t]{0.3\textwidth}
		\vspace{0.6cm}
        \centering
        \includegraphics[width=\textwidth]{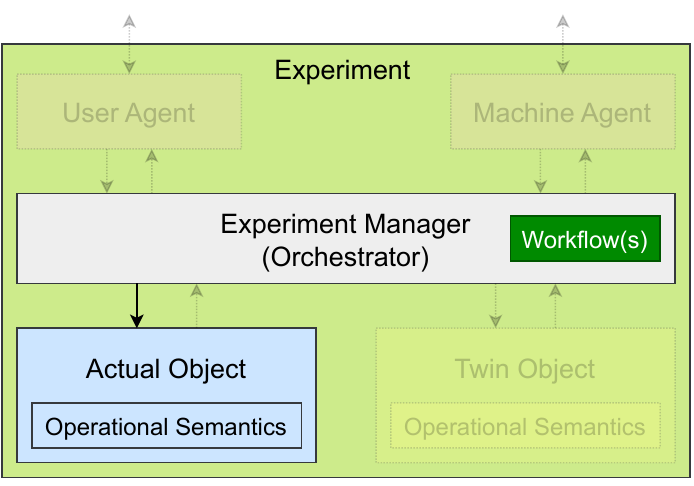}
        \caption{Reference architecture, variant 5, without Twin Object, User Agent and Machine Agent.}
        \label{fig:var-4}
    \end{minipage}\hfill
	\begin{minipage}[t]{0.3\textwidth}
		\vspace{0.6cm}
        \centering
        \includegraphics[width=\textwidth]{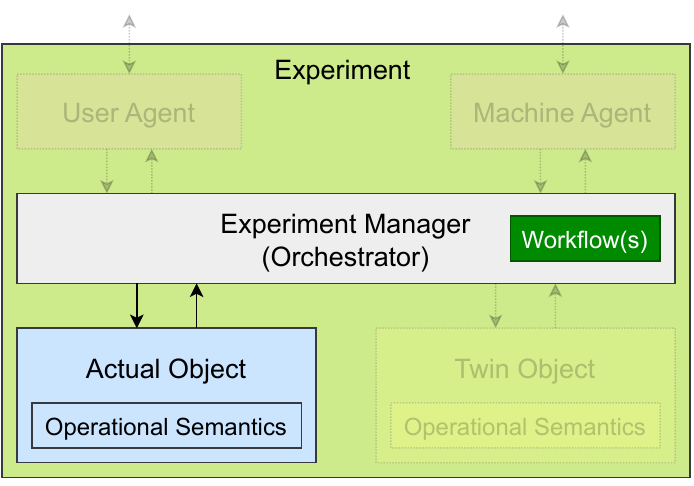}
        \caption{Reference architecture, variant 6, without Twin Object, User Agent and Machine Agent.}
        \label{fig:var-5}
    \end{minipage}\\

	\begin{minipage}[t]{0.3\textwidth}
		\vspace{0.6cm}
        \centering
        \includegraphics[width=\textwidth]{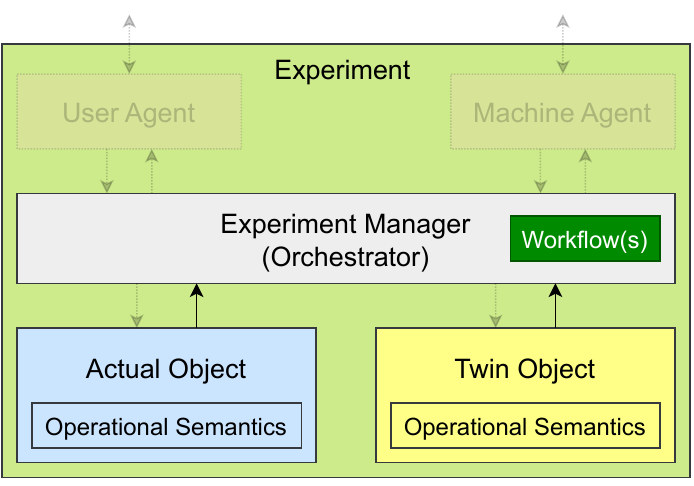}
        \caption{Reference architecture, variant 7, without User Agent and Machine Agent.}
        \label{fig:var-6}
    \end{minipage}\hfill
	\begin{minipage}[t]{0.3\textwidth}
		\vspace{0.6cm}
        \centering
        \includegraphics[width=\textwidth]{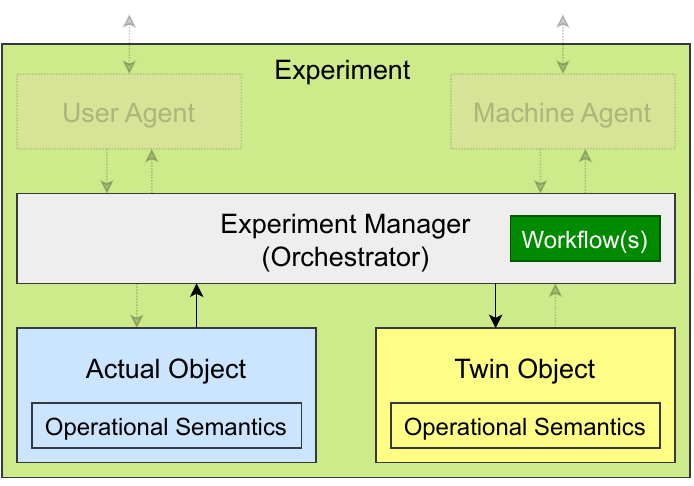}
        \caption{Reference architecture, variant 8, without User Agent and Machine Agent.}
        \label{fig:var-7}
    \end{minipage}\hfill
	\begin{minipage}[t]{0.3\textwidth}
		\vspace{0.6cm}
        \centering
        \includegraphics[width=\textwidth]{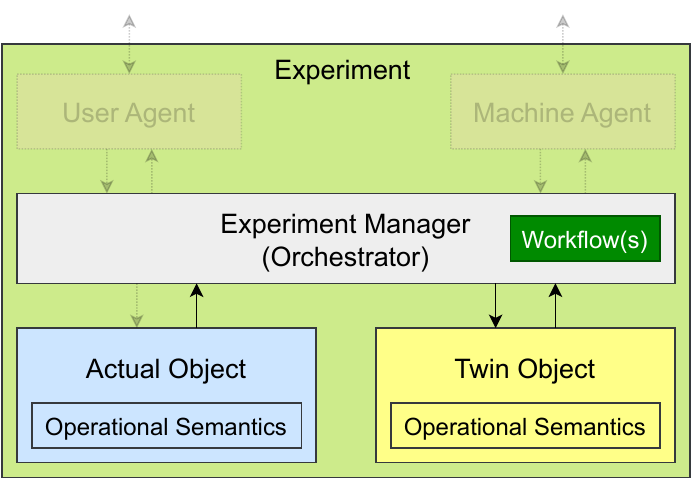}
        \caption{Reference architecture, variant 9, without User Agent and Machine Agent.}
        \label{fig:var-8}
    \end{minipage}\\

\end{figure}
\begin{figure}[htb]
	\centering
	
\end{figure}
\begin{figure}[p]
	\centering
	\begin{minipage}[t]{0.3\textwidth}
		\vspace{0.6cm}
        \centering
        \includegraphics[width=\textwidth]{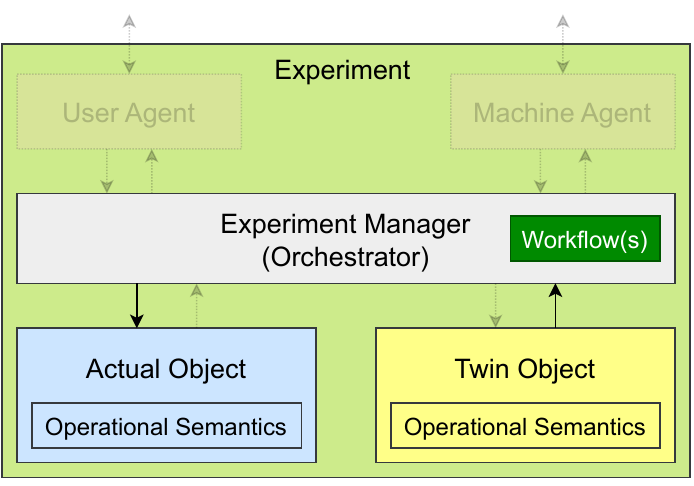}
        \caption{Reference architecture, variant 10, without User Agent and Machine Agent.}
        \label{fig:var-9}
    \end{minipage}\hfill
	\begin{minipage}[t]{0.3\textwidth}
		\vspace{0.6cm}
        \centering
        \includegraphics[width=\textwidth]{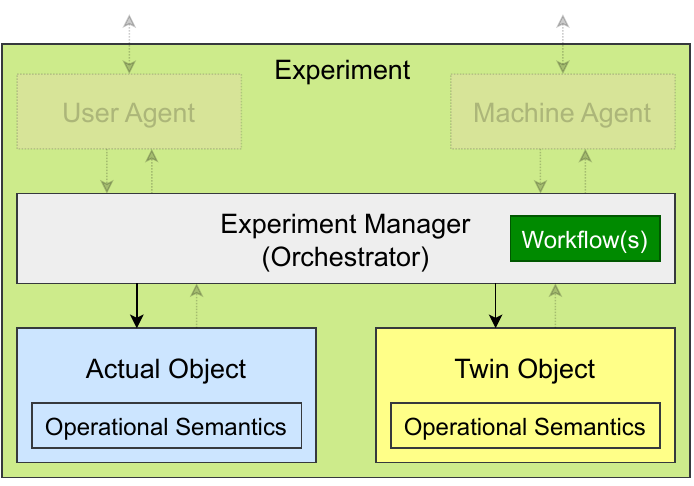}
        \caption{Reference architecture, variant 11, without User Agent and Machine Agent.}
        \label{fig:var-10}
    \end{minipage}\hfill
	\begin{minipage}[t]{0.3\textwidth}
		\vspace{0.6cm}
        \centering
        \includegraphics[width=\textwidth]{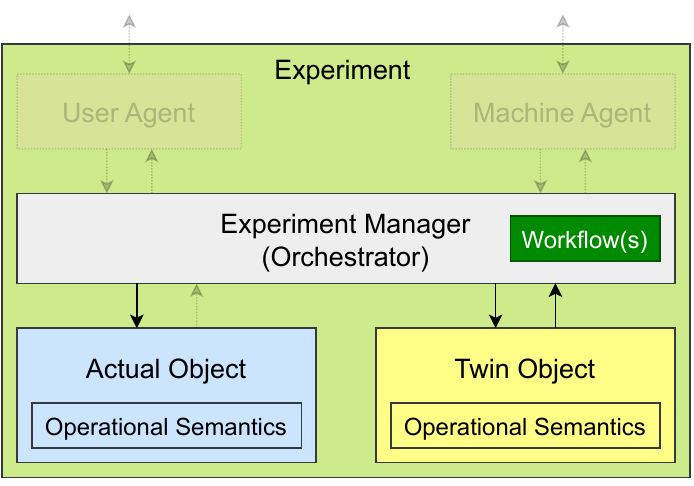}
        \caption{Reference architecture, variant 12, without User Agent and Machine Agent.}
        \label{fig:var-11}
    \end{minipage}\\

	\begin{minipage}[t]{0.3\textwidth}
		\vspace{0.6cm}
        \centering
        \includegraphics[width=\textwidth]{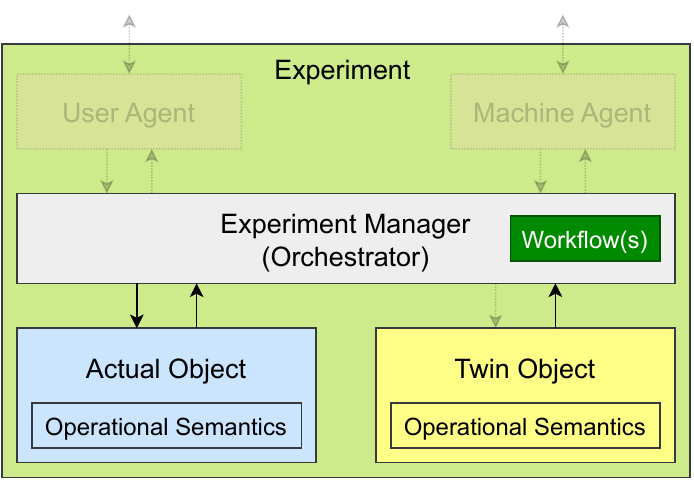}
        \caption{Reference architecture, variant 13, without User Agent and Machine Agent.}
        \label{fig:var-12}
    \end{minipage}\hfill
	\begin{minipage}[t]{0.3\textwidth}
		\vspace{0.6cm}
        \centering
        \includegraphics[width=\textwidth]{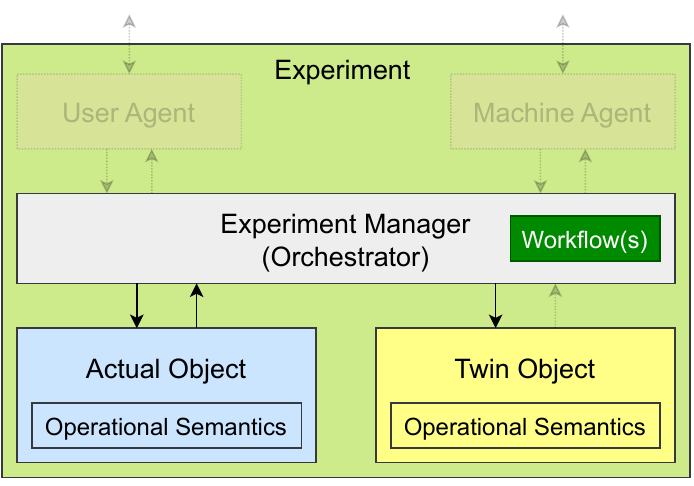}
        \caption{Reference architecture, variant 14, without User Agent and Machine Agent.}
        \label{fig:var-13}
    \end{minipage}\hfill
	\begin{minipage}[t]{0.3\textwidth}
		\vspace{0.6cm}
        \centering
        \includegraphics[width=\textwidth]{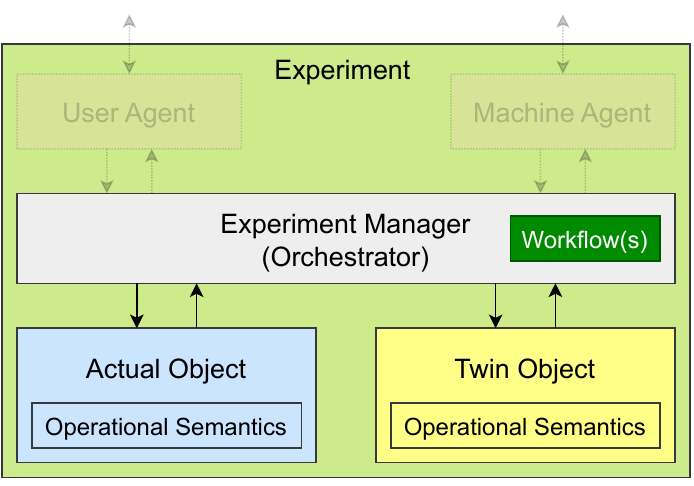}
        \caption{Reference architecture, variant 15, without User Agent and Machine Agent.}
        \label{fig:var-14}
    \end{minipage}\\

	\begin{minipage}[t]{0.3\textwidth}
		\vspace{0.6cm}
        \centering
        \includegraphics[width=\textwidth]{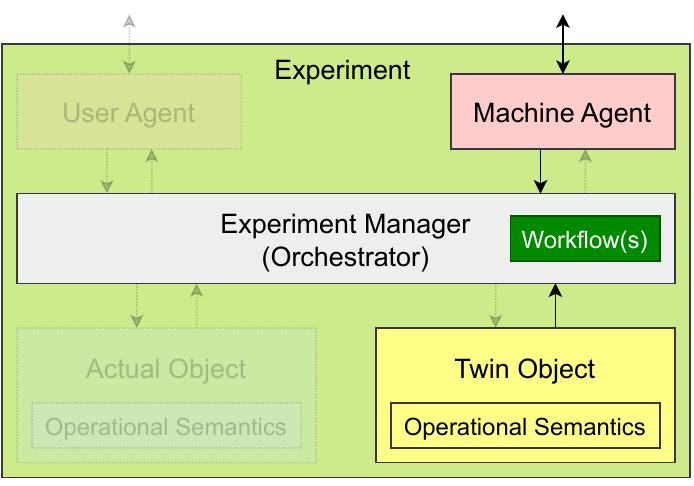}
        \caption{Reference architecture, variant 16, without Actual Object and User Agent.}
        \label{fig:var-15}
    \end{minipage}\hfill
	\begin{minipage}[t]{0.3\textwidth}
		\vspace{0.6cm}
        \centering
        \includegraphics[width=\textwidth]{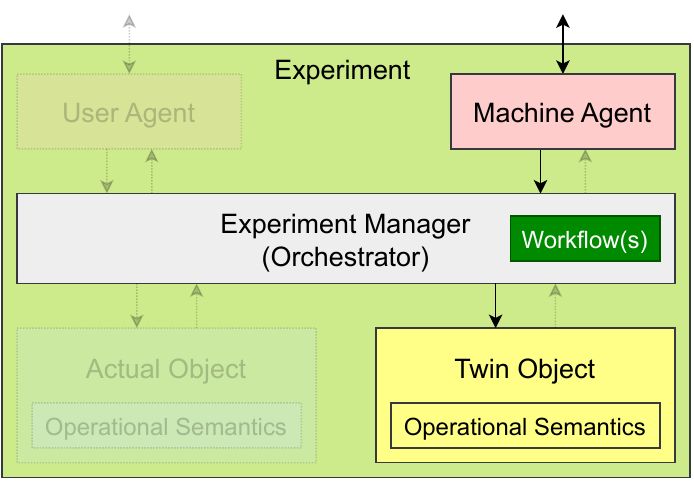}
        \caption{Reference architecture, variant 17, without Actual Object and User Agent.}
        \label{fig:var-16}
    \end{minipage}\hfill
	\begin{minipage}[t]{0.3\textwidth}
		\vspace{0.6cm}
        \centering
        \includegraphics[width=\textwidth]{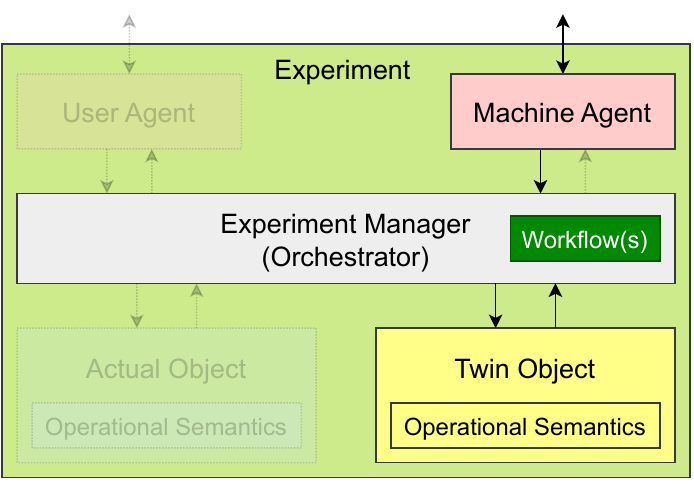}
        \caption{Reference architecture, variant 18, without Actual Object and User Agent.}
        \label{fig:var-17}
    \end{minipage}\\

	\begin{minipage}[t]{0.3\textwidth}
		\vspace{0.6cm}
        \centering
        \includegraphics[width=\textwidth]{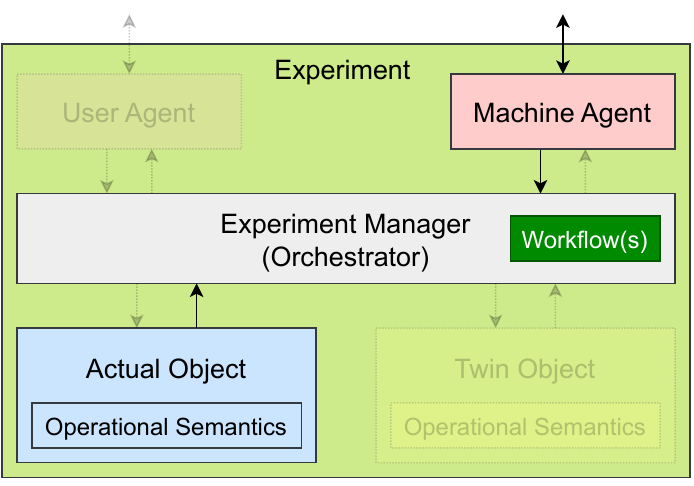}
        \caption{Reference architecture, variant 19, without Twin Object and User Agent.}
        \label{fig:var-18}
    \end{minipage}\hfill
	\begin{minipage}[t]{0.3\textwidth}
		\vspace{0.6cm}
        \centering
        \includegraphics[width=\textwidth]{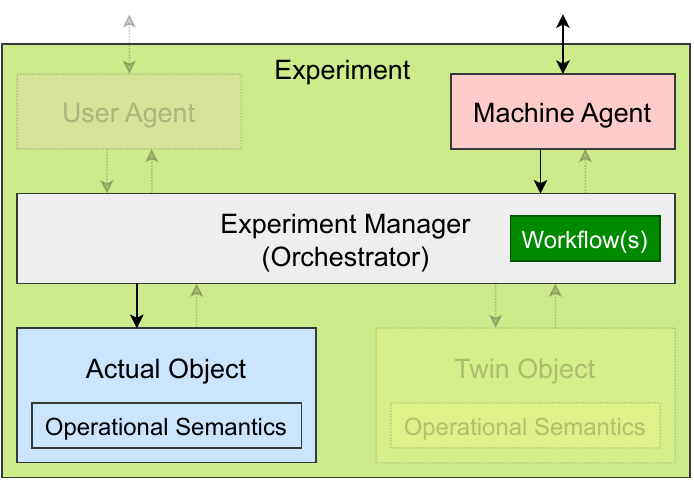}
        \caption{Reference architecture, variant 20, without Twin Object and User Agent.}
        \label{fig:var-19}
    \end{minipage}\hfill
	\begin{minipage}[t]{0.3\textwidth}
		\vspace{0.6cm}
        \centering
        \includegraphics[width=\textwidth]{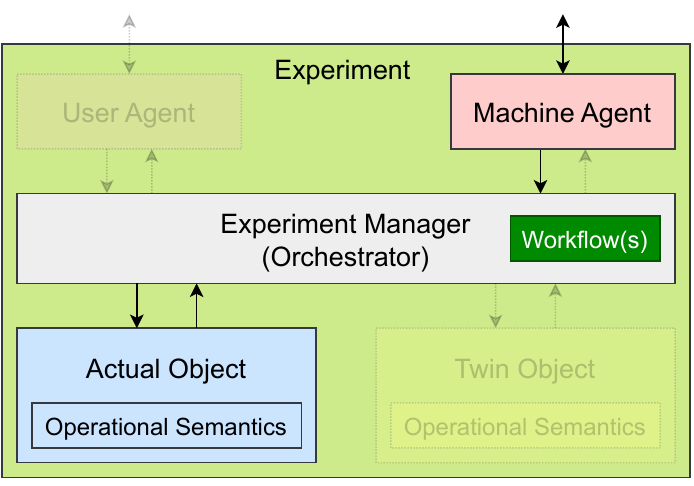}
        \caption{Reference architecture, variant 21, without Twin Object and User Agent.}
        \label{fig:var-20}
    \end{minipage}\\

\end{figure}
\begin{figure}[p]
	\centering
	\begin{minipage}[t]{0.3\textwidth}
		\vspace{0.6cm}
        \centering
        \includegraphics[width=\textwidth]{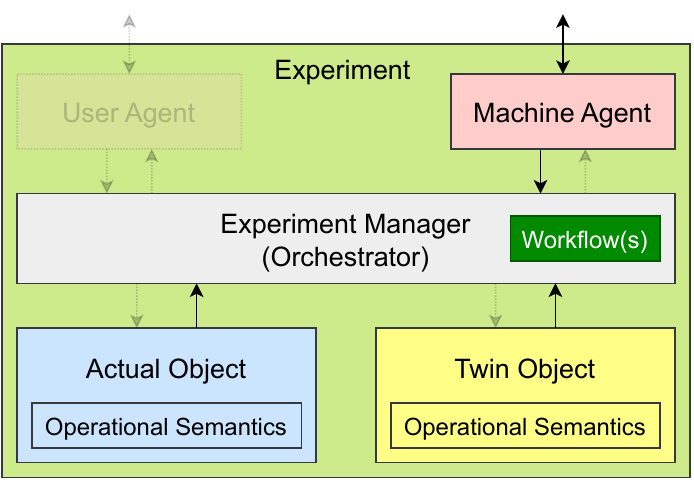}
        \caption{Reference architecture, variant 22, without User Agent.}
        \label{fig:var-21}
    \end{minipage}\hfill
	\begin{minipage}[t]{0.3\textwidth}
		\vspace{0.6cm}
        \centering
        \includegraphics[width=\textwidth]{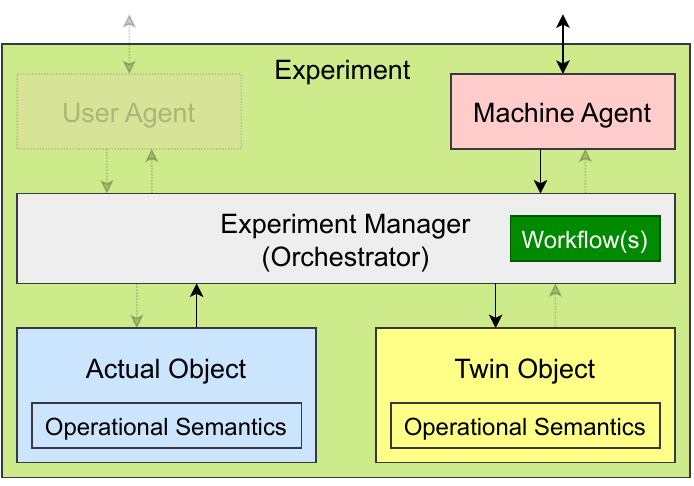}
        \caption{Reference architecture, variant 23, without User Agent.}
        \label{fig:var-22}
    \end{minipage}\hfill
	\begin{minipage}[t]{0.3\textwidth}
		\vspace{0.6cm}
        \centering
        \includegraphics[width=\textwidth]{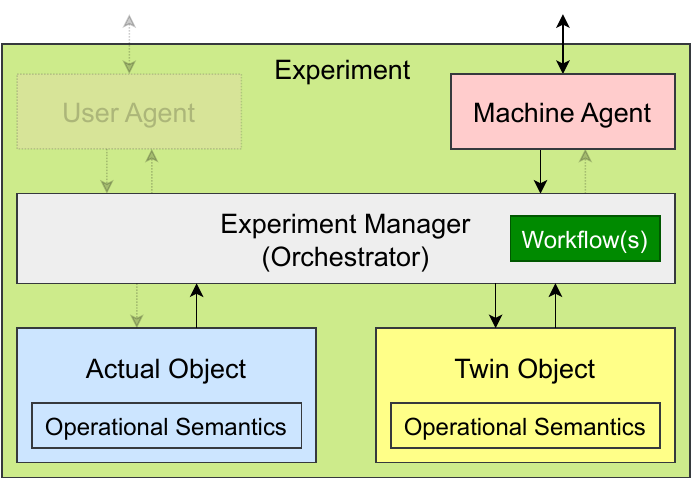}
        \caption{Reference architecture, variant 24, without User Agent.}
        \label{fig:var-23}
    \end{minipage}\\

	\begin{minipage}[t]{0.3\textwidth}
		\vspace{0.6cm}
        \centering
        \includegraphics[width=\textwidth]{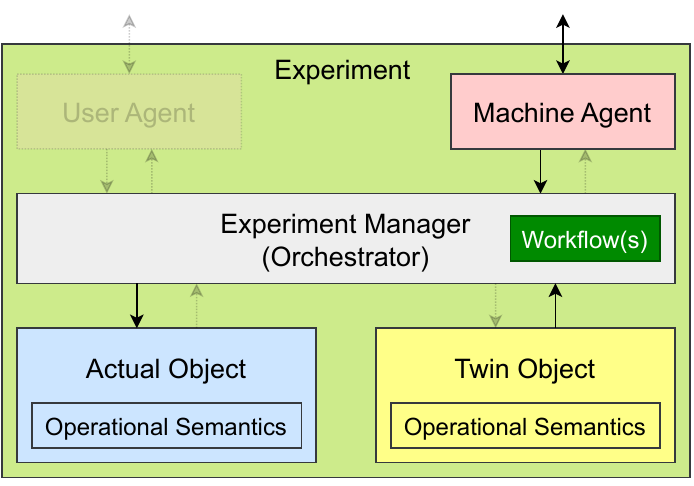}
        \caption{Reference architecture, variant 25, without User Agent.}
        \label{fig:var-24}
    \end{minipage}\hfill
	\begin{minipage}[t]{0.3\textwidth}
		\vspace{0.6cm}
        \centering
        \includegraphics[width=\textwidth]{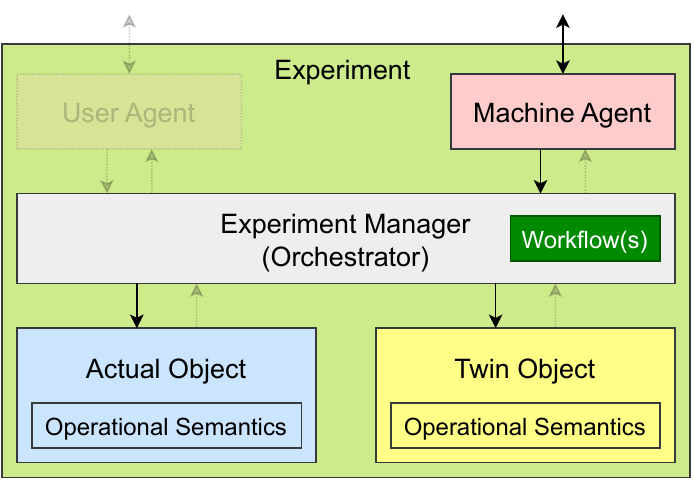}
        \caption{Reference architecture, variant 26, without User Agent.}
        \label{fig:var-25}
    \end{minipage}\hfill
	\begin{minipage}[t]{0.3\textwidth}
		\vspace{0.6cm}
        \centering
        \includegraphics[width=\textwidth]{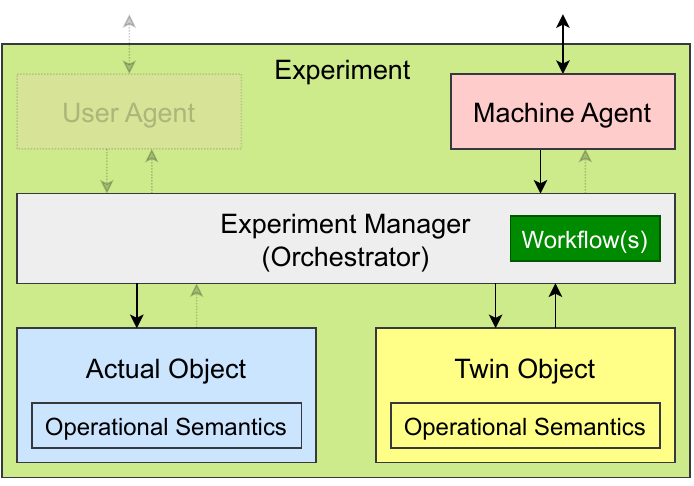}
        \caption{Reference architecture, variant 27, without User Agent.}
        \label{fig:var-26}
    \end{minipage}\\

	\begin{minipage}[t]{0.3\textwidth}
		\vspace{0.6cm}
        \centering
        \includegraphics[width=\textwidth]{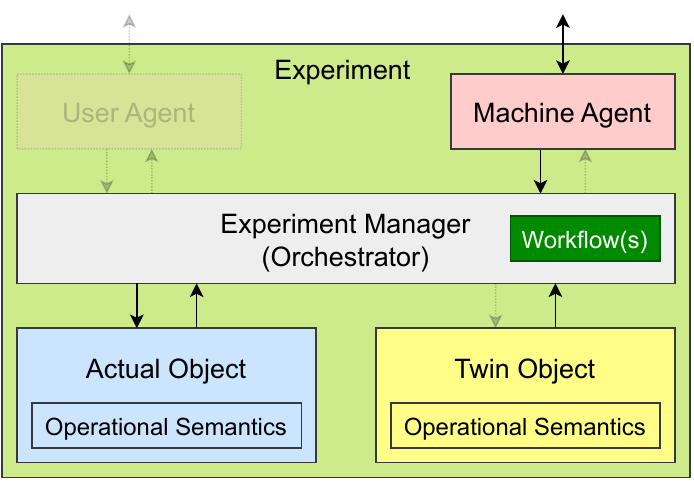}
        \caption{Reference architecture, variant 28, without User Agent.}
        \label{fig:var-27}
    \end{minipage}\hfill
	\begin{minipage}[t]{0.3\textwidth}
		\vspace{0.6cm}
        \centering
        \includegraphics[width=\textwidth]{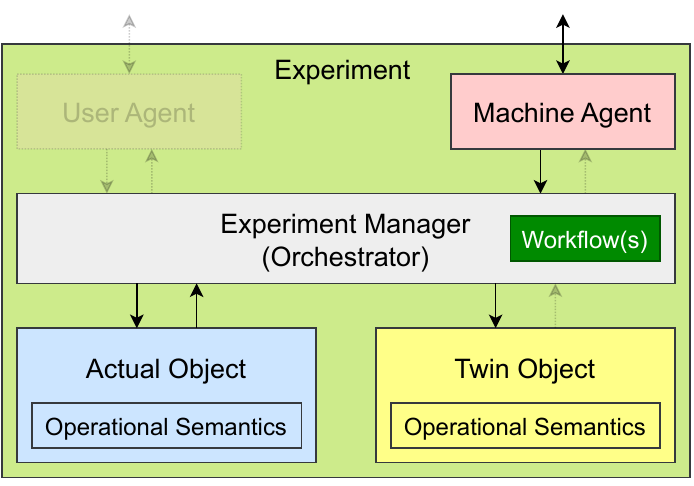}
        \caption{Reference architecture, variant 29, without User Agent.}
        \label{fig:var-28}
    \end{minipage}\hfill
	\begin{minipage}[t]{0.3\textwidth}
		\vspace{0.6cm}
        \centering
        \includegraphics[width=\textwidth]{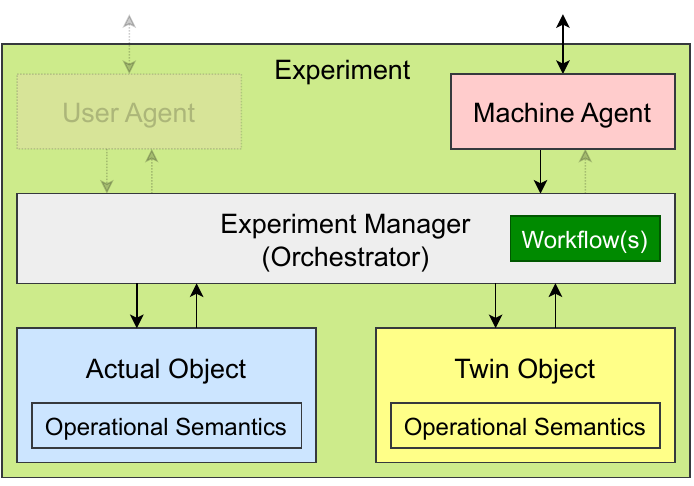}
        \caption{Reference architecture, variant 30, without User Agent.}
        \label{fig:var-29}
    \end{minipage}\\

	\begin{minipage}[t]{0.3\textwidth}
		\vspace{0.6cm}
        \centering
        \includegraphics[width=\textwidth]{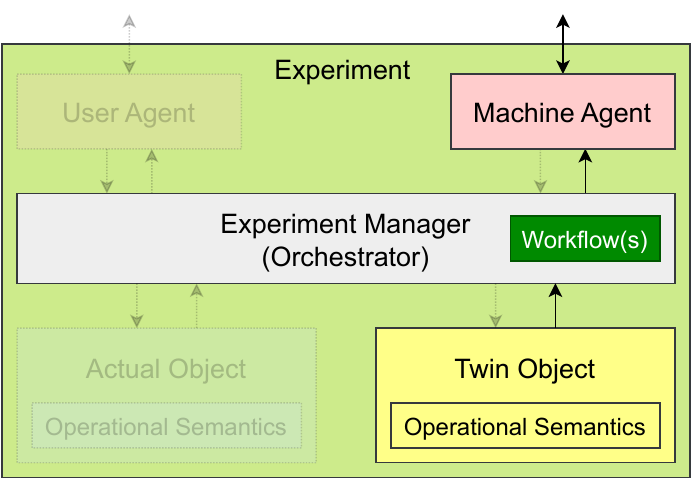}
        \caption{Reference architecture, variant 31, without Actual Object and User Agent.}
        \label{fig:var-30}
    \end{minipage}\hfill
	\begin{minipage}[t]{0.3\textwidth}
		\vspace{0.6cm}
        \centering
        \includegraphics[width=\textwidth]{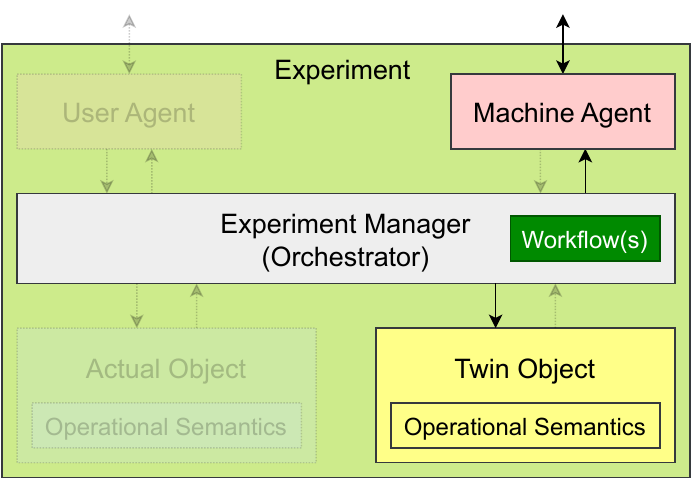}
        \caption{Reference architecture, variant 32, without Actual Object and User Agent.}
        \label{fig:var-31}
    \end{minipage}\hfill
	\begin{minipage}[t]{0.3\textwidth}
		\vspace{0.6cm}
        \centering
        \includegraphics[width=\textwidth]{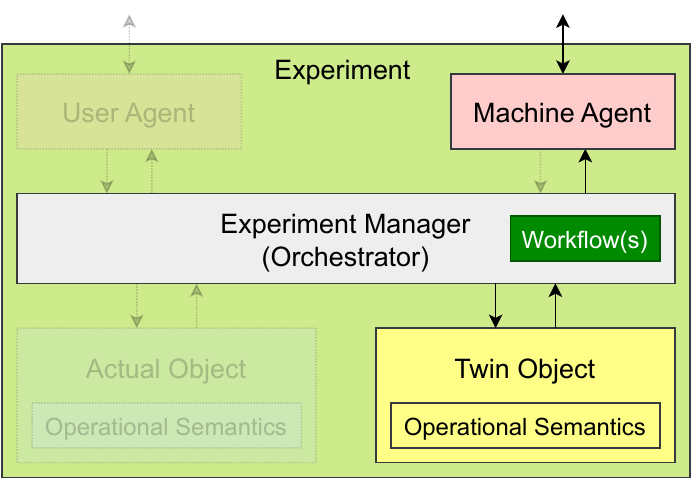}
        \caption{Reference architecture, variant 33, without Actual Object and User Agent.}
        \label{fig:var-32}
    \end{minipage}\\

\end{figure}
\begin{figure}[p]
	\centering
	\begin{minipage}[t]{0.3\textwidth}
		\vspace{0.6cm}
        \centering
        \includegraphics[width=\textwidth]{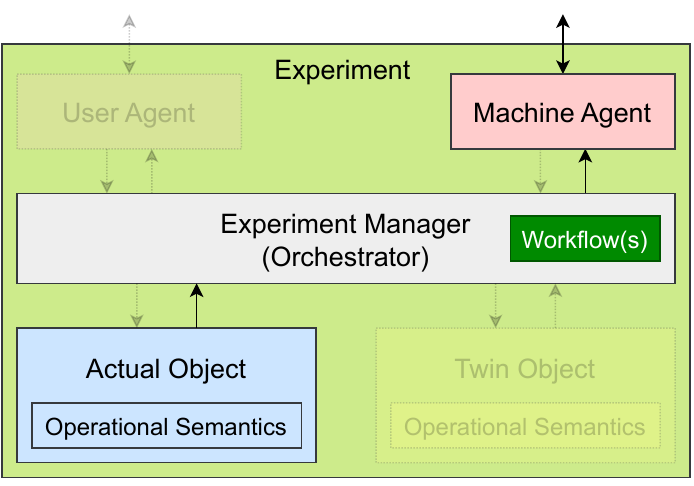}
        \caption{Reference architecture, variant 34, without Twin Object and User Agent.}
        \label{fig:var-33}
    \end{minipage}\hfill
	\begin{minipage}[t]{0.3\textwidth}
		\vspace{0.6cm}
        \centering
        \includegraphics[width=\textwidth]{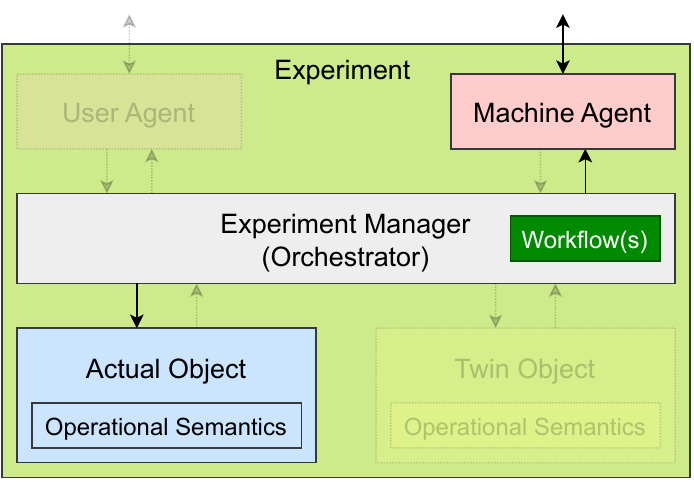}
        \caption{Reference architecture, variant 35, without Twin Object and User Agent.}
        \label{fig:var-34}
    \end{minipage}\hfill
	\begin{minipage}[t]{0.3\textwidth}
		\vspace{0.6cm}
        \centering
        \includegraphics[width=\textwidth]{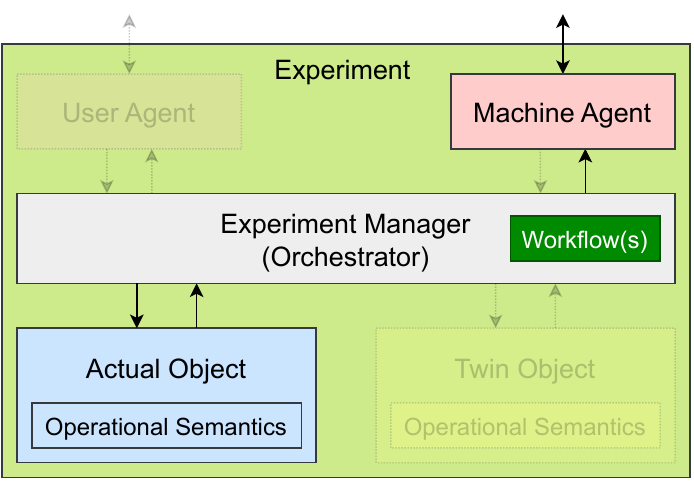}
        \caption{Reference architecture, variant 36, without Twin Object and User Agent.}
        \label{fig:var-35}
    \end{minipage}\\

	\begin{minipage}[t]{0.3\textwidth}
		\vspace{0.6cm}
        \centering
        \includegraphics[width=\textwidth]{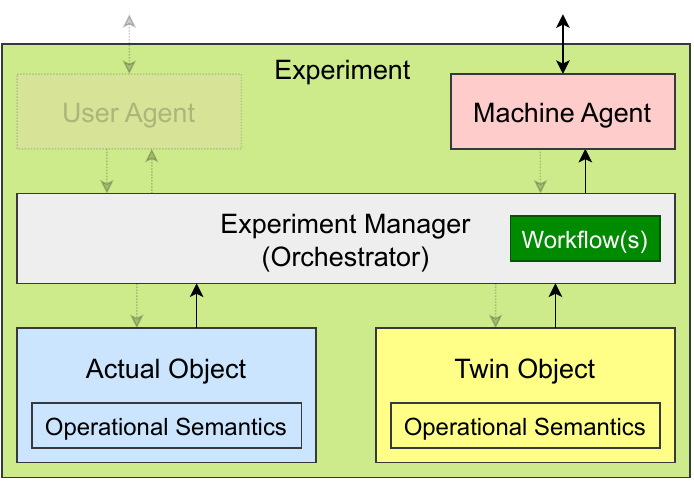}
        \caption{Reference architecture, variant 37, without User Agent.}
        \label{fig:var-36}
    \end{minipage}\hfill
	\begin{minipage}[t]{0.3\textwidth}
		\vspace{0.6cm}
        \centering
        \includegraphics[width=\textwidth]{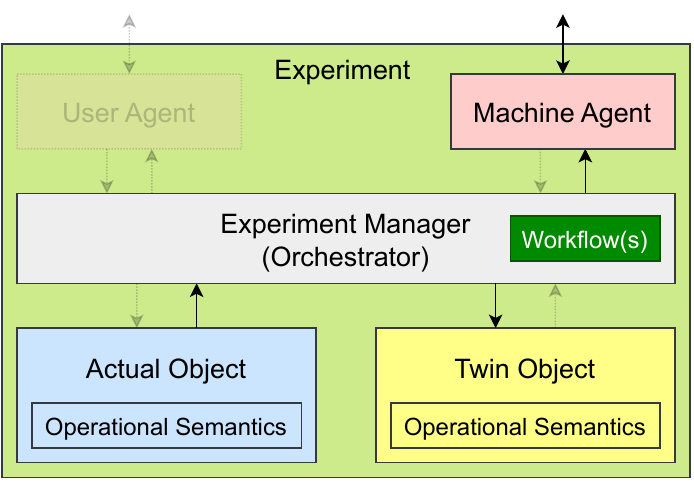}
        \caption{Reference architecture, variant 38, without User Agent.}
        \label{fig:var-37}
    \end{minipage}\hfill
	\begin{minipage}[t]{0.3\textwidth}
		\vspace{0.6cm}
        \centering
        \includegraphics[width=\textwidth]{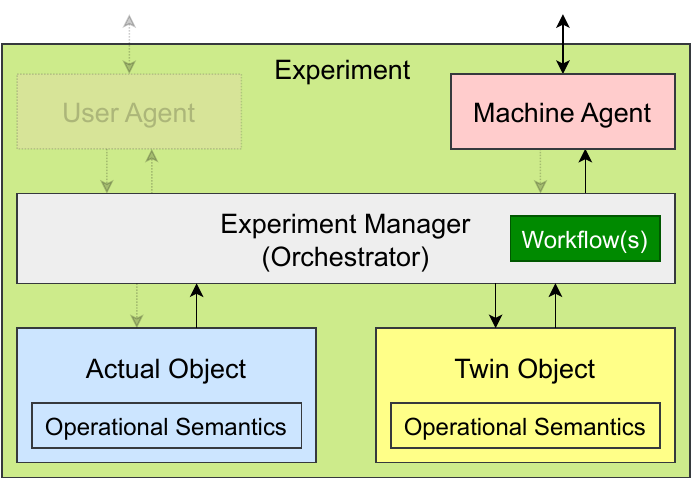}
        \caption{Reference architecture, variant 39, without User Agent.}
        \label{fig:var-38}
    \end{minipage}\\

	\begin{minipage}[t]{0.3\textwidth}
		\vspace{0.6cm}
        \centering
        \includegraphics[width=\textwidth]{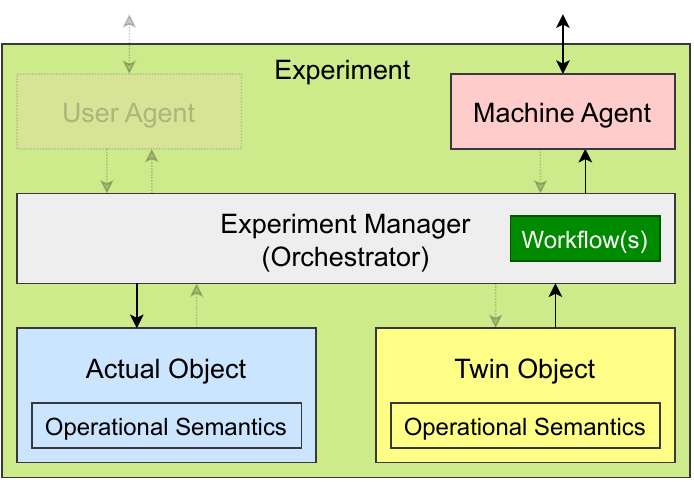}
        \caption{Reference architecture, variant 40, without User Agent.}
        \label{fig:var-39}
    \end{minipage}\hfill
	\begin{minipage}[t]{0.3\textwidth}
		\vspace{0.6cm}
        \centering
        \includegraphics[width=\textwidth]{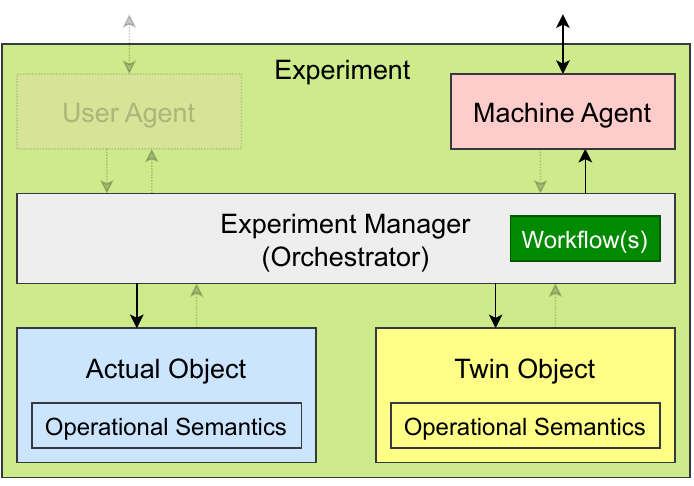}
        \caption{Reference architecture, variant 41, without User Agent.}
        \label{fig:var-40}
    \end{minipage}\hfill
	\begin{minipage}[t]{0.3\textwidth}
		\vspace{0.6cm}
        \centering
        \includegraphics[width=\textwidth]{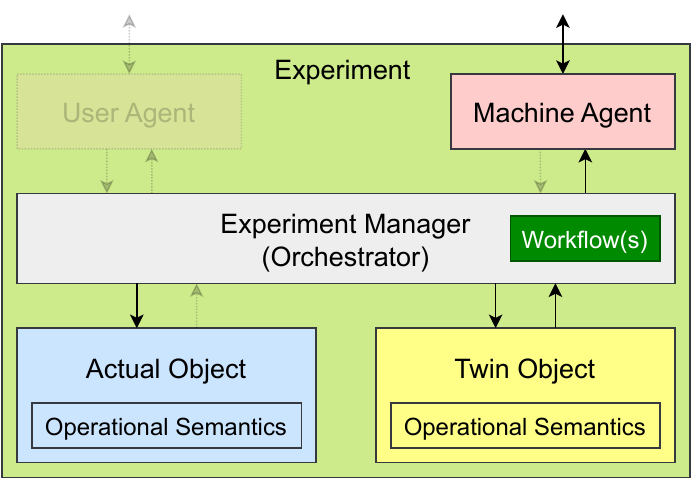}
        \caption{Reference architecture, variant 42, without User Agent.}
        \label{fig:var-41}
    \end{minipage}\\

	\begin{minipage}[t]{0.3\textwidth}
		\vspace{0.6cm}
        \centering
        \includegraphics[width=\textwidth]{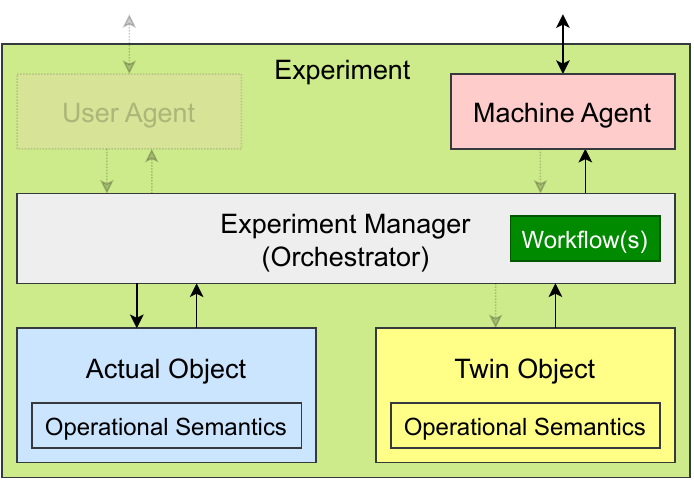}
        \caption{Reference architecture, variant 43, without User Agent.}
        \label{fig:var-42}
    \end{minipage}\hfill
	\begin{minipage}[t]{0.3\textwidth}
		\vspace{0.6cm}
        \centering
        \includegraphics[width=\textwidth]{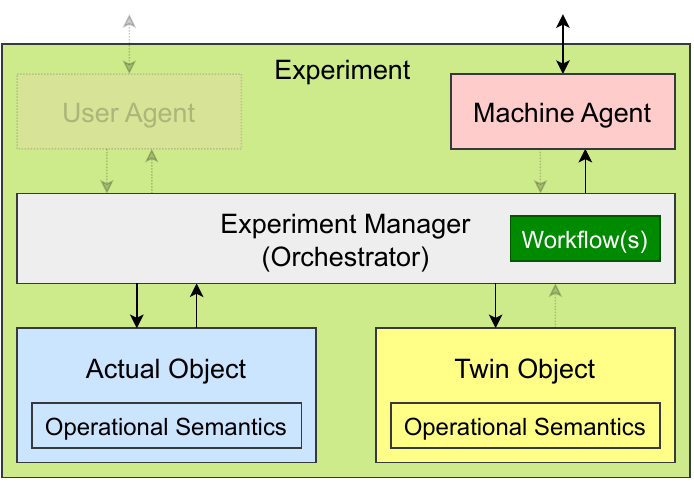}
        \caption{Reference architecture, variant 44, without User Agent.}
        \label{fig:var-43}
    \end{minipage}\hfill
	\begin{minipage}[t]{0.3\textwidth}
		\vspace{0.6cm}
        \centering
        \includegraphics[width=\textwidth]{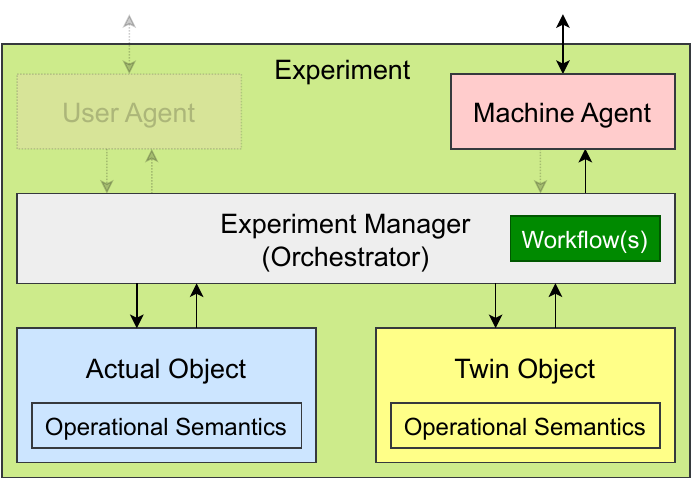}
        \caption{Reference architecture, variant 45, without User Agent.}
        \label{fig:var-44}
    \end{minipage}\\

\end{figure}
\begin{figure}[p]
	\centering
	\begin{minipage}[t]{0.3\textwidth}
		\vspace{0.6cm}
        \centering
        \includegraphics[width=\textwidth]{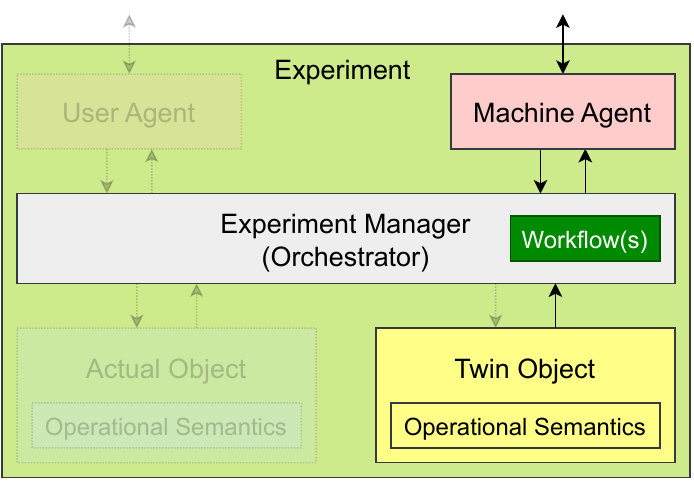}
        \caption{Reference architecture, variant 46, without Actual Object and User Agent.}
        \label{fig:var-45}
    \end{minipage}\hfill
	\begin{minipage}[t]{0.3\textwidth}
		\vspace{0.6cm}
        \centering
        \includegraphics[width=\textwidth]{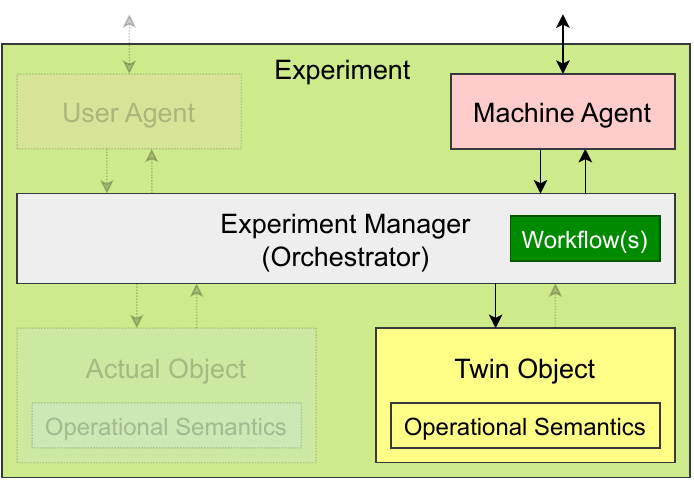}
        \caption{Reference architecture, variant 47, without Actual Object and User Agent.}
        \label{fig:var-46}
    \end{minipage}\hfill
	\begin{minipage}[t]{0.3\textwidth}
		\vspace{0.6cm}
        \centering
        \includegraphics[width=\textwidth]{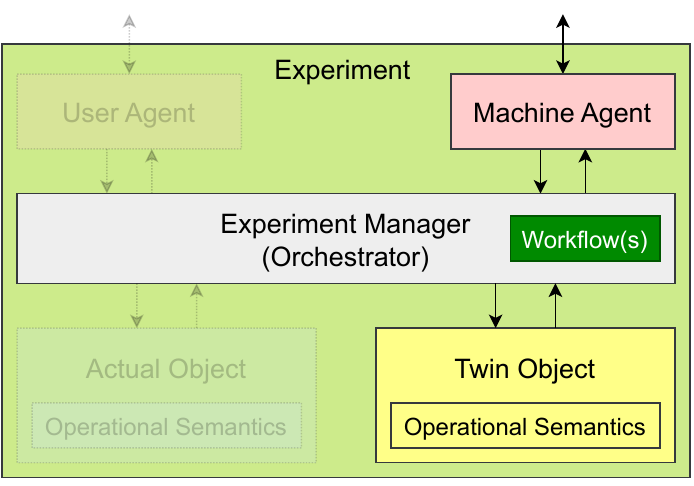}
        \caption{Reference architecture, variant 48, without Actual Object and User Agent.}
        \label{fig:var-47}
    \end{minipage}\\

	\begin{minipage}[t]{0.3\textwidth}
		\vspace{0.6cm}
        \centering
        \includegraphics[width=\textwidth]{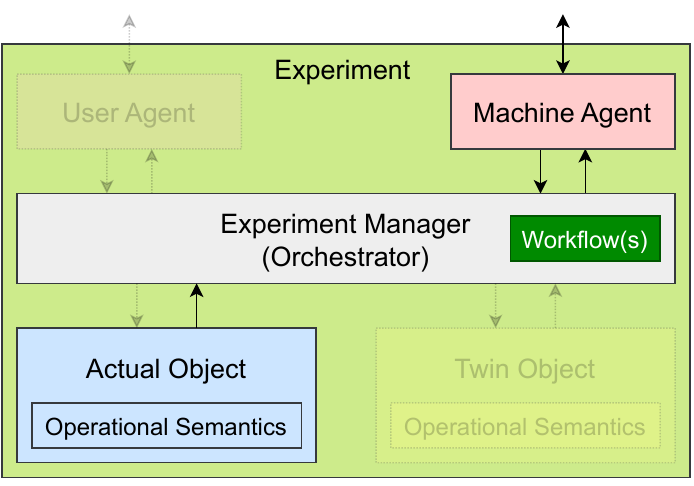}
        \caption{Reference architecture, variant 49, without Twin Object and User Agent.}
        \label{fig:var-48}
    \end{minipage}\hfill
	\begin{minipage}[t]{0.3\textwidth}
		\vspace{0.6cm}
        \centering
        \includegraphics[width=\textwidth]{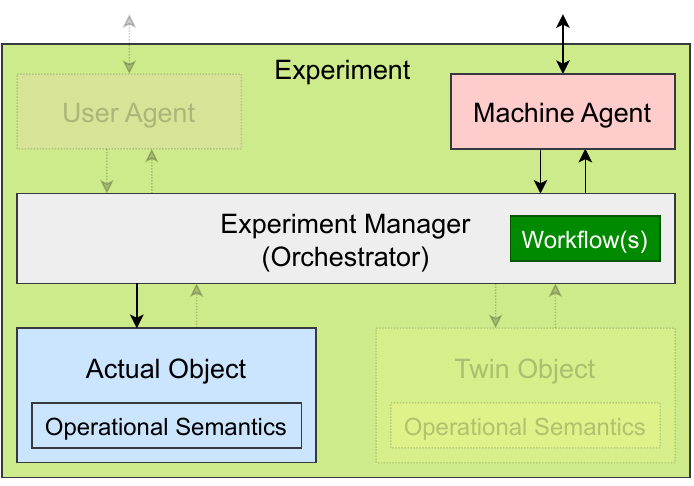}
        \caption{Reference architecture, variant 50, without Twin Object and User Agent.}
        \label{fig:var-49}
    \end{minipage}\hfill
	\begin{minipage}[t]{0.3\textwidth}
		\vspace{0.6cm}
        \centering
        \includegraphics[width=\textwidth]{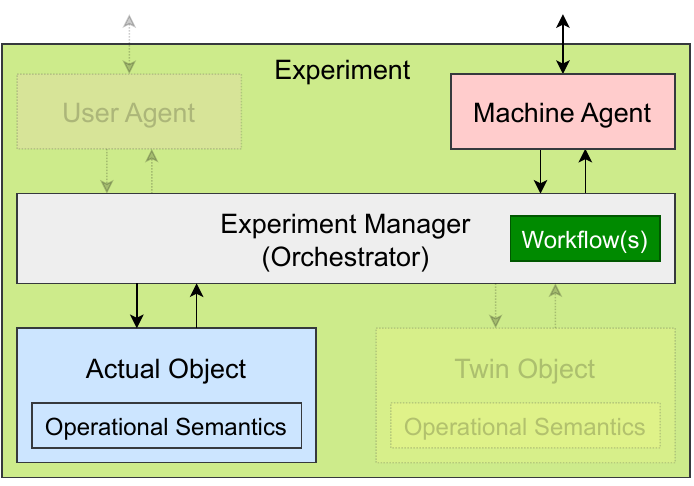}
        \caption{Reference architecture, variant 51, without Twin Object and User Agent.}
        \label{fig:var-50}
    \end{minipage}\\

	\begin{minipage}[t]{0.3\textwidth}
		\vspace{0.6cm}
        \centering
        \includegraphics[width=\textwidth]{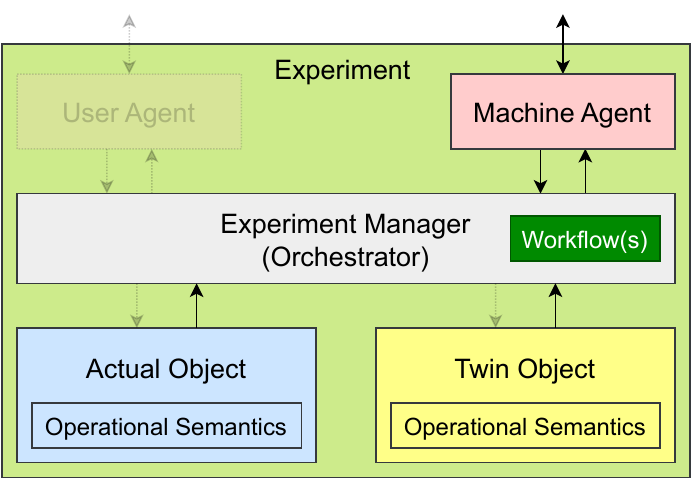}
        \caption{Reference architecture, variant 52, without User Agent.}
        \label{fig:var-51}
    \end{minipage}\hfill
	\begin{minipage}[t]{0.3\textwidth}
		\vspace{0.6cm}
        \centering
        \includegraphics[width=\textwidth]{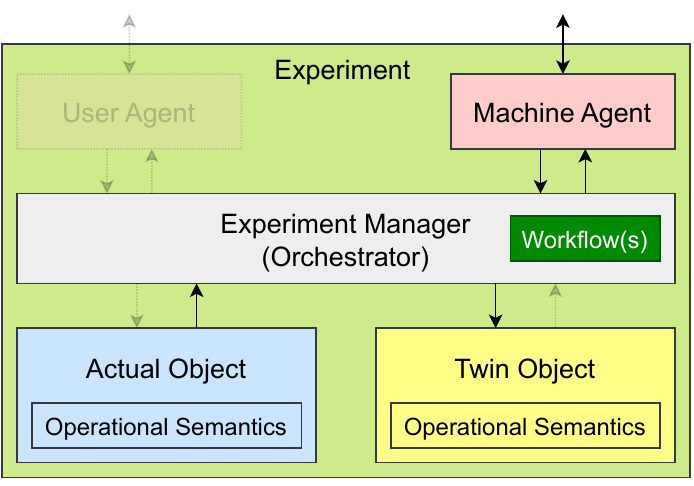}
        \caption{Reference architecture, variant 53, without User Agent.}
        \label{fig:var-52}
    \end{minipage}\hfill
	\begin{minipage}[t]{0.3\textwidth}
		\vspace{0.6cm}
        \centering
        \includegraphics[width=\textwidth]{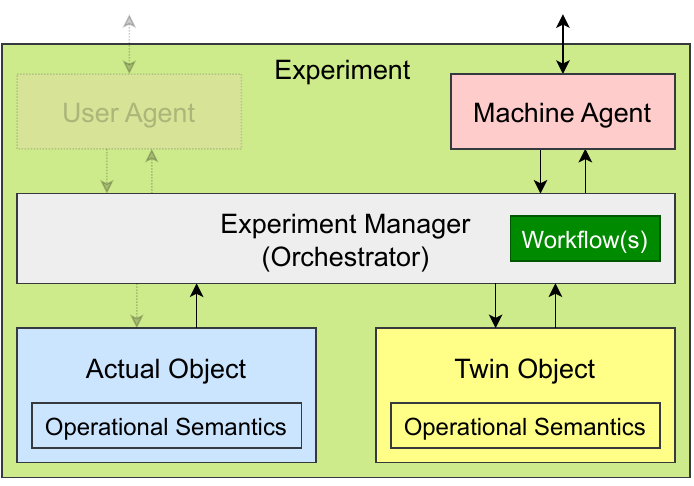}
        \caption{Reference architecture, variant 54, without User Agent.}
        \label{fig:var-53}
    \end{minipage}\\

	\begin{minipage}[t]{0.3\textwidth}
		\vspace{0.6cm}
        \centering
        \includegraphics[width=\textwidth]{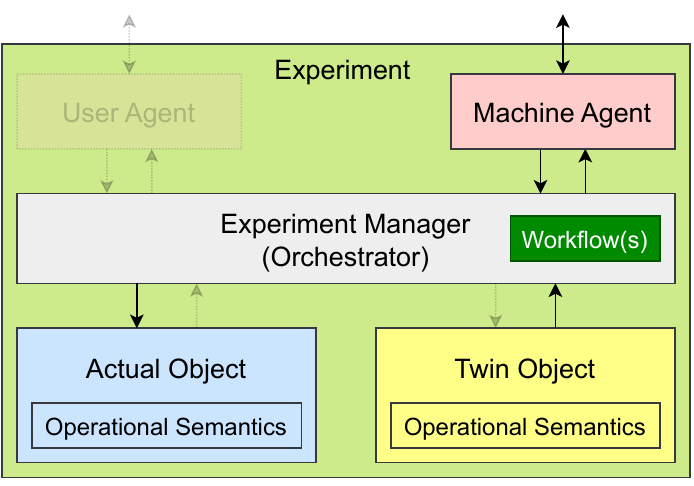}
        \caption{Reference architecture, variant 55, without User Agent.}
        \label{fig:var-54}
    \end{minipage}\hfill
	\begin{minipage}[t]{0.3\textwidth}
		\vspace{0.6cm}
        \centering
        \includegraphics[width=\textwidth]{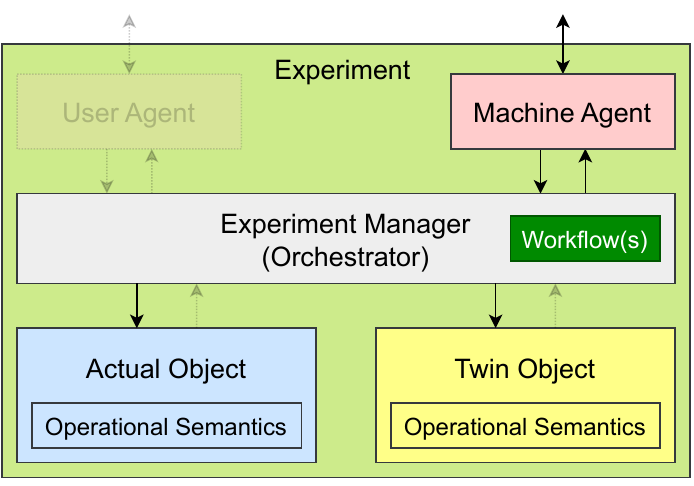}
        \caption{Reference architecture, variant 56, without User Agent.}
        \label{fig:var-55}
    \end{minipage}\hfill
	\begin{minipage}[t]{0.3\textwidth}
		\vspace{0.6cm}
        \centering
        \includegraphics[width=\textwidth]{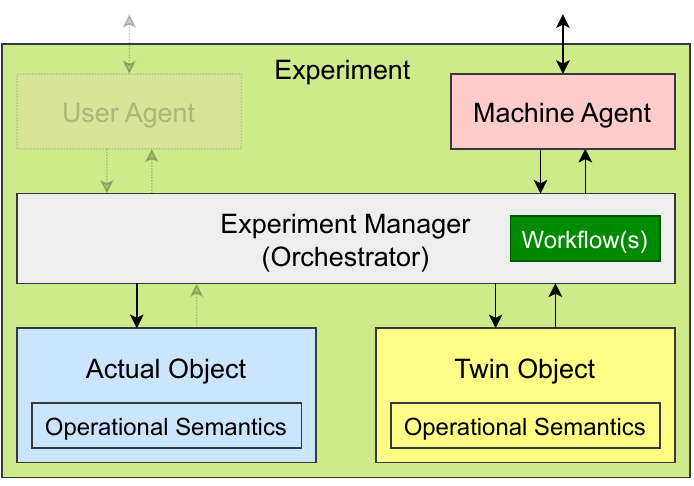}
        \caption{Reference architecture, variant 57, without User Agent.}
        \label{fig:var-56}
    \end{minipage}\\

\end{figure}
\begin{figure}[p]
	\centering
	\begin{minipage}[t]{0.3\textwidth}
		\vspace{0.6cm}
        \centering
        \includegraphics[width=\textwidth]{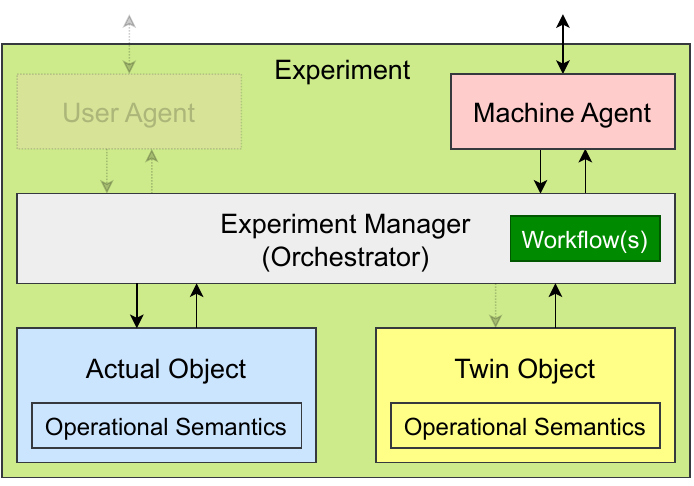}
        \caption{Reference architecture, variant 58, without User Agent.}
        \label{fig:var-57}
    \end{minipage}\hfill
	\begin{minipage}[t]{0.3\textwidth}
		\vspace{0.6cm}
        \centering
        \includegraphics[width=\textwidth]{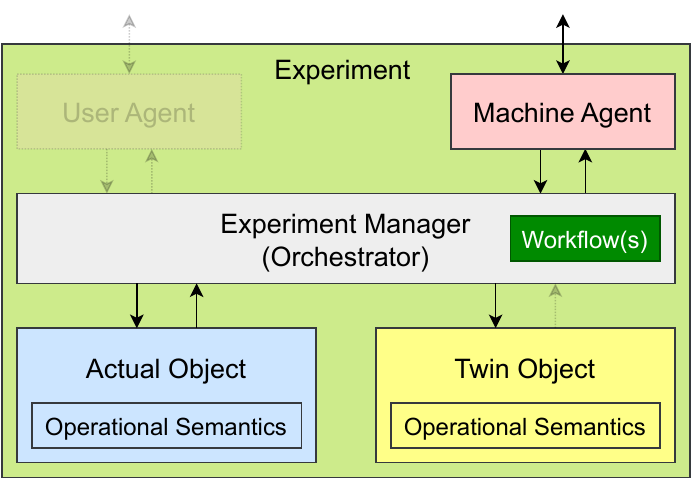}
        \caption{Reference architecture, variant 59, without User Agent.}
        \label{fig:var-58}
    \end{minipage}\hfill
	\begin{minipage}[t]{0.3\textwidth}
		\vspace{0.6cm}
        \centering
        \includegraphics[width=\textwidth]{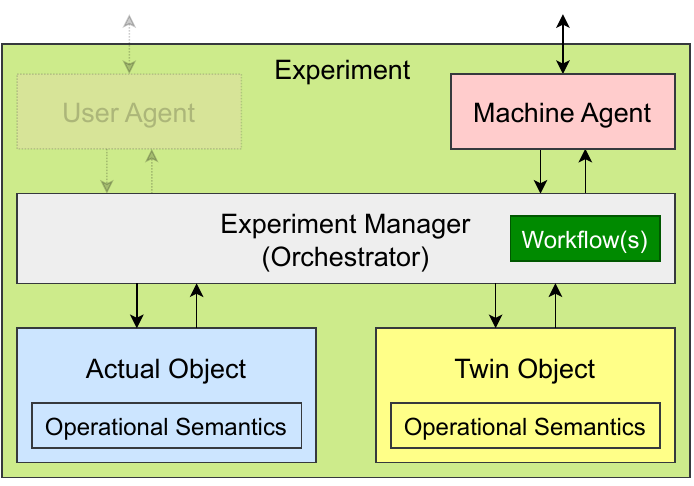}
        \caption{Reference architecture, variant 60, without User Agent.}
        \label{fig:var-59}
    \end{minipage}\\

	\begin{minipage}[t]{0.3\textwidth}
		\vspace{0.6cm}
        \centering
        \includegraphics[width=\textwidth]{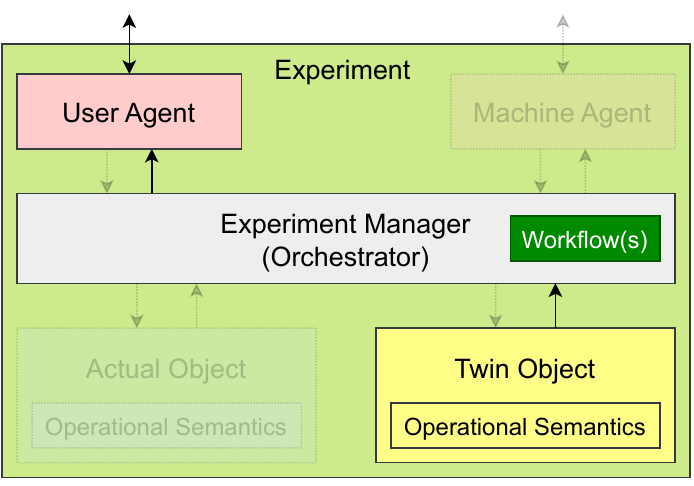}
        \caption{Reference architecture, variant 61, without Actual Object and Machine Agent.}
        \label{fig:var-60}
    \end{minipage}\hfill
	\begin{minipage}[t]{0.3\textwidth}
		\vspace{0.6cm}
        \centering
        \includegraphics[width=\textwidth]{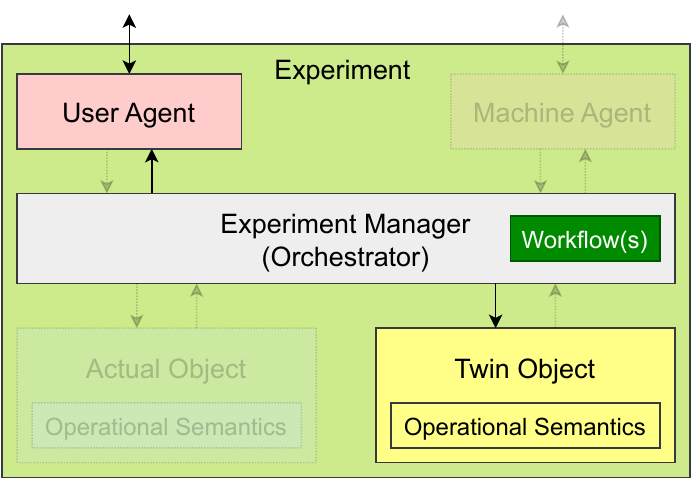}
        \caption{Reference architecture, variant 62, without Actual Object and Machine Agent.}
        \label{fig:var-61}
    \end{minipage}\hfill
	\begin{minipage}[t]{0.3\textwidth}
		\vspace{0.6cm}
        \centering
        \includegraphics[width=\textwidth]{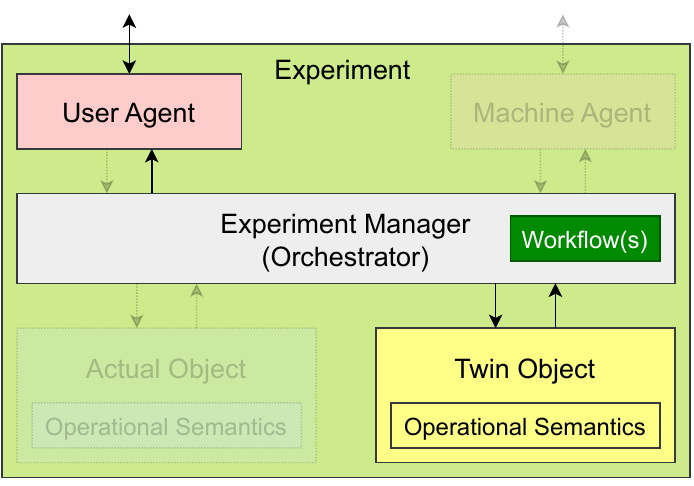}
        \caption{Reference architecture, variant 63, without Actual Object and Machine Agent.}
        \label{fig:var-62}
    \end{minipage}\\

	\begin{minipage}[t]{0.3\textwidth}
		\vspace{0.6cm}
        \centering
        \includegraphics[width=\textwidth]{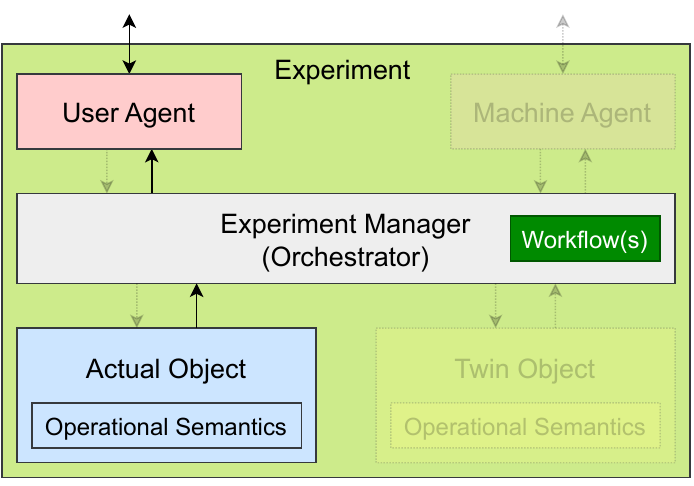}
        \caption{Reference architecture, variant 64, without Twin Object and Machine Agent.}
        \label{fig:var-63}
    \end{minipage}\hfill
	\begin{minipage}[t]{0.3\textwidth}
		\vspace{0.6cm}
        \centering
        \includegraphics[width=\textwidth]{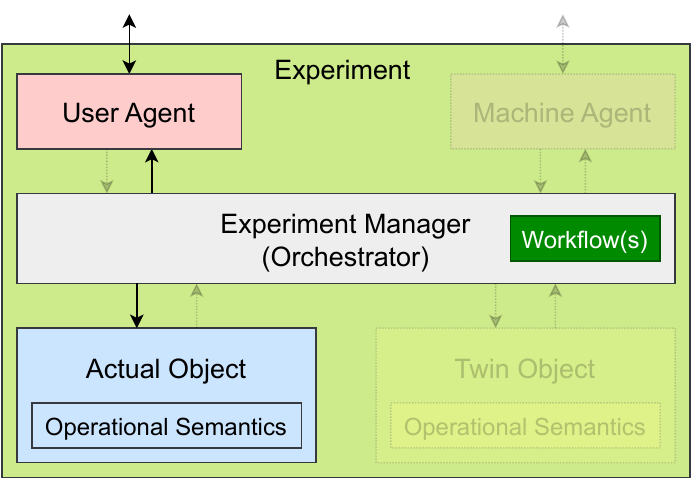}
        \caption{Reference architecture, variant 65, without Twin Object and Machine Agent.}
        \label{fig:var-64}
    \end{minipage}\hfill
	\begin{minipage}[t]{0.3\textwidth}
		\vspace{0.6cm}
        \centering
        \includegraphics[width=\textwidth]{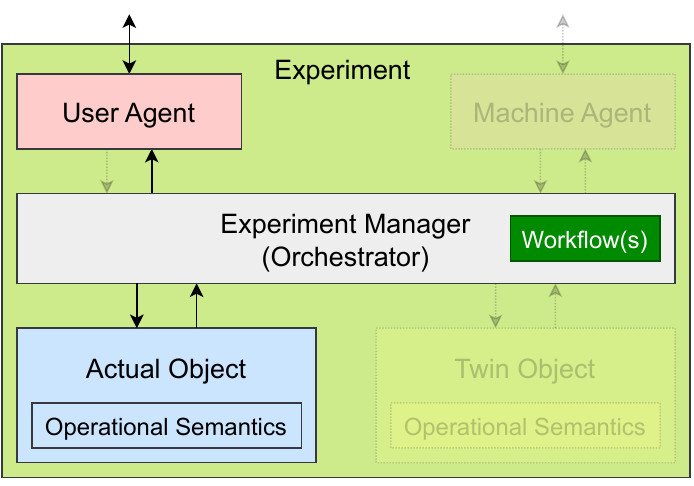}
        \caption{Reference architecture, variant 66, without Twin Object and Machine Agent.}
        \label{fig:var-65}
    \end{minipage}\\

	\begin{minipage}[t]{0.3\textwidth}
		\vspace{0.6cm}
        \centering
        \includegraphics[width=\textwidth]{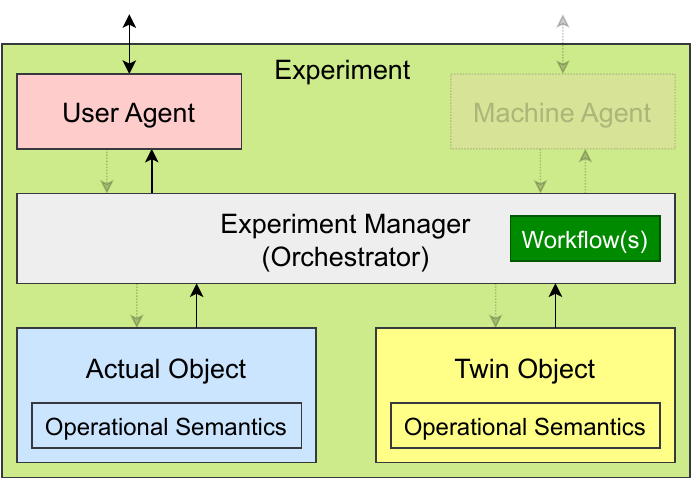}
        \caption{Reference architecture, variant 67, without Machine Agent.}
        \label{fig:var-66}
    \end{minipage}\hfill
	\begin{minipage}[t]{0.3\textwidth}
		\vspace{0.6cm}
        \centering
        \includegraphics[width=\textwidth]{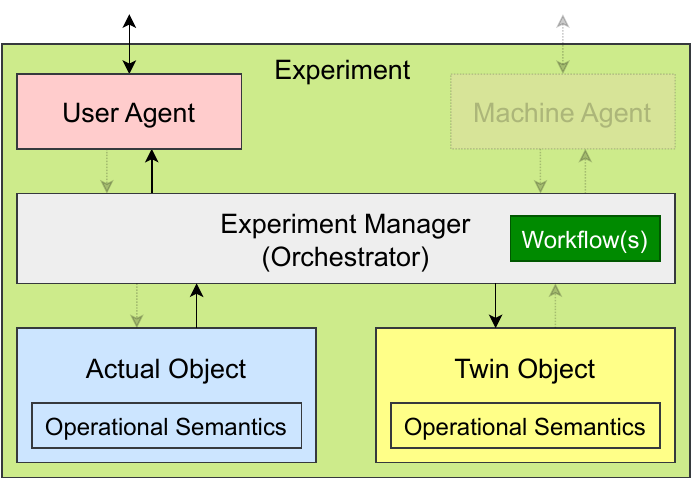}
        \caption{Reference architecture, variant 68, without Machine Agent.}
        \label{fig:var-67}
    \end{minipage}\hfill
	\begin{minipage}[t]{0.3\textwidth}
		\vspace{0.6cm}
        \centering
        \includegraphics[width=\textwidth]{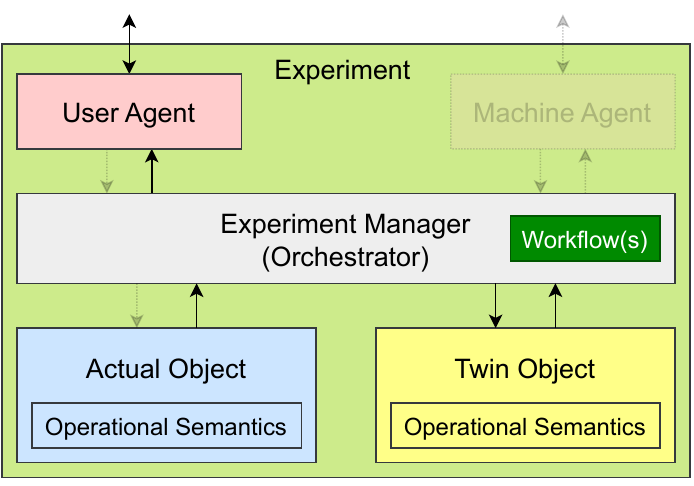}
        \caption{Reference architecture, variant 69, without Machine Agent.}
        \label{fig:var-68}
    \end{minipage}\\

\end{figure}
\begin{figure}[p]
	\centering
	\begin{minipage}[t]{0.3\textwidth}
		\vspace{0.6cm}
        \centering
        \includegraphics[width=\textwidth]{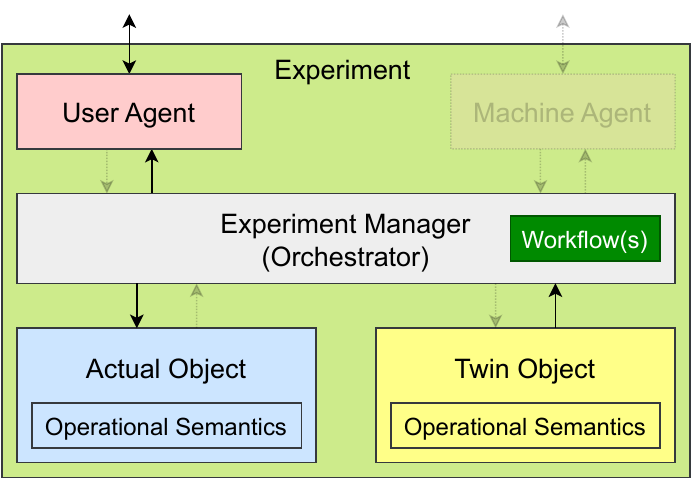}
        \caption{Reference architecture, variant 70, without Machine Agent.}
        \label{fig:var-69}
    \end{minipage}\hfill
	\begin{minipage}[t]{0.3\textwidth}
		\vspace{0.6cm}
        \centering
        \includegraphics[width=\textwidth]{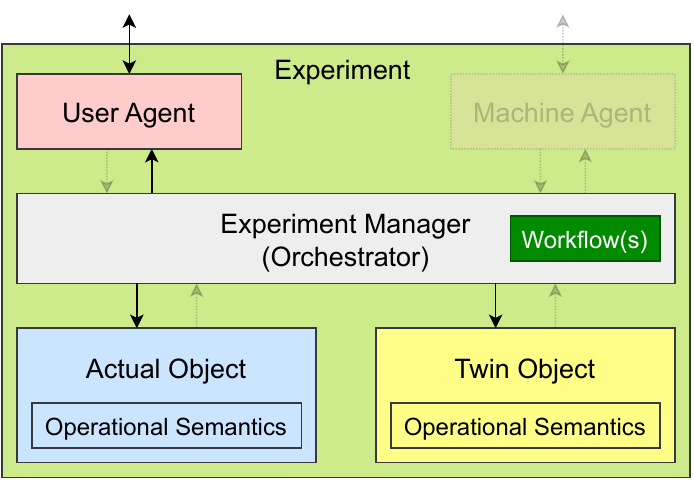}
        \caption{Reference architecture, variant 71, without Machine Agent.}
        \label{fig:var-70}
    \end{minipage}\hfill
	\begin{minipage}[t]{0.3\textwidth}
		\vspace{0.6cm}
        \centering
        \includegraphics[width=\textwidth]{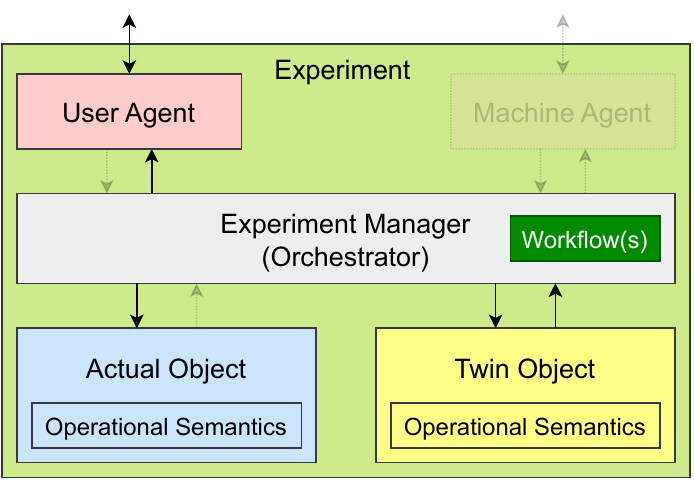}
        \caption{Reference architecture, variant 72, without Machine Agent.}
        \label{fig:var-71}
    \end{minipage}\\

	\begin{minipage}[t]{0.3\textwidth}
		\vspace{0.6cm}
        \centering
        \includegraphics[width=\textwidth]{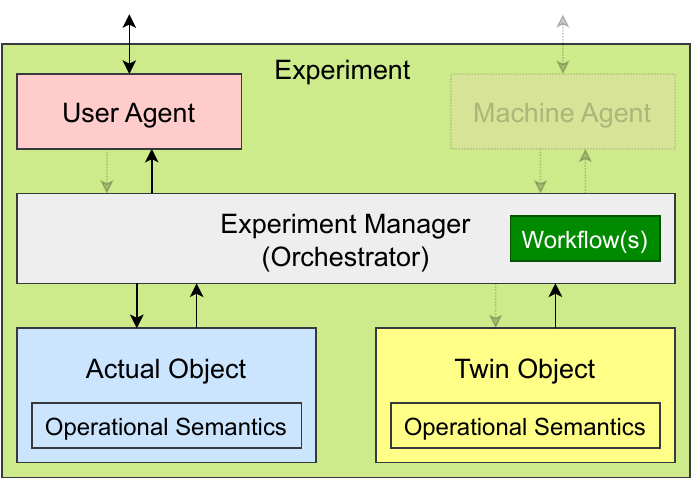}
        \caption{Reference architecture, variant 73, without Machine Agent.}
        \label{fig:var-72}
    \end{minipage}\hfill
	\begin{minipage}[t]{0.3\textwidth}
		\vspace{0.6cm}
        \centering
        \includegraphics[width=\textwidth]{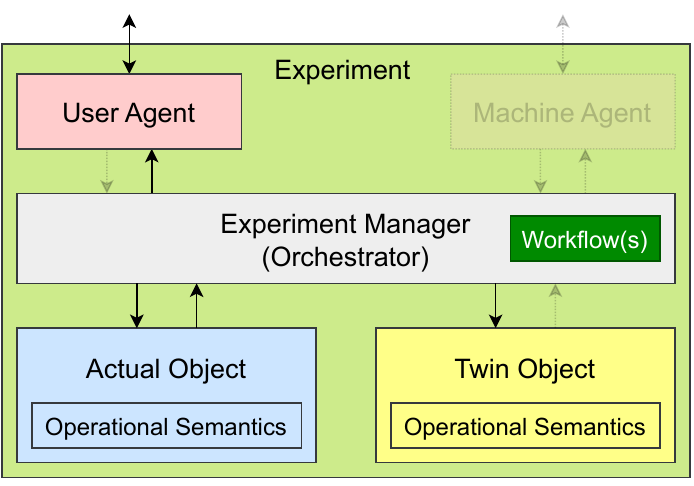}
        \caption{Reference architecture, variant 74, without Machine Agent.}
        \label{fig:var-73}
    \end{minipage}\hfill
	\begin{minipage}[t]{0.3\textwidth}
		\vspace{0.6cm}
        \centering
        \includegraphics[width=\textwidth]{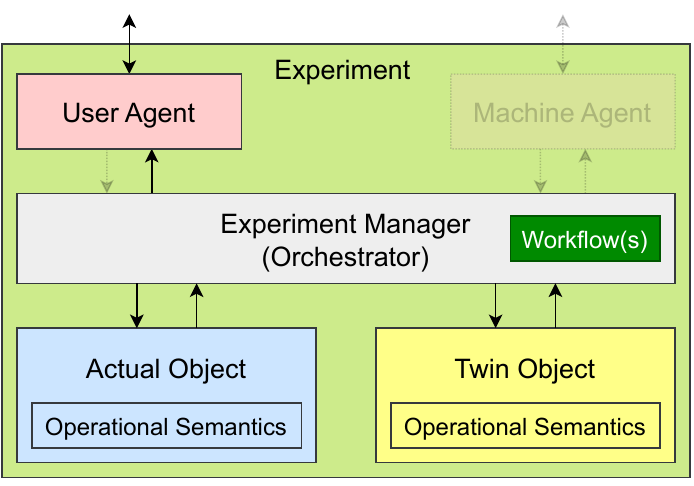}
        \caption{Reference architecture, variant 75, without Machine Agent.}
        \label{fig:var-74}
    \end{minipage}\\

	\begin{minipage}[t]{0.3\textwidth}
		\vspace{0.6cm}
        \centering
        \includegraphics[width=\textwidth]{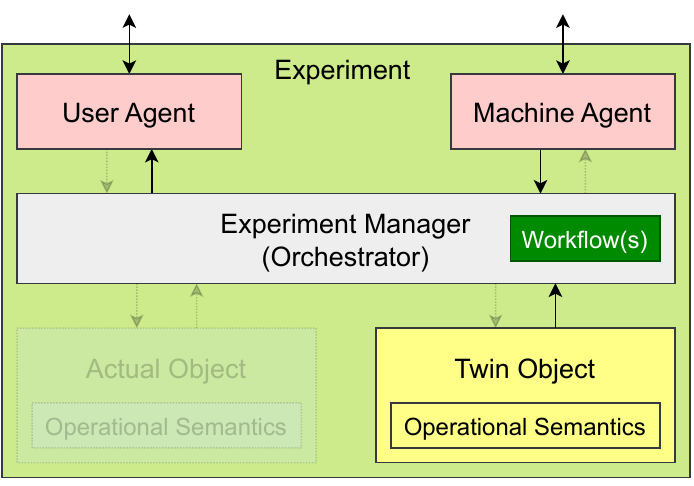}
        \caption{Reference architecture, variant 76, without Actual Object.}
        \label{fig:var-75}
    \end{minipage}\hfill
	\begin{minipage}[t]{0.3\textwidth}
		\vspace{0.6cm}
        \centering
        \includegraphics[width=\textwidth]{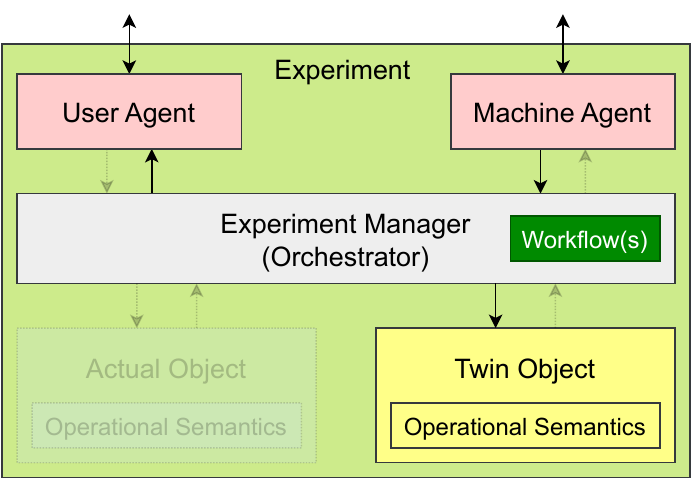}
        \caption{Reference architecture, variant 77, without Actual Object.}
        \label{fig:var-76}
    \end{minipage}\hfill
	\begin{minipage}[t]{0.3\textwidth}
		\vspace{0.6cm}
        \centering
        \includegraphics[width=\textwidth]{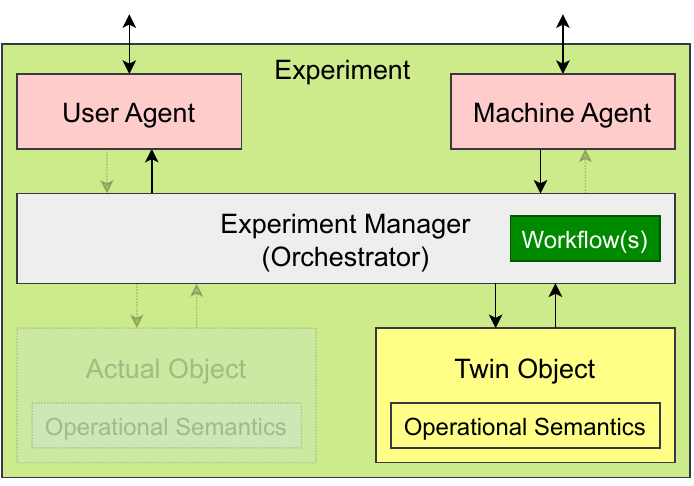}
        \caption{Reference architecture, variant 78, without Actual Object.}
        \label{fig:var-77}
    \end{minipage}\\

	\begin{minipage}[t]{0.3\textwidth}
		\vspace{0.6cm}
        \centering
        \includegraphics[width=\textwidth]{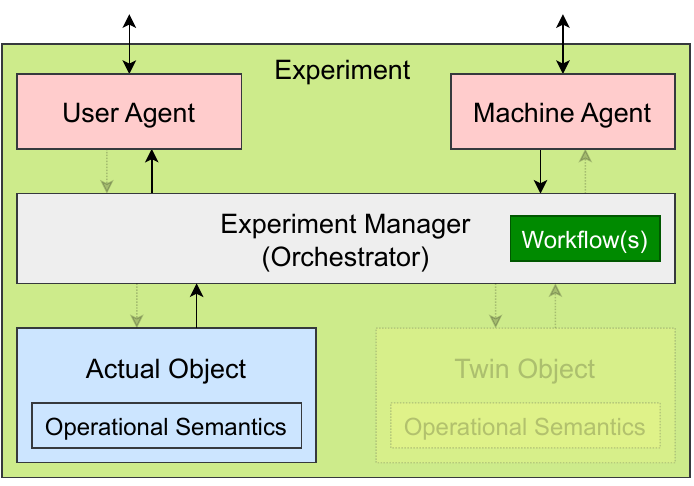}
        \caption{Reference architecture, variant 79, without Twin Object.}
        \label{fig:var-78}
    \end{minipage}\hfill
	\begin{minipage}[t]{0.3\textwidth}
		\vspace{0.6cm}
        \centering
        \includegraphics[width=\textwidth]{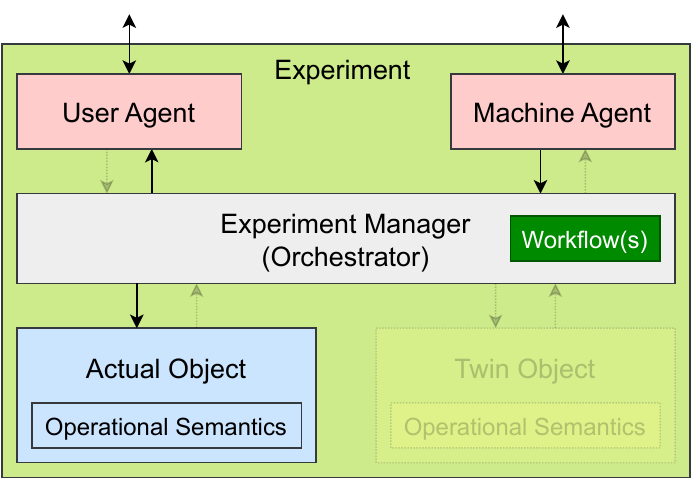}
        \caption{Reference architecture, variant 80, without Twin Object.}
        \label{fig:var-79}
    \end{minipage}\hfill
	\begin{minipage}[t]{0.3\textwidth}
		\vspace{0.6cm}
        \centering
        \includegraphics[width=\textwidth]{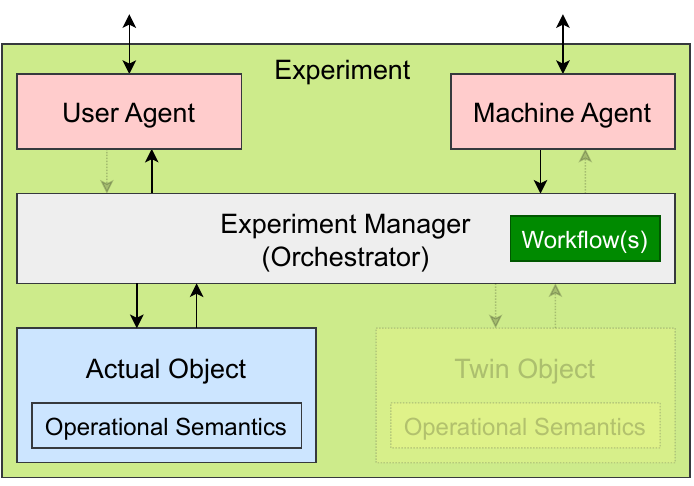}
        \caption{Reference architecture, variant 81, without Twin Object.}
        \label{fig:var-80}
    \end{minipage}\\

\end{figure}
\begin{figure}[p]
	\centering
	\begin{minipage}[t]{0.3\textwidth}
		\vspace{0.6cm}
        \centering
        \includegraphics[width=\textwidth]{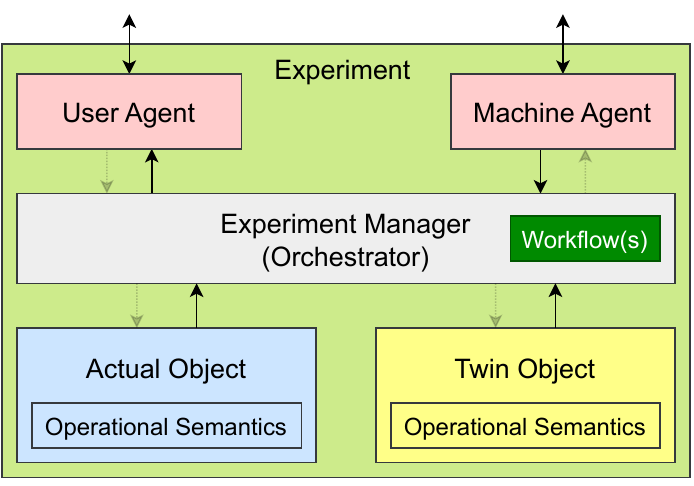}
        \caption{Reference architecture, variant 82.}
        \label{fig:var-81}
    \end{minipage}\hfill
	\begin{minipage}[t]{0.3\textwidth}
		\vspace{0.6cm}
        \centering
        \includegraphics[width=\textwidth]{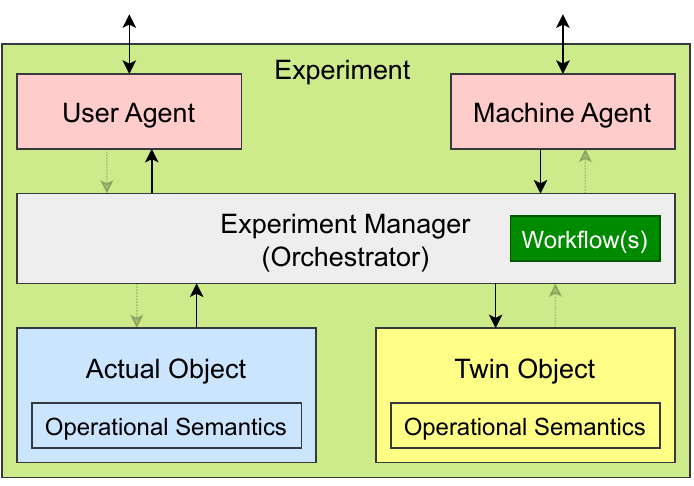}
        \caption{Reference architecture, variant 83.}
        \label{fig:var-82}
    \end{minipage}\hfill
	\begin{minipage}[t]{0.3\textwidth}
		\vspace{0.6cm}
        \centering
        \includegraphics[width=\textwidth]{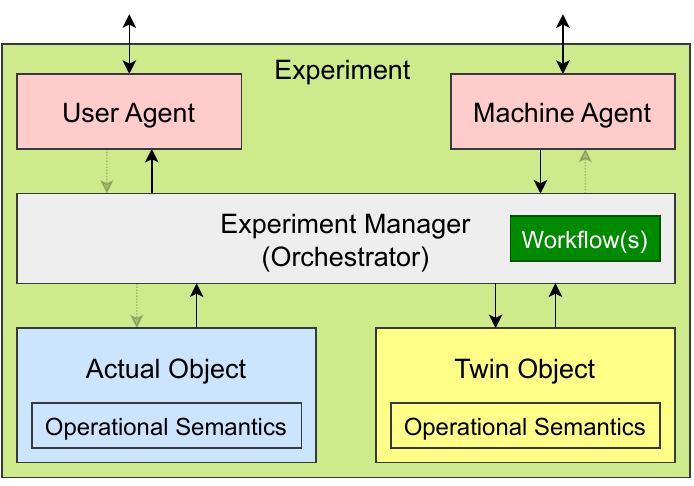}
        \caption{Reference architecture, variant 84.}
        \label{fig:var-83}
    \end{minipage}\\

	\begin{minipage}[t]{0.3\textwidth}
		\vspace{0.6cm}
        \centering
        \includegraphics[width=\textwidth]{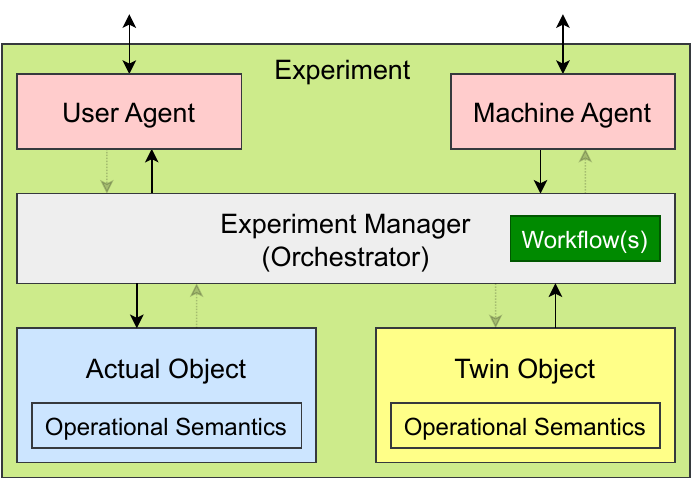}
        \caption{Reference architecture, variant 85.}
        \label{fig:var-84}
    \end{minipage}\hfill
	\begin{minipage}[t]{0.3\textwidth}
		\vspace{0.6cm}
        \centering
        \includegraphics[width=\textwidth]{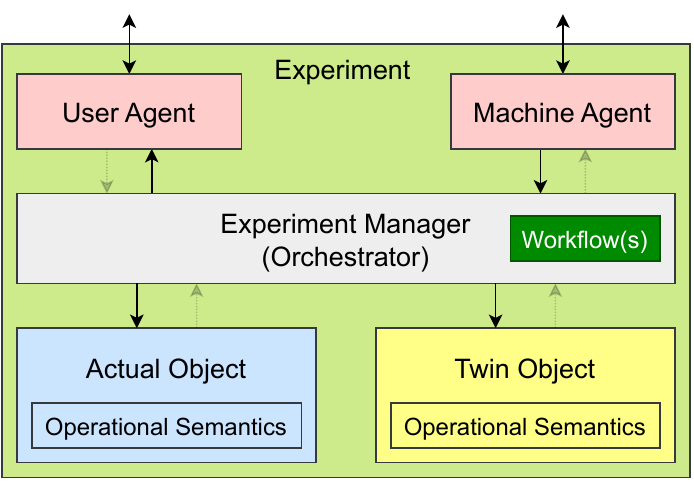}
        \caption{Reference architecture, variant 86.}
        \label{fig:var-85}
    \end{minipage}\hfill
	\begin{minipage}[t]{0.3\textwidth}
		\vspace{0.6cm}
        \centering
        \includegraphics[width=\textwidth]{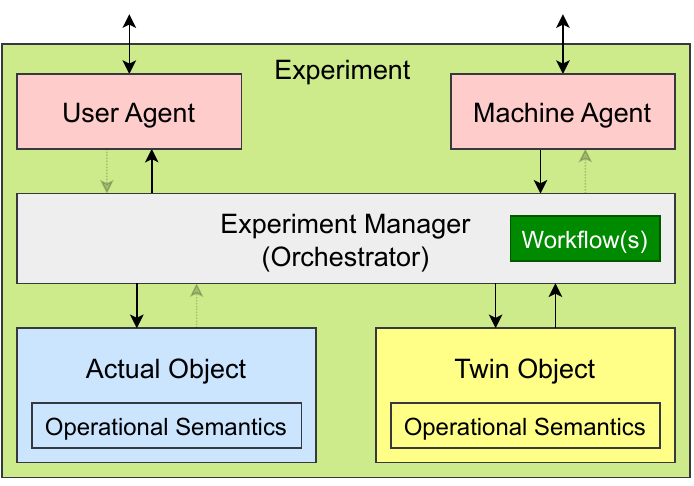}
        \caption{Reference architecture, variant 87.}
        \label{fig:var-86}
    \end{minipage}\\

	\begin{minipage}[t]{0.3\textwidth}
		\vspace{0.6cm}
        \centering
        \includegraphics[width=\textwidth]{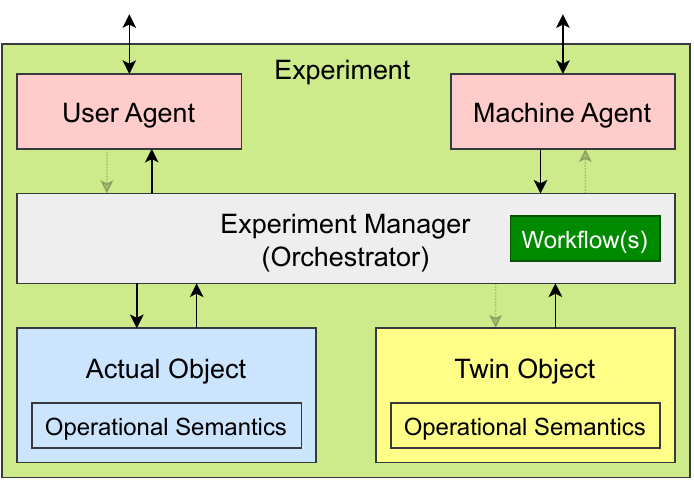}
        \caption{Reference architecture, variant 88.}
        \label{fig:var-87}
    \end{minipage}\hfill
	\begin{minipage}[t]{0.3\textwidth}
		\vspace{0.6cm}
        \centering
        \includegraphics[width=\textwidth]{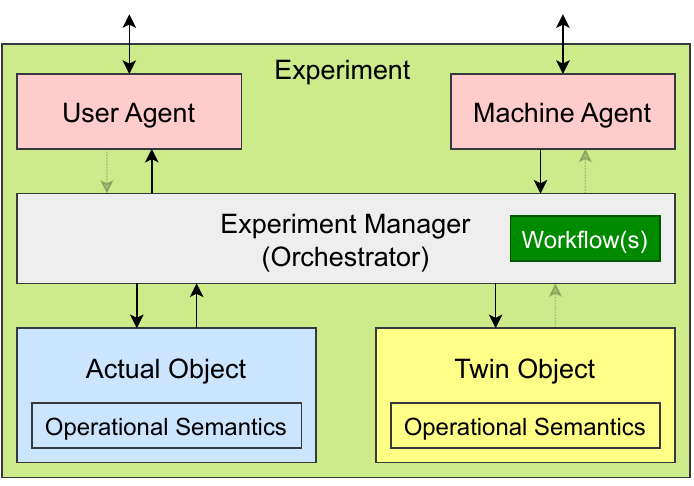}
        \caption{Reference architecture, variant 89.}
        \label{fig:var-88}
    \end{minipage}\hfill
	\begin{minipage}[t]{0.3\textwidth}
		\vspace{0.6cm}
        \centering
        \includegraphics[width=\textwidth]{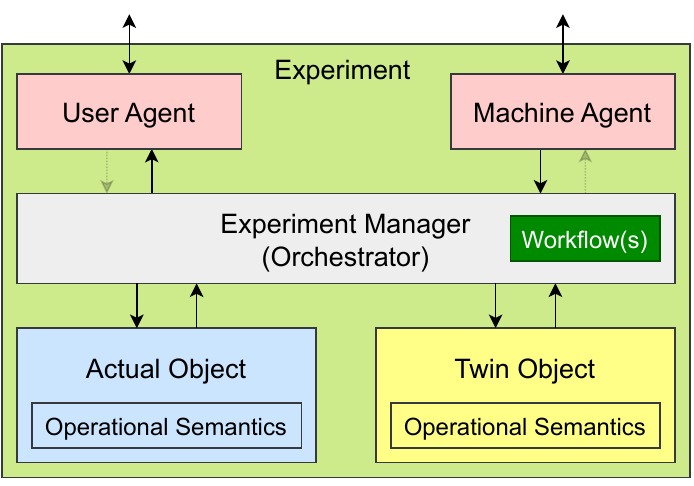}
        \caption{Reference architecture, variant 90.}
        \label{fig:var-89}
    \end{minipage}\\

	\begin{minipage}[t]{0.3\textwidth}
		\vspace{0.6cm}
        \centering
        \includegraphics[width=\textwidth]{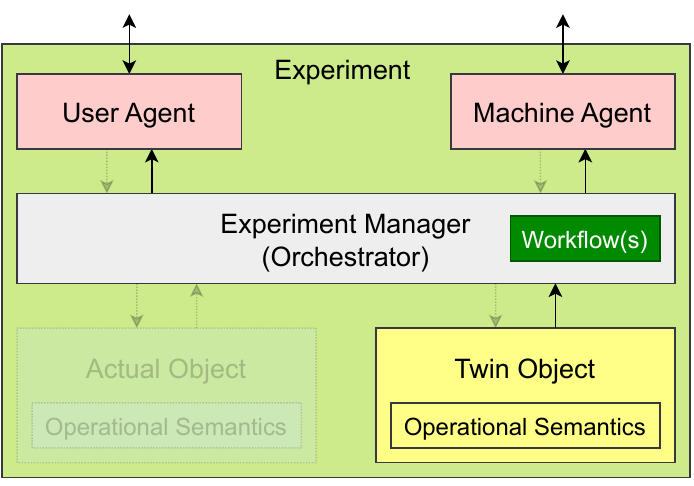}
        \caption{Reference architecture, variant 91, without Actual Object.}
        \label{fig:var-90}
    \end{minipage}\hfill
	\begin{minipage}[t]{0.3\textwidth}
		\vspace{0.6cm}
        \centering
        \includegraphics[width=\textwidth]{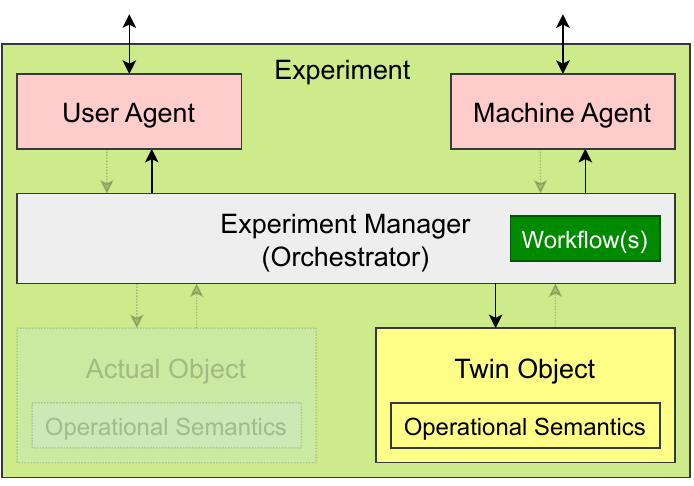}
        \caption{Reference architecture, variant 92, without Actual Object.}
        \label{fig:var-91}
    \end{minipage}\hfill
	\begin{minipage}[t]{0.3\textwidth}
		\vspace{0.6cm}
        \centering
        \includegraphics[width=\textwidth]{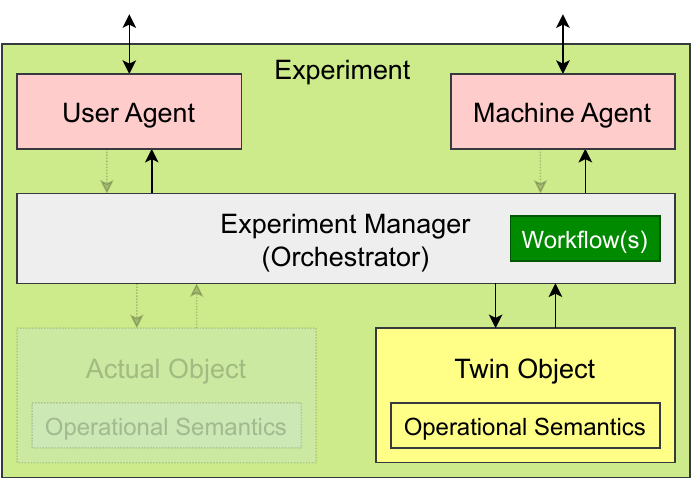}
        \caption{Reference architecture, variant 93, without Actual Object.}
        \label{fig:var-92}
    \end{minipage}\\

\end{figure}
\begin{figure}[p]
	\centering
	\begin{minipage}[t]{0.3\textwidth}
		\vspace{0.6cm}
        \centering
        \includegraphics[width=\textwidth]{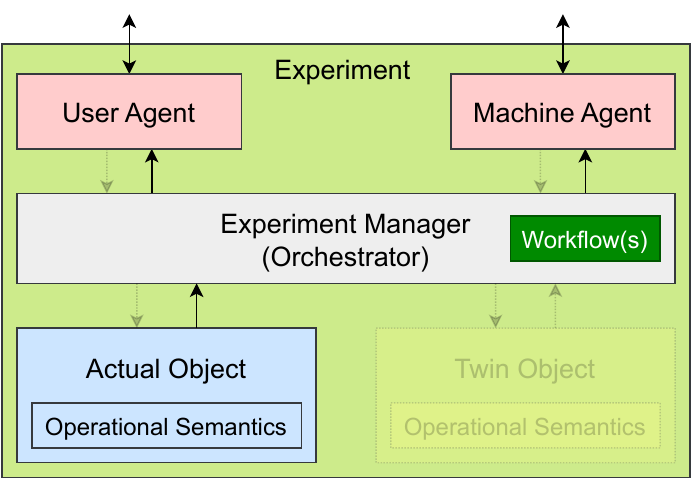}
        \caption{Reference architecture, variant 94, without Twin Object.}
        \label{fig:var-93}
    \end{minipage}\hfill
	\begin{minipage}[t]{0.3\textwidth}
		\vspace{0.6cm}
        \centering
        \includegraphics[width=\textwidth]{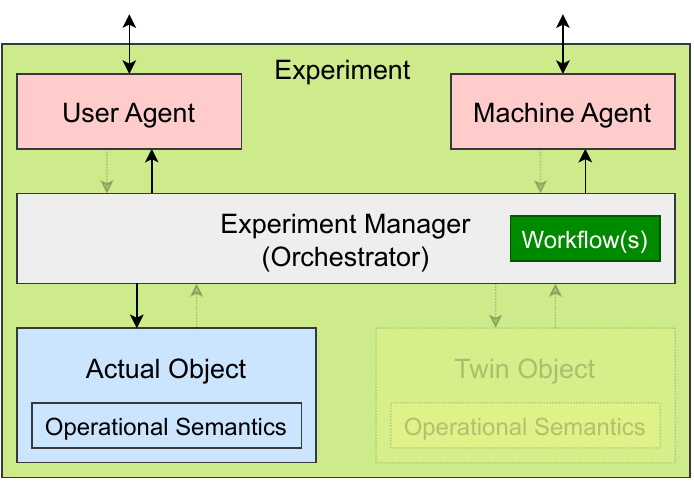}
        \caption{Reference architecture, variant 95, without Twin Object.}
        \label{fig:var-94}
    \end{minipage}\hfill
	\begin{minipage}[t]{0.3\textwidth}
		\vspace{0.6cm}
        \centering
        \includegraphics[width=\textwidth]{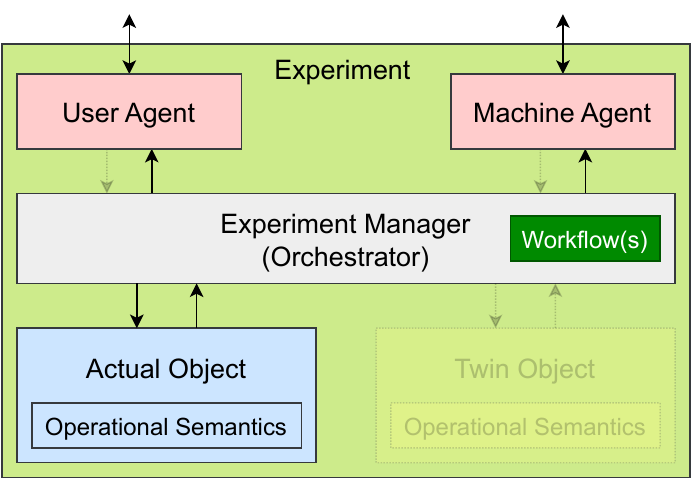}
        \caption{Reference architecture, variant 96, without Twin Object.}
        \label{fig:var-95}
    \end{minipage}\\

	\begin{minipage}[t]{0.3\textwidth}
		\vspace{0.6cm}
        \centering
        \includegraphics[width=\textwidth]{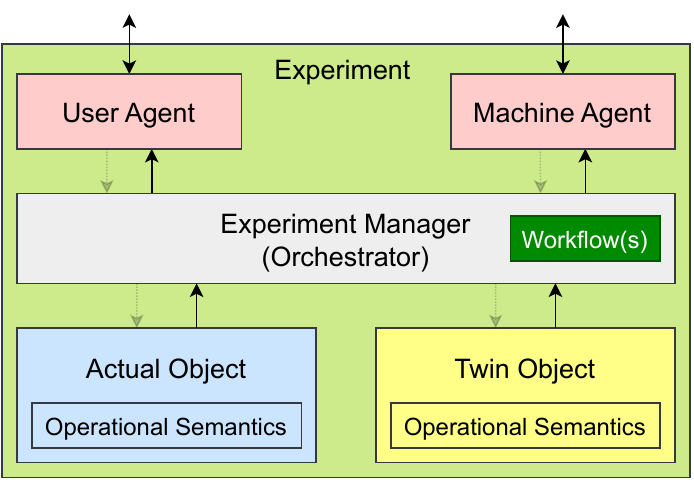}
        \caption{Reference architecture, variant 97.}
        \label{fig:var-96}
    \end{minipage}\hfill
	\begin{minipage}[t]{0.3\textwidth}
		\vspace{0.6cm}
        \centering
        \includegraphics[width=\textwidth]{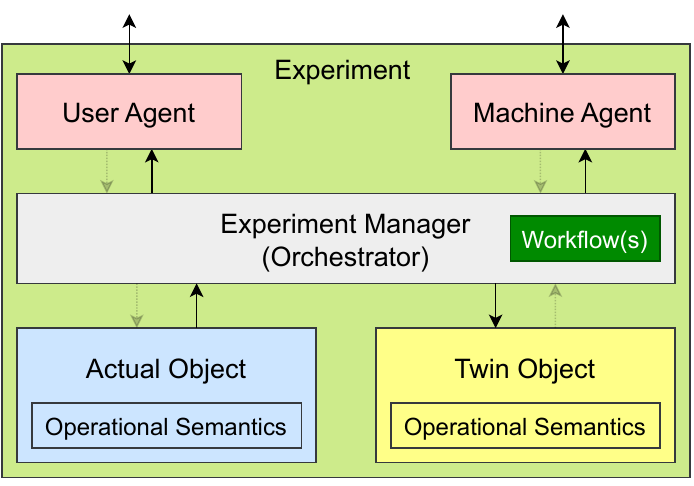}
        \caption{Reference architecture, variant 98.}
        \label{fig:var-97}
    \end{minipage}\hfill
	\begin{minipage}[t]{0.3\textwidth}
		\vspace{0.6cm}
        \centering
        \includegraphics[width=\textwidth]{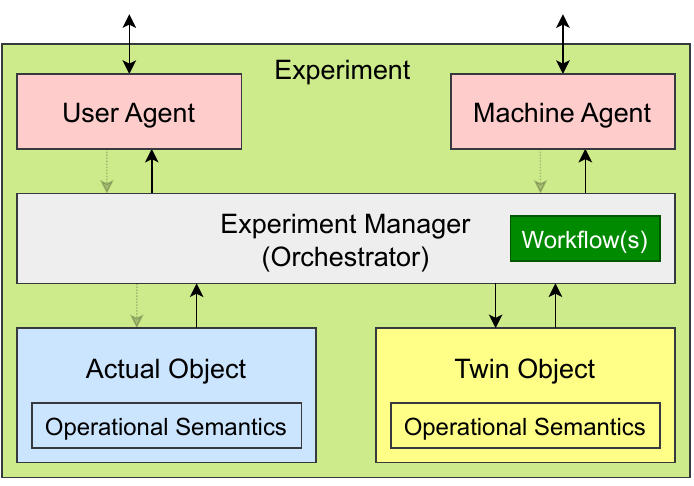}
        \caption{Reference architecture, variant 99.}
        \label{fig:var-98}
    \end{minipage}\\

	\begin{minipage}[t]{0.3\textwidth}
		\vspace{0.6cm}
        \centering
        \includegraphics[width=\textwidth]{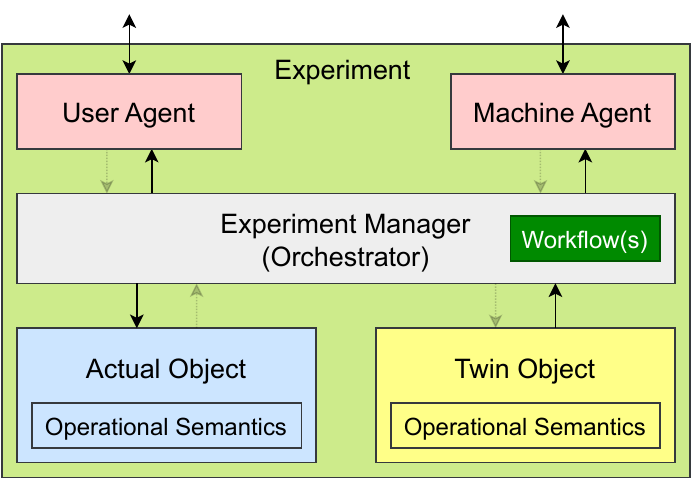}
        \caption{Reference architecture, variant 100.}
        \label{fig:var-99}
    \end{minipage}\hfill
	\begin{minipage}[t]{0.3\textwidth}
		\vspace{0.6cm}
        \centering
        \includegraphics[width=\textwidth]{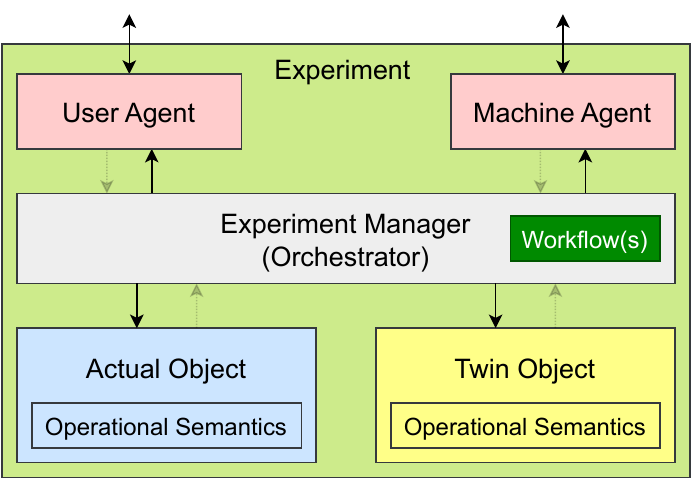}
        \caption{Reference architecture, variant 101.}
        \label{fig:var-100}
    \end{minipage}\hfill
	\begin{minipage}[t]{0.3\textwidth}
		\vspace{0.6cm}
        \centering
        \includegraphics[width=\textwidth]{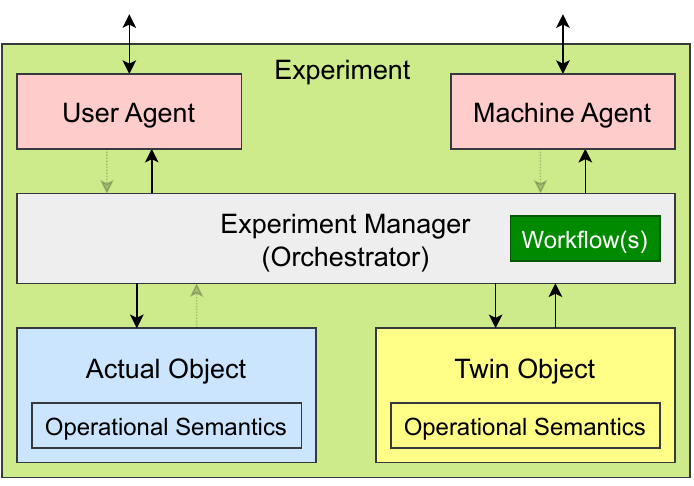}
        \caption{Reference architecture, variant 102.}
        \label{fig:var-101}
    \end{minipage}\\

	\begin{minipage}[t]{0.3\textwidth}
		\vspace{0.6cm}
        \centering
        \includegraphics[width=\textwidth]{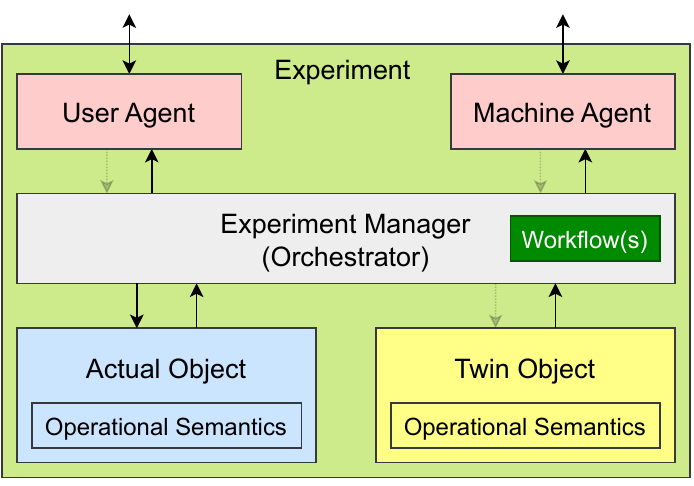}
        \caption{Reference architecture, variant 103.}
        \label{fig:var-102}
    \end{minipage}\hfill
	\begin{minipage}[t]{0.3\textwidth}
		\vspace{0.6cm}
        \centering
        \includegraphics[width=\textwidth]{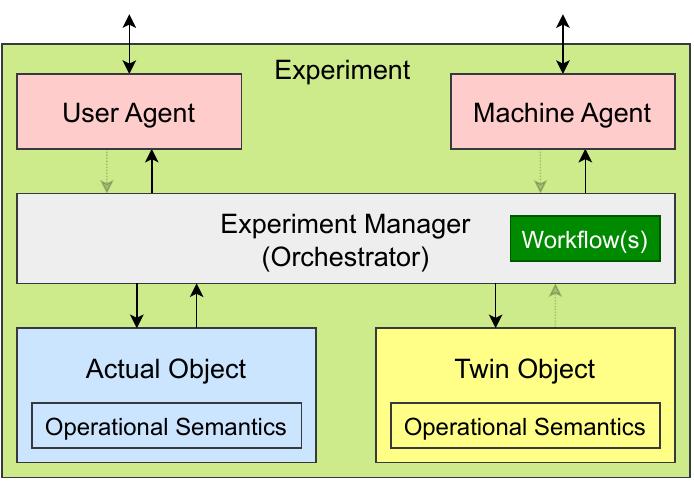}
        \caption{Reference architecture, variant 104.}
        \label{fig:var-103}
    \end{minipage}\hfill
	\begin{minipage}[t]{0.3\textwidth}
		\vspace{0.6cm}
        \centering
        \includegraphics[width=\textwidth]{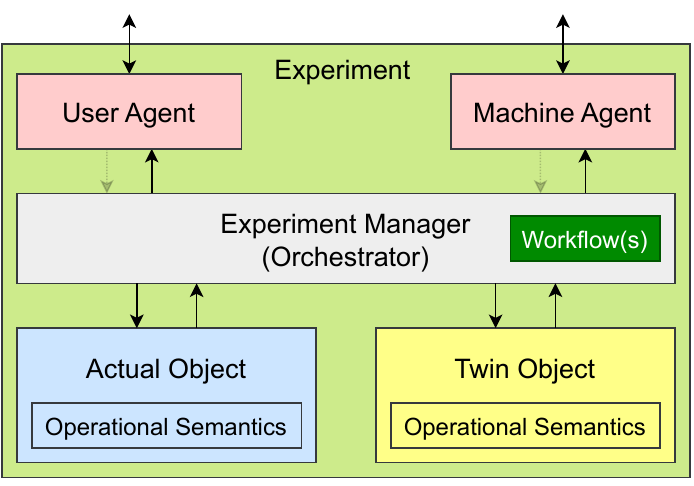}
        \caption{Reference architecture, variant 105.}
        \label{fig:var-104}
    \end{minipage}\\

\end{figure}
\begin{figure}[p]
	\centering
	\begin{minipage}[t]{0.3\textwidth}
		\vspace{0.6cm}
        \centering
        \includegraphics[width=\textwidth]{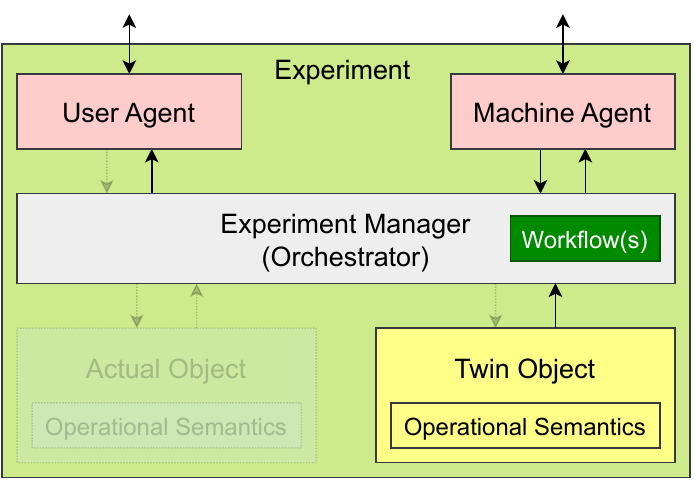}
        \caption{Reference architecture, variant 106, without Actual Object.}
        \label{fig:var-105}
    \end{minipage}\hfill
	\begin{minipage}[t]{0.3\textwidth}
		\vspace{0.6cm}
        \centering
        \includegraphics[width=\textwidth]{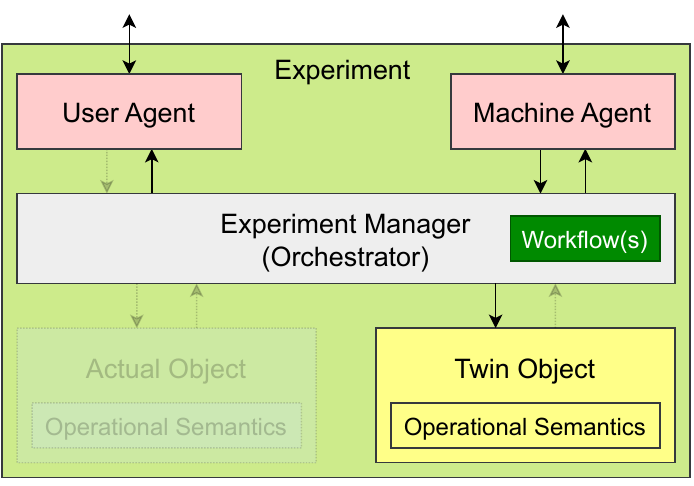}
        \caption{Reference architecture, variant 107, without Actual Object.}
        \label{fig:var-106}
    \end{minipage}\hfill
	\begin{minipage}[t]{0.3\textwidth}
		\vspace{0.6cm}
        \centering
        \includegraphics[width=\textwidth]{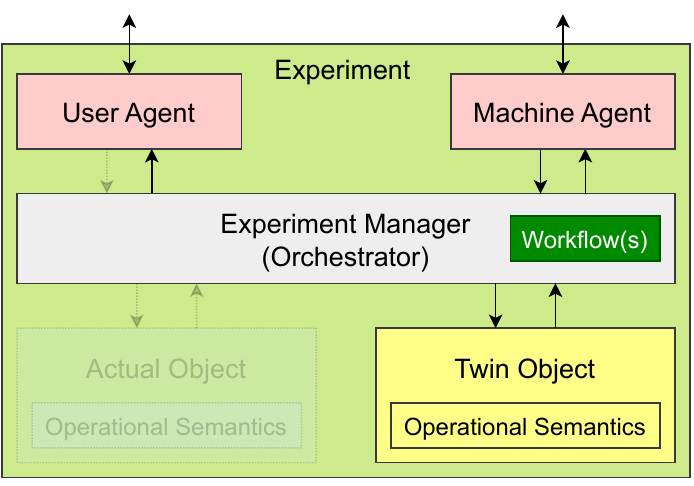}
        \caption{Reference architecture, variant 108, without Actual Object.}
        \label{fig:var-107}
    \end{minipage}\\

	\begin{minipage}[t]{0.3\textwidth}
		\vspace{0.6cm}
        \centering
        \includegraphics[width=\textwidth]{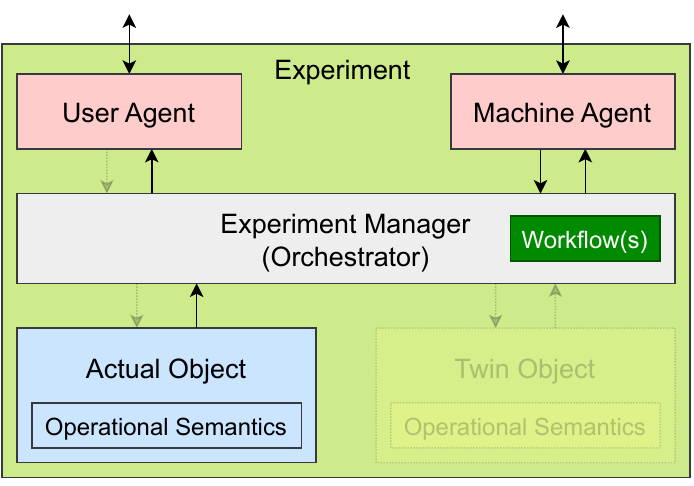}
        \caption{Reference architecture, variant 109, without Twin Object.}
        \label{fig:var-108}
    \end{minipage}\hfill
	\begin{minipage}[t]{0.3\textwidth}
		\vspace{0.6cm}
        \centering
        \includegraphics[width=\textwidth]{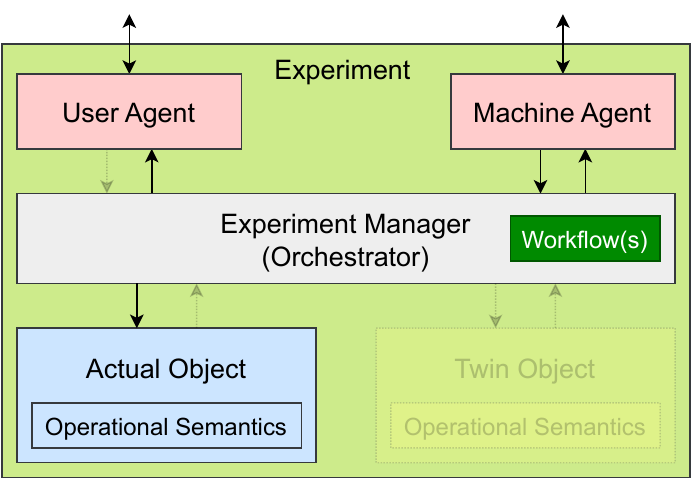}
        \caption{Reference architecture, variant 110, without Twin Object.}
        \label{fig:var-109}
    \end{minipage}\hfill
	\begin{minipage}[t]{0.3\textwidth}
		\vspace{0.6cm}
        \centering
        \includegraphics[width=\textwidth]{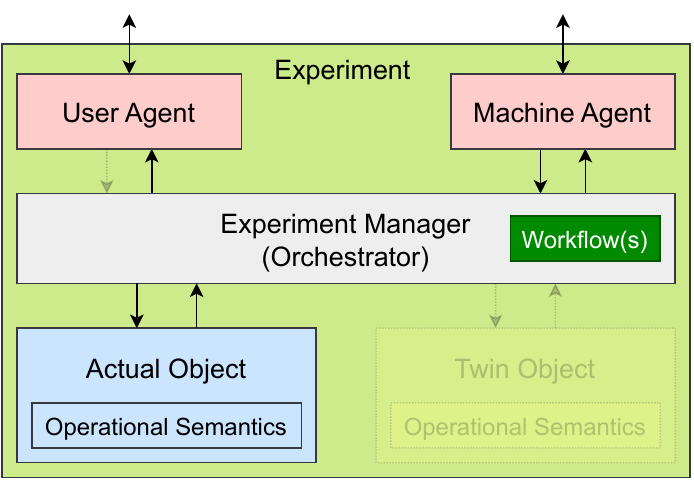}
        \caption{Reference architecture, variant 111, without Twin Object.}
        \label{fig:var-110}
    \end{minipage}\\

	\begin{minipage}[t]{0.3\textwidth}
		\vspace{0.6cm}
        \centering
        \includegraphics[width=\textwidth]{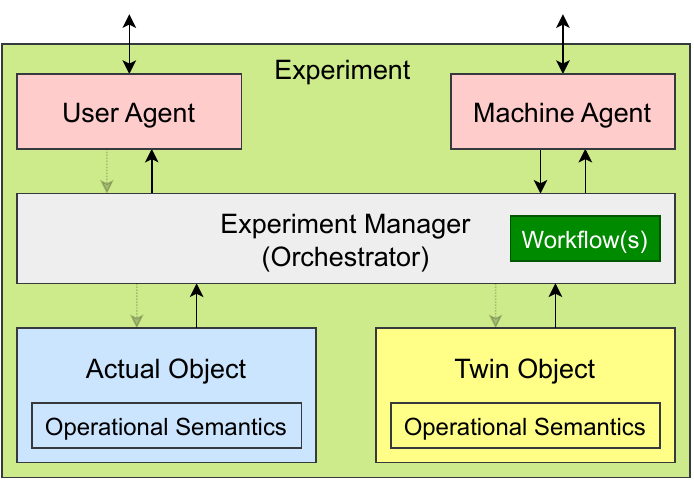}
        \caption{Reference architecture, variant 112.}
        \label{fig:var-111}
    \end{minipage}\hfill
	\begin{minipage}[t]{0.3\textwidth}
		\vspace{0.6cm}
        \centering
        \includegraphics[width=\textwidth]{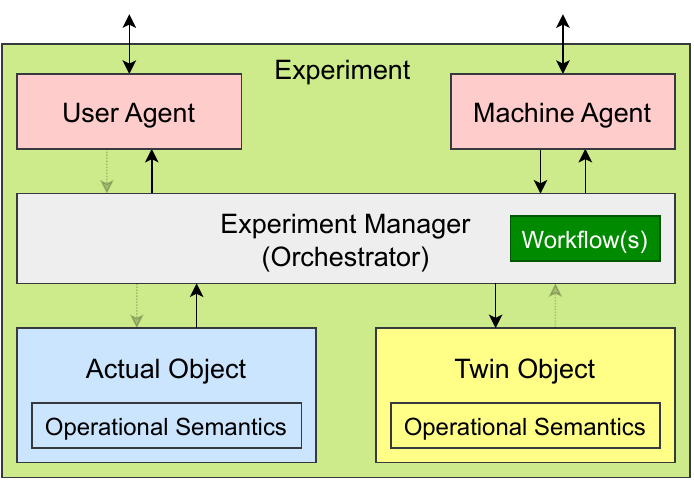}
        \caption{Reference architecture, variant 113.}
        \label{fig:var-112}
    \end{minipage}\hfill
	\begin{minipage}[t]{0.3\textwidth}
		\vspace{0.6cm}
        \centering
        \includegraphics[width=\textwidth]{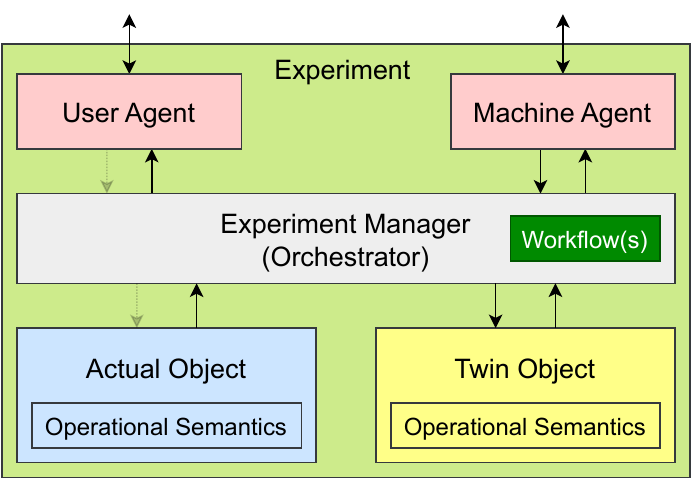}
        \caption{Reference architecture, variant 114.}
        \label{fig:var-113}
    \end{minipage}\\

	\begin{minipage}[t]{0.3\textwidth}
		\vspace{0.6cm}
        \centering
        \includegraphics[width=\textwidth]{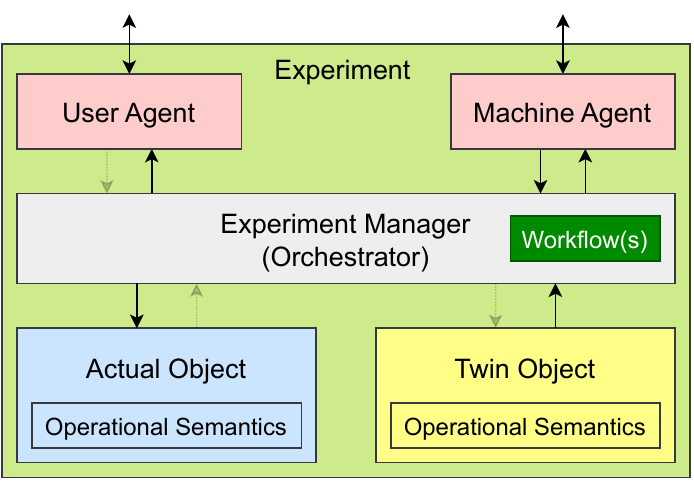}
        \caption{Reference architecture, variant 115.}
        \label{fig:var-114}
    \end{minipage}\hfill
	\begin{minipage}[t]{0.3\textwidth}
		\vspace{0.6cm}
        \centering
        \includegraphics[width=\textwidth]{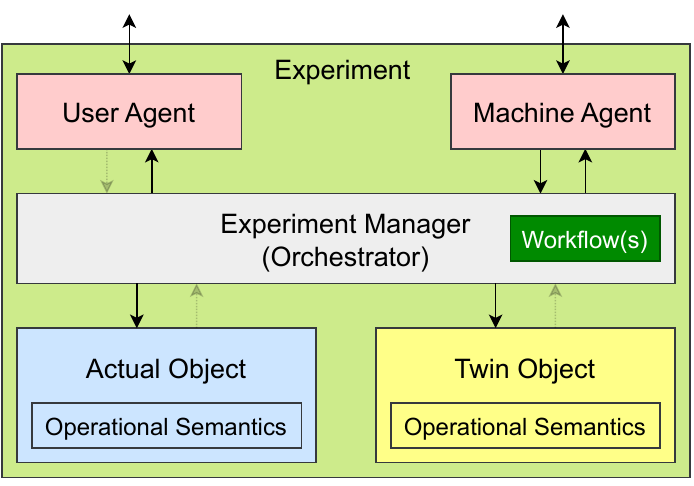}
        \caption{Reference architecture, variant 116.}
        \label{fig:var-115}
    \end{minipage}\hfill
	\begin{minipage}[t]{0.3\textwidth}
		\vspace{0.6cm}
        \centering
        \includegraphics[width=\textwidth]{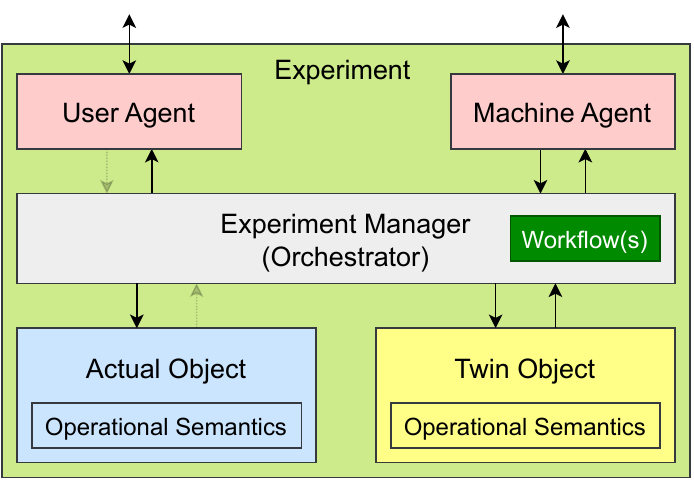}
        \caption{Reference architecture, variant 117.}
        \label{fig:var-116}
    \end{minipage}\\

\end{figure}
\begin{figure}[p]
	\centering
	\begin{minipage}[t]{0.3\textwidth}
		\vspace{0.6cm}
        \centering
        \includegraphics[width=\textwidth]{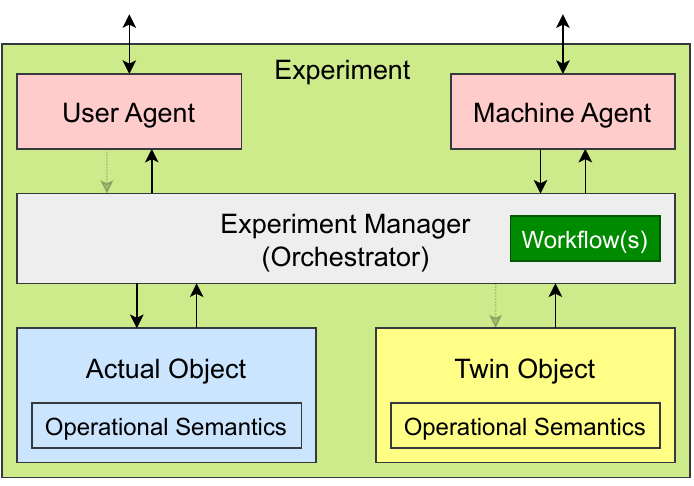}
        \caption{Reference architecture, variant 118.}
        \label{fig:var-117}
    \end{minipage}\hfill
	\begin{minipage}[t]{0.3\textwidth}
		\vspace{0.6cm}
        \centering
        \includegraphics[width=\textwidth]{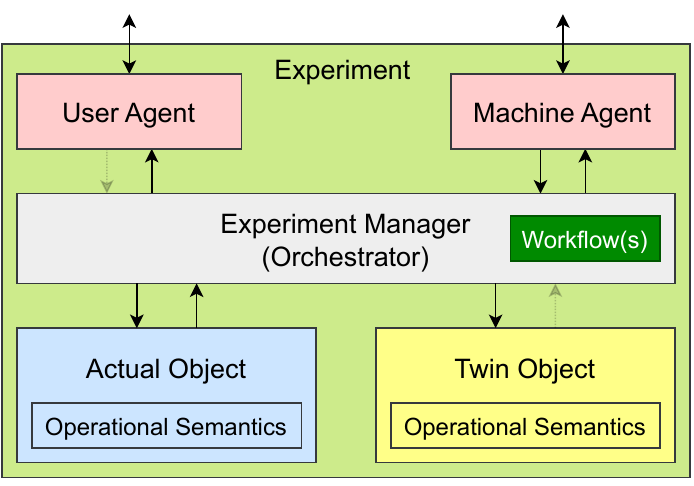}
        \caption{Reference architecture, variant 119.}
        \label{fig:var-118}
    \end{minipage}\hfill
	\begin{minipage}[t]{0.3\textwidth}
		\vspace{0.6cm}
        \centering
        \includegraphics[width=\textwidth]{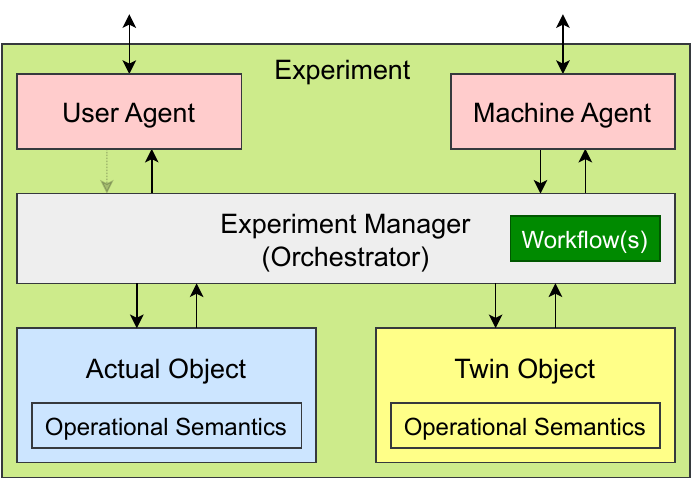}
        \caption{Reference architecture, variant 120.}
        \label{fig:var-119}
    \end{minipage}\\

	\begin{minipage}[t]{0.3\textwidth}
		\vspace{0.6cm}
        \centering
        \includegraphics[width=\textwidth]{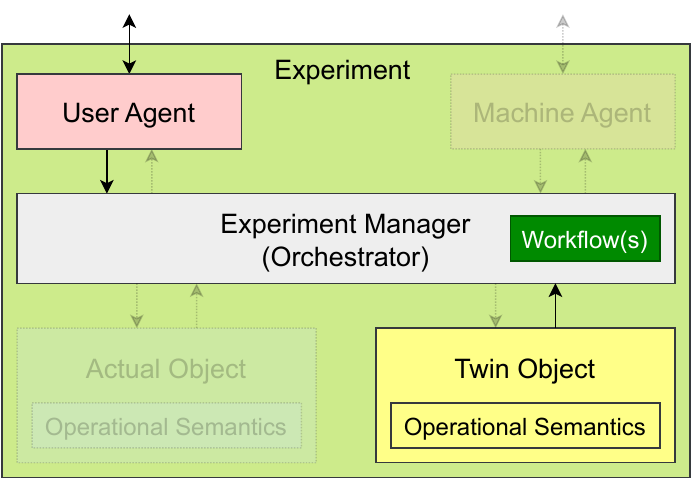}
        \caption{Reference architecture, variant 121, without Actual Object and Machine Agent.}
        \label{fig:var-120}
    \end{minipage}\hfill
	\begin{minipage}[t]{0.3\textwidth}
		\vspace{0.6cm}
        \centering
        \includegraphics[width=\textwidth]{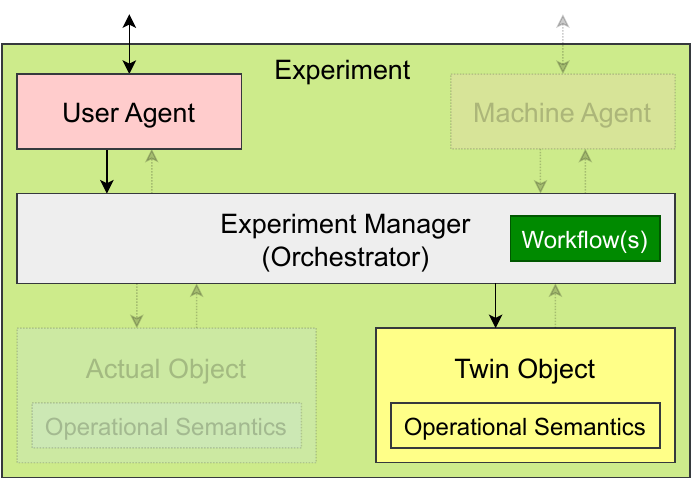}
        \caption{Reference architecture, variant 122, without Actual Object and Machine Agent.}
        \label{fig:var-121}
    \end{minipage}\hfill
	\begin{minipage}[t]{0.3\textwidth}
		\vspace{0.6cm}
        \centering
        \includegraphics[width=\textwidth]{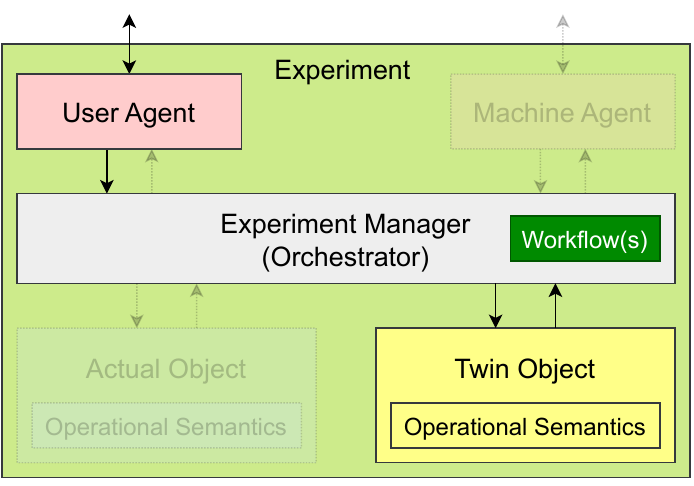}
        \caption{Reference architecture, variant 123, without Actual Object and Machine Agent.}
        \label{fig:var-122}
    \end{minipage}\\

	\begin{minipage}[t]{0.3\textwidth}
		\vspace{0.6cm}
        \centering
        \includegraphics[width=\textwidth]{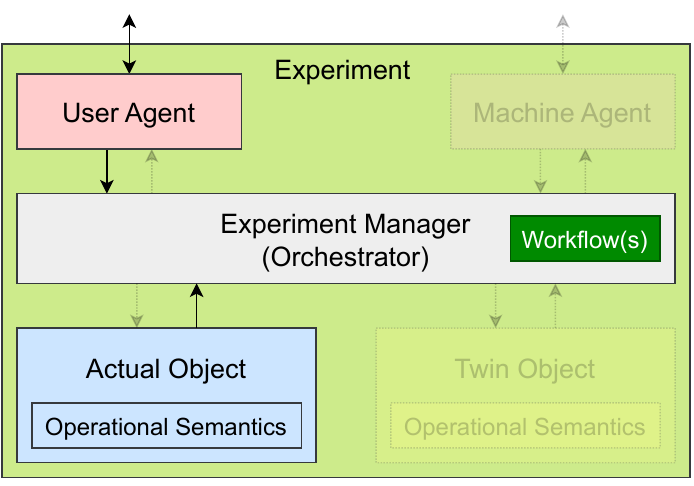}
        \caption{Reference architecture, variant 124, without Twin Object and Machine Agent.}
        \label{fig:var-123}
    \end{minipage}\hfill
	\begin{minipage}[t]{0.3\textwidth}
		\vspace{0.6cm}
        \centering
        \includegraphics[width=\textwidth]{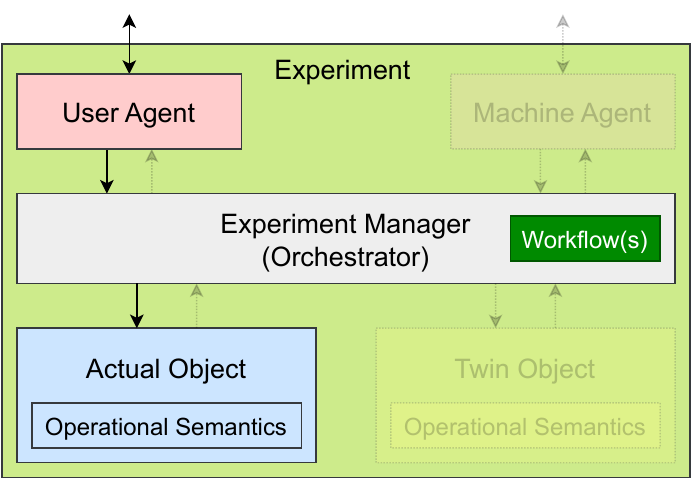}
        \caption{Reference architecture, variant 125, without Twin Object and Machine Agent.}
        \label{fig:var-124}
    \end{minipage}\hfill
	\begin{minipage}[t]{0.3\textwidth}
		\vspace{0.6cm}
        \centering
        \includegraphics[width=\textwidth]{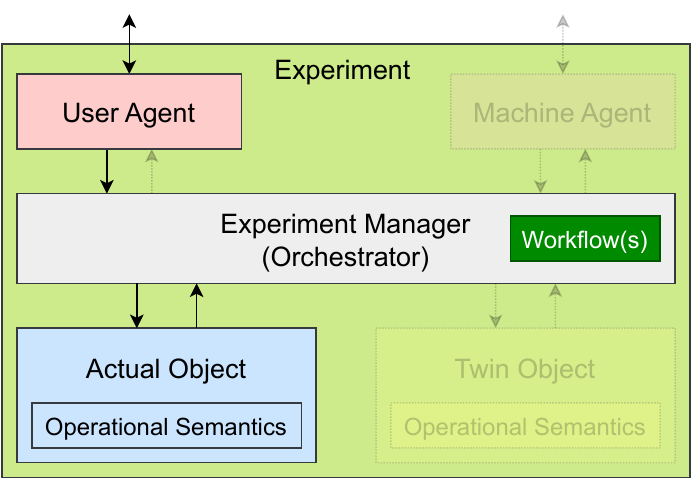}
        \caption{Reference architecture, variant 126, without Twin Object and Machine Agent.}
        \label{fig:var-125}
    \end{minipage}\\

	\begin{minipage}[t]{0.3\textwidth}
		\vspace{0.6cm}
        \centering
        \includegraphics[width=\textwidth]{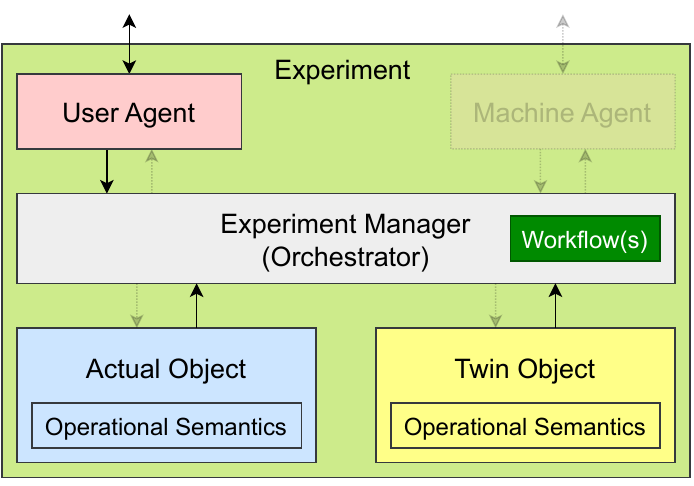}
        \caption{Reference architecture, variant 127, without Machine Agent.}
        \label{fig:var-126}
    \end{minipage}\hfill
	\begin{minipage}[t]{0.3\textwidth}
		\vspace{0.6cm}
        \centering
        \includegraphics[width=\textwidth]{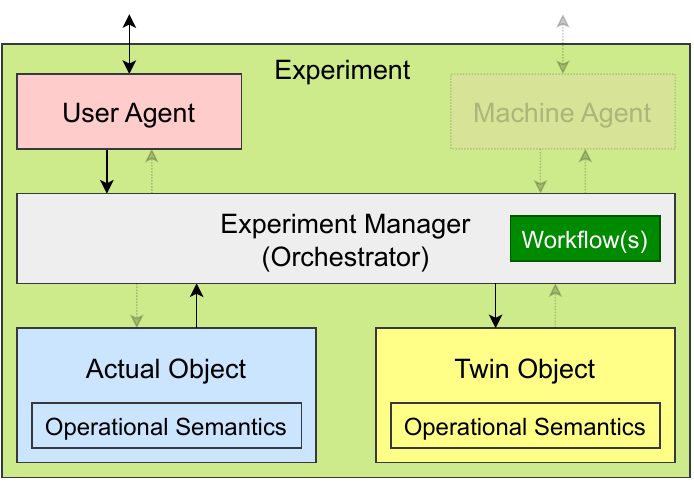}
        \caption{Reference architecture, variant 128, without Machine Agent.}
        \label{fig:var-127}
    \end{minipage}\hfill
	\begin{minipage}[t]{0.3\textwidth}
		\vspace{0.6cm}
        \centering
        \includegraphics[width=\textwidth]{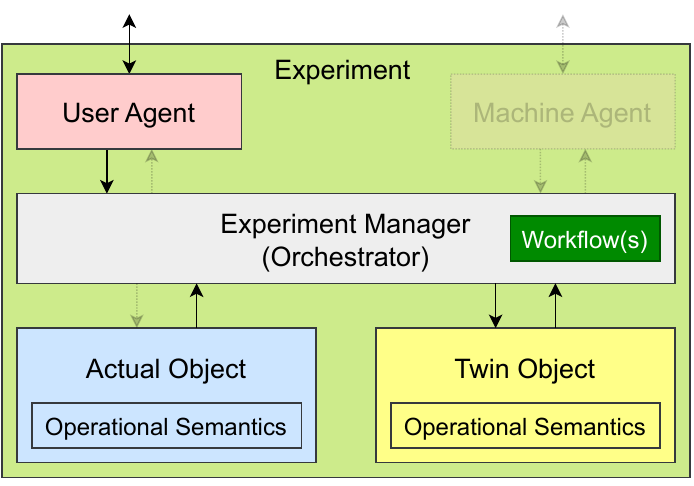}
        \caption{Reference architecture, variant 129, without Machine Agent.}
        \label{fig:var-128}
    \end{minipage}\\

\end{figure}
\begin{figure}[p]
	\centering
	\begin{minipage}[t]{0.3\textwidth}
		\vspace{0.6cm}
        \centering
        \includegraphics[width=\textwidth]{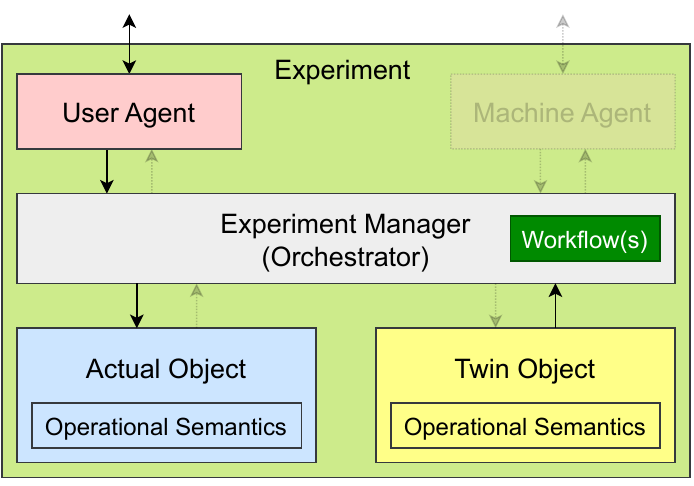}
        \caption{Reference architecture, variant 130, without Machine Agent.}
        \label{fig:var-129}
    \end{minipage}\hfill
	\begin{minipage}[t]{0.3\textwidth}
		\vspace{0.6cm}
        \centering
        \includegraphics[width=\textwidth]{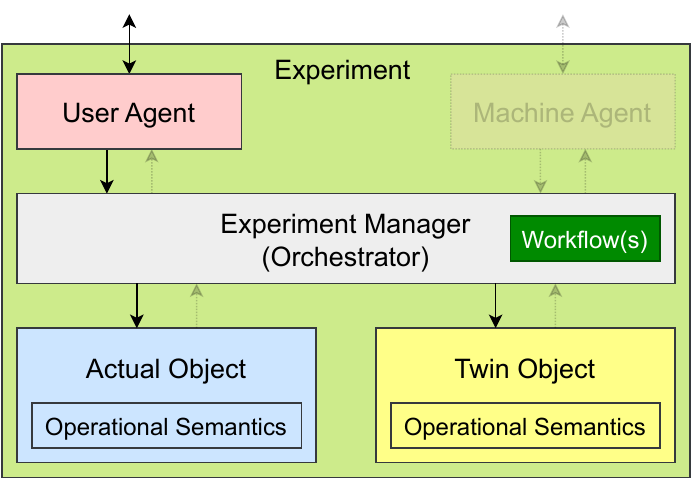}
        \caption{Reference architecture, variant 131, without Machine Agent.}
        \label{fig:var-130}
    \end{minipage}\hfill
	\begin{minipage}[t]{0.3\textwidth}
		\vspace{0.6cm}
        \centering
        \includegraphics[width=\textwidth]{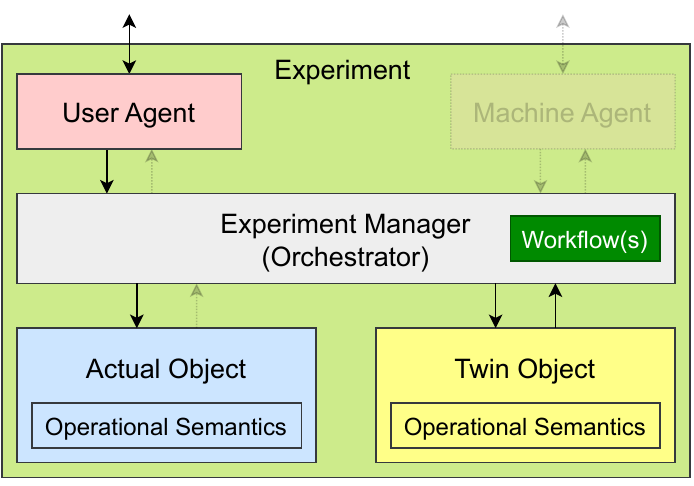}
        \caption{Reference architecture, variant 132, without Machine Agent.}
        \label{fig:var-131}
    \end{minipage}\\

	\begin{minipage}[t]{0.3\textwidth}
		\vspace{0.6cm}
        \centering
        \includegraphics[width=\textwidth]{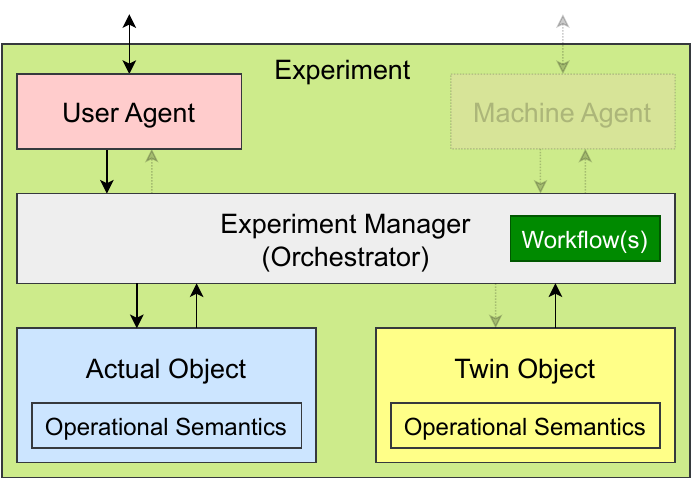}
        \caption{Reference architecture, variant 133, without Machine Agent.}
        \label{fig:var-132}
    \end{minipage}\hfill
	\begin{minipage}[t]{0.3\textwidth}
		\vspace{0.6cm}
        \centering
        \includegraphics[width=\textwidth]{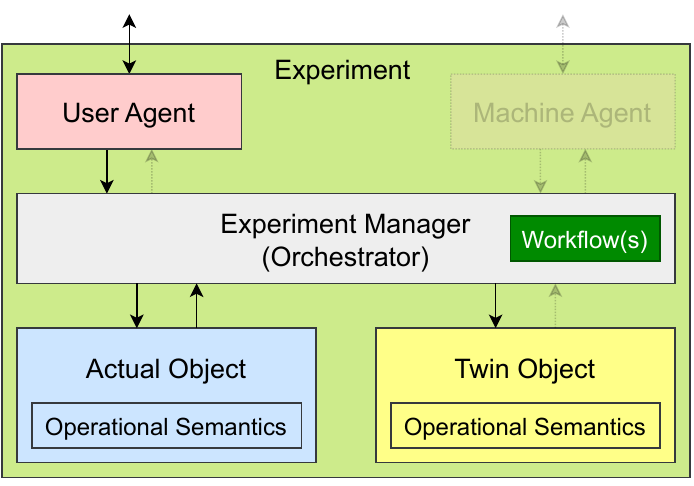}
        \caption{Reference architecture, variant 134, without Machine Agent.}
        \label{fig:var-133}
    \end{minipage}\hfill
	\begin{minipage}[t]{0.3\textwidth}
		\vspace{0.6cm}
        \centering
        \includegraphics[width=\textwidth]{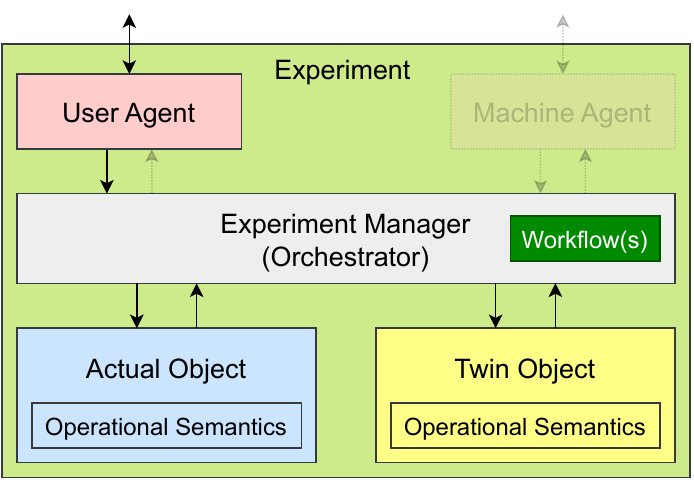}
        \caption{Reference architecture, variant 135, without Machine Agent.}
        \label{fig:var-134}
    \end{minipage}\\

	\begin{minipage}[t]{0.3\textwidth}
		\vspace{0.6cm}
        \centering
        \includegraphics[width=\textwidth]{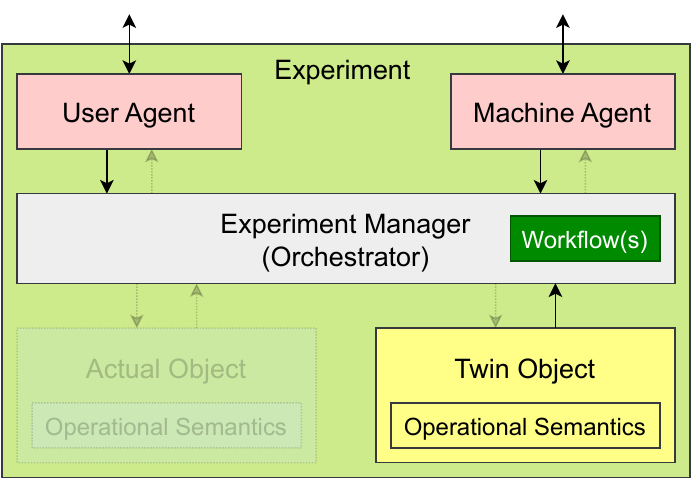}
        \caption{Reference architecture, variant 136, without Actual Object.}
        \label{fig:var-135}
    \end{minipage}\hfill
	\begin{minipage}[t]{0.3\textwidth}
		\vspace{0.6cm}
        \centering
        \includegraphics[width=\textwidth]{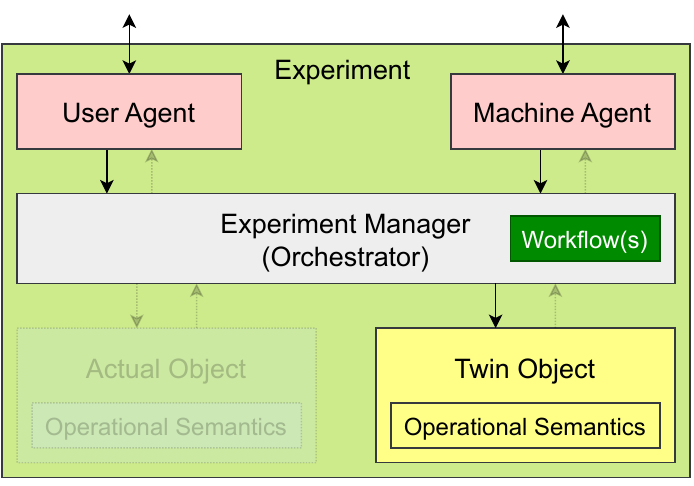}
        \caption{Reference architecture, variant 137, without Actual Object.}
        \label{fig:var-136}
    \end{minipage}\hfill
	\begin{minipage}[t]{0.3\textwidth}
		\vspace{0.6cm}
        \centering
        \includegraphics[width=\textwidth]{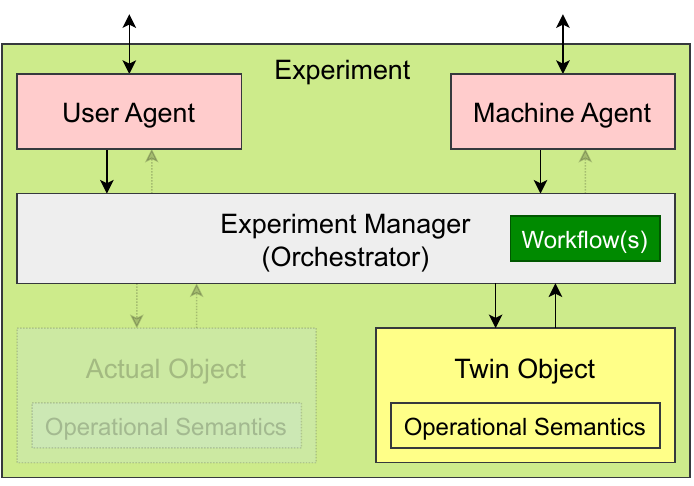}
        \caption{Reference architecture, variant 138, without Actual Object.}
        \label{fig:var-137}
    \end{minipage}\\

	\begin{minipage}[t]{0.3\textwidth}
		\vspace{0.6cm}
        \centering
        \includegraphics[width=\textwidth]{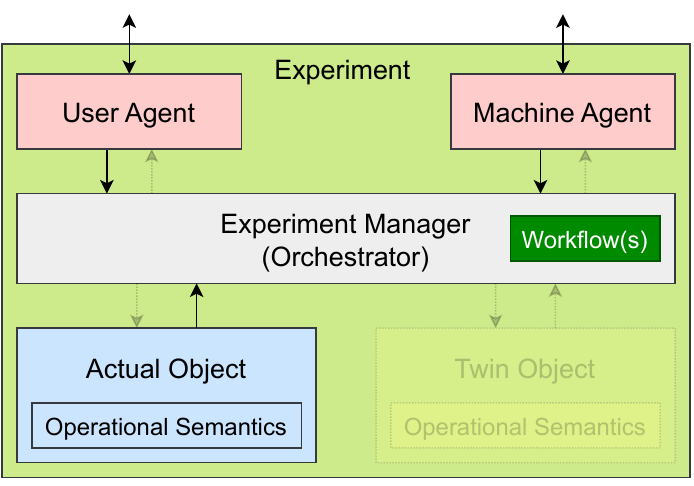}
        \caption{Reference architecture, variant 139, without Twin Object.}
        \label{fig:var-138}
    \end{minipage}\hfill
	\begin{minipage}[t]{0.3\textwidth}
		\vspace{0.6cm}
        \centering
        \includegraphics[width=\textwidth]{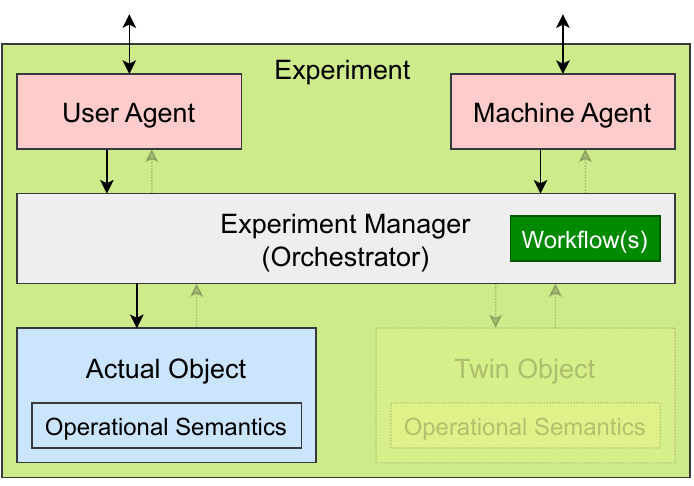}
        \caption{Reference architecture, variant 140, without Twin Object.}
        \label{fig:var-139}
    \end{minipage}\hfill
	\begin{minipage}[t]{0.3\textwidth}
		\vspace{0.6cm}
        \centering
        \includegraphics[width=\textwidth]{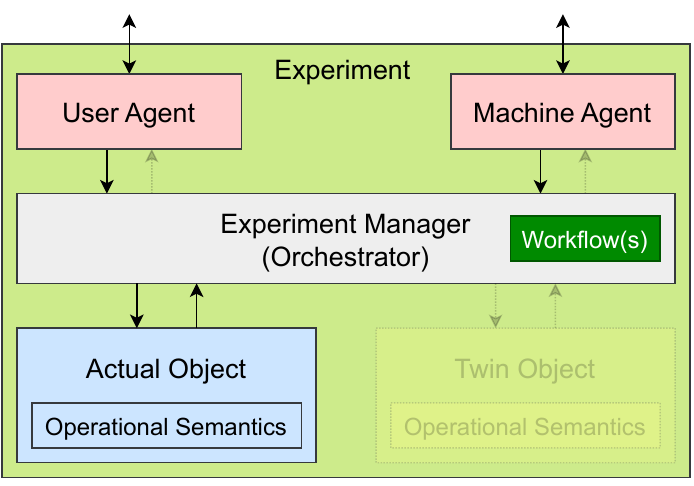}
        \caption{Reference architecture, variant 141, without Twin Object.}
        \label{fig:var-140}
    \end{minipage}\\

\end{figure}
\begin{figure}[p]
	\centering
	\begin{minipage}[t]{0.3\textwidth}
		\vspace{0.6cm}
        \centering
        \includegraphics[width=\textwidth]{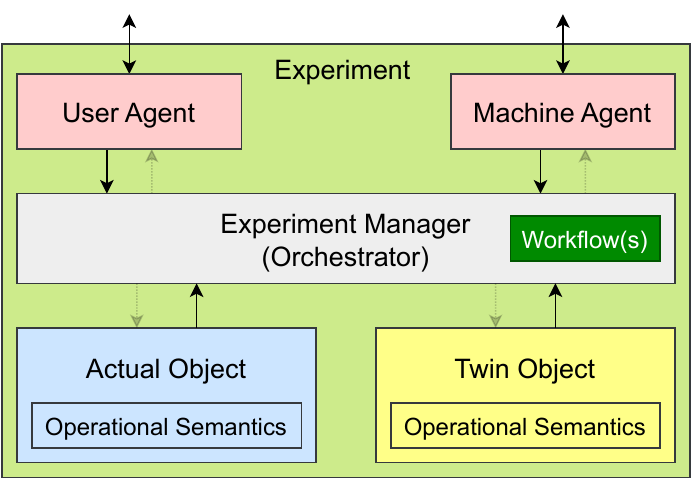}
        \caption{Reference architecture, variant 142.}
        \label{fig:var-141}
    \end{minipage}\hfill
	\begin{minipage}[t]{0.3\textwidth}
		\vspace{0.6cm}
        \centering
        \includegraphics[width=\textwidth]{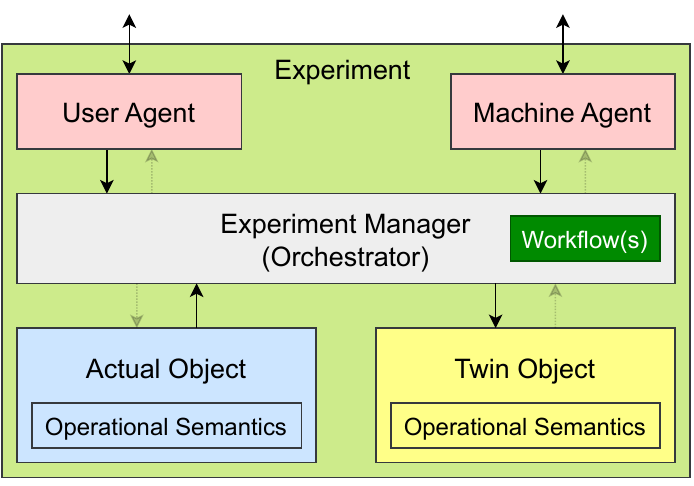}
        \caption{Reference architecture, variant 143.}
        \label{fig:var-142}
    \end{minipage}\hfill
	\begin{minipage}[t]{0.3\textwidth}
		\vspace{0.6cm}
        \centering
        \includegraphics[width=\textwidth]{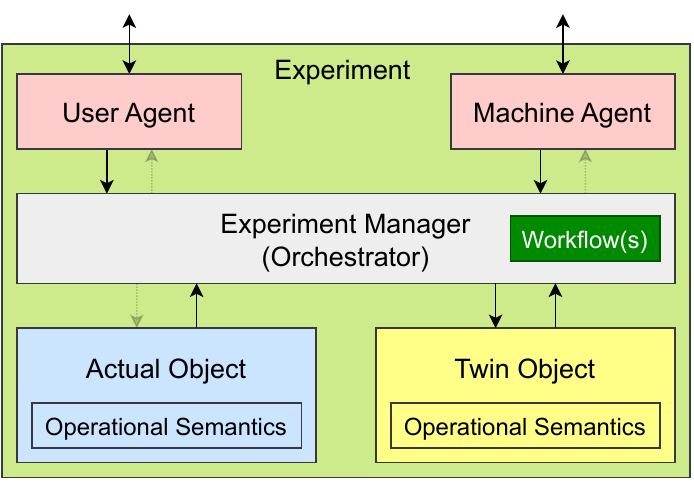}
        \caption{Reference architecture, variant 144.}
        \label{fig:var-143}
    \end{minipage}\\

	\begin{minipage}[t]{0.3\textwidth}
		\vspace{0.6cm}
        \centering
        \includegraphics[width=\textwidth]{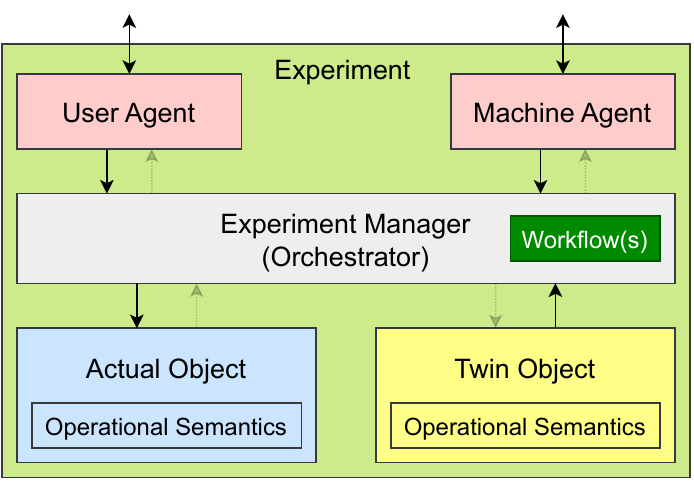}
        \caption{Reference architecture, variant 145.}
        \label{fig:var-144}
    \end{minipage}\hfill
	\begin{minipage}[t]{0.3\textwidth}
		\vspace{0.6cm}
        \centering
        \includegraphics[width=\textwidth]{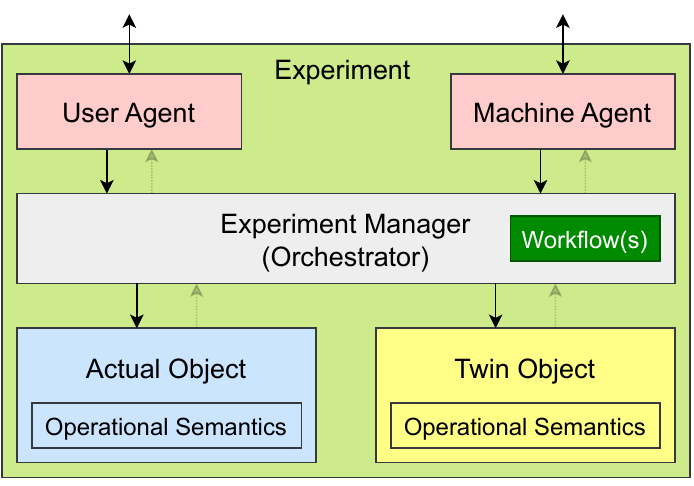}
        \caption{Reference architecture, variant 146.}
        \label{fig:var-145}
    \end{minipage}\hfill
	\begin{minipage}[t]{0.3\textwidth}
		\vspace{0.6cm}
        \centering
        \includegraphics[width=\textwidth]{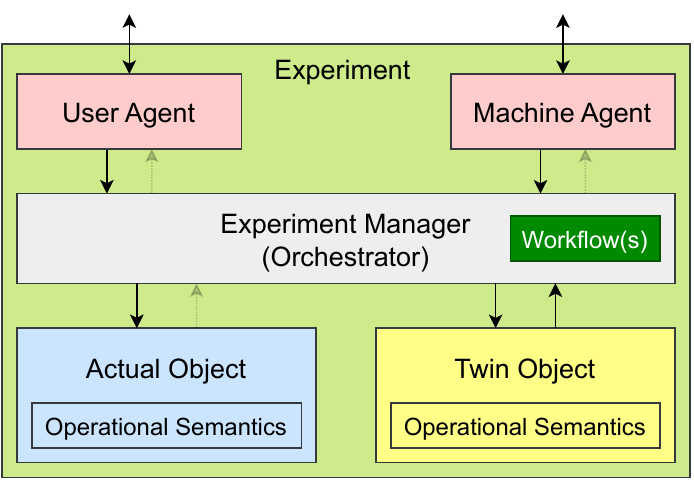}
        \caption{Reference architecture, variant 147.}
        \label{fig:var-146}
    \end{minipage}\\

	\begin{minipage}[t]{0.3\textwidth}
		\vspace{0.6cm}
        \centering
        \includegraphics[width=\textwidth]{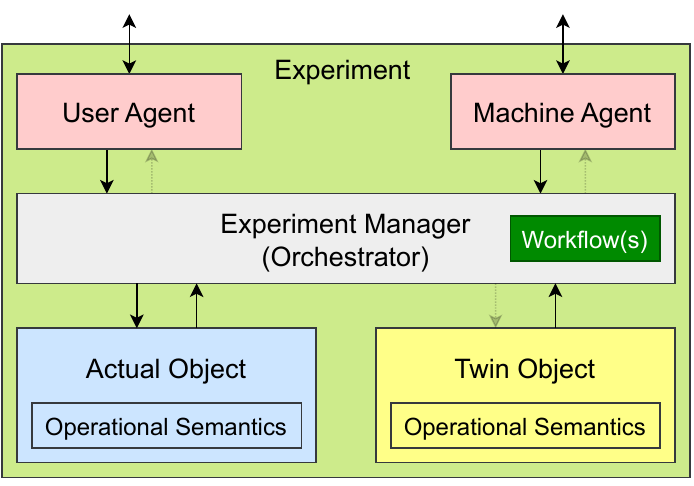}
        \caption{Reference architecture, variant 148.}
        \label{fig:var-147}
    \end{minipage}\hfill
	\begin{minipage}[t]{0.3\textwidth}
		\vspace{0.6cm}
        \centering
        \includegraphics[width=\textwidth]{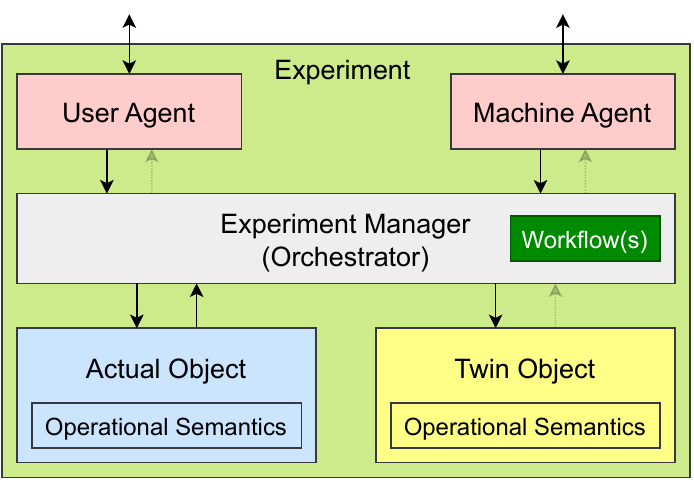}
        \caption{Reference architecture, variant 149.}
        \label{fig:var-148}
    \end{minipage}\hfill
	\begin{minipage}[t]{0.3\textwidth}
		\vspace{0.6cm}
        \centering
        \includegraphics[width=\textwidth]{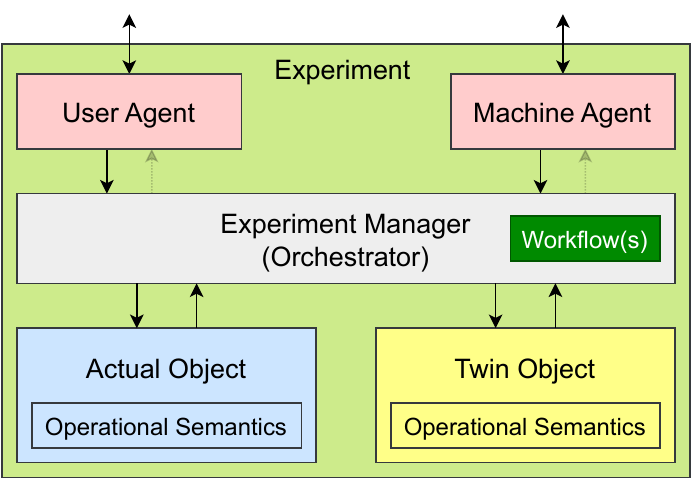}
        \caption{Reference architecture, variant 150.}
        \label{fig:var-149}
    \end{minipage}\\

	\begin{minipage}[t]{0.3\textwidth}
		\vspace{0.6cm}
        \centering
        \includegraphics[width=\textwidth]{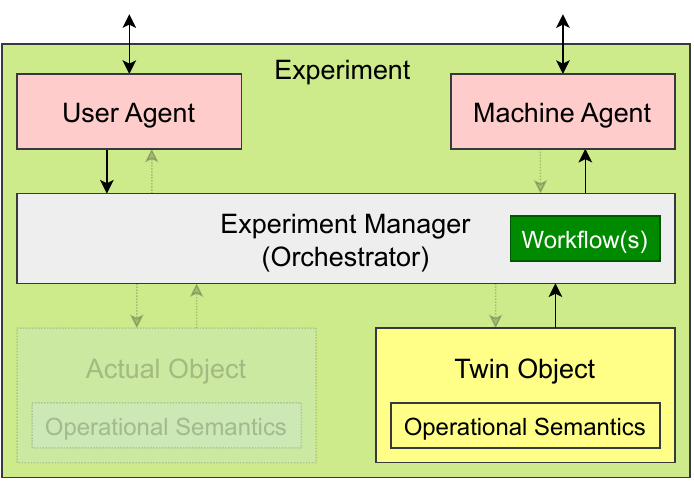}
        \caption{Reference architecture, variant 151, without Actual Object.}
        \label{fig:var-150}
    \end{minipage}\hfill
	\begin{minipage}[t]{0.3\textwidth}
		\vspace{0.6cm}
        \centering
        \includegraphics[width=\textwidth]{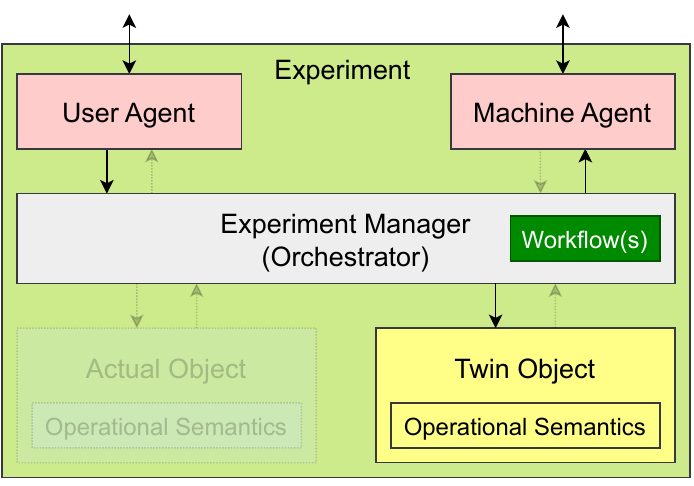}
        \caption{Reference architecture, variant 152, without Actual Object.}
        \label{fig:var-151}
    \end{minipage}\hfill
	\begin{minipage}[t]{0.3\textwidth}
		\vspace{0.6cm}
        \centering
        \includegraphics[width=\textwidth]{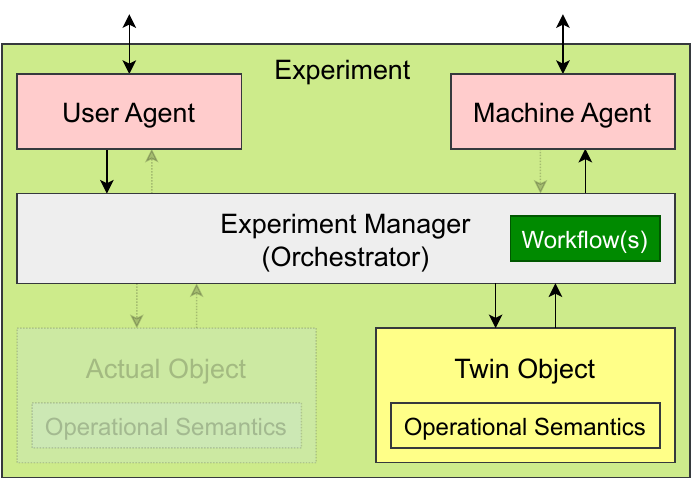}
        \caption{Reference architecture, variant 153, without Actual Object.}
        \label{fig:var-152}
    \end{minipage}\\

\end{figure}
\begin{figure}[p]
	\centering
	\begin{minipage}[t]{0.3\textwidth}
		\vspace{0.6cm}
        \centering
        \includegraphics[width=\textwidth]{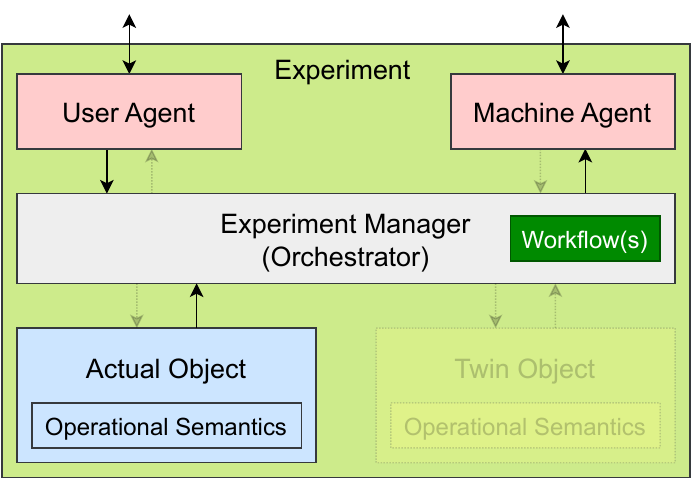}
        \caption{Reference architecture, variant 154, without Twin Object.}
        \label{fig:var-153}
    \end{minipage}\hfill
	\begin{minipage}[t]{0.3\textwidth}
		\vspace{0.6cm}
        \centering
        \includegraphics[width=\textwidth]{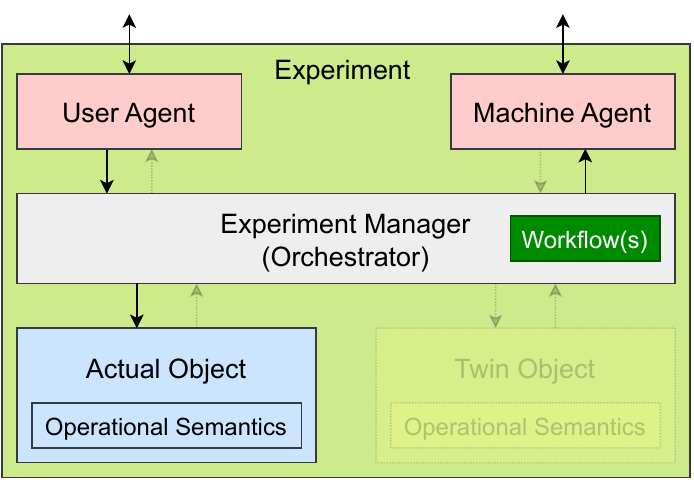}
        \caption{Reference architecture, variant 155, without Twin Object.}
        \label{fig:var-154}
    \end{minipage}\hfill
	\begin{minipage}[t]{0.3\textwidth}
		\vspace{0.6cm}
        \centering
        \includegraphics[width=\textwidth]{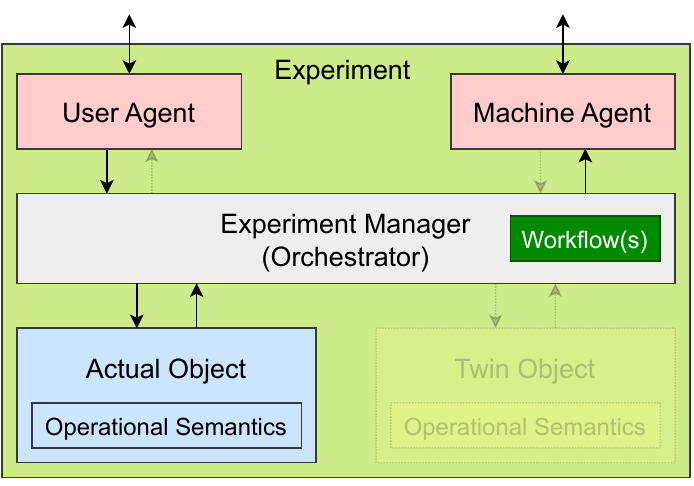}
        \caption{Reference architecture, variant 156, without Twin Object.}
        \label{fig:var-155}
    \end{minipage}\\

	\begin{minipage}[t]{0.3\textwidth}
		\vspace{0.6cm}
        \centering
        \includegraphics[width=\textwidth]{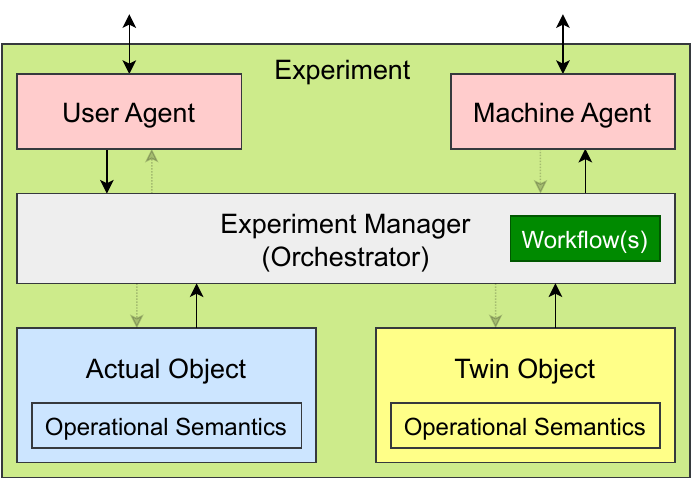}
        \caption{Reference architecture, variant 157.}
        \label{fig:var-156}
    \end{minipage}\hfill
	\begin{minipage}[t]{0.3\textwidth}
		\vspace{0.6cm}
        \centering
        \includegraphics[width=\textwidth]{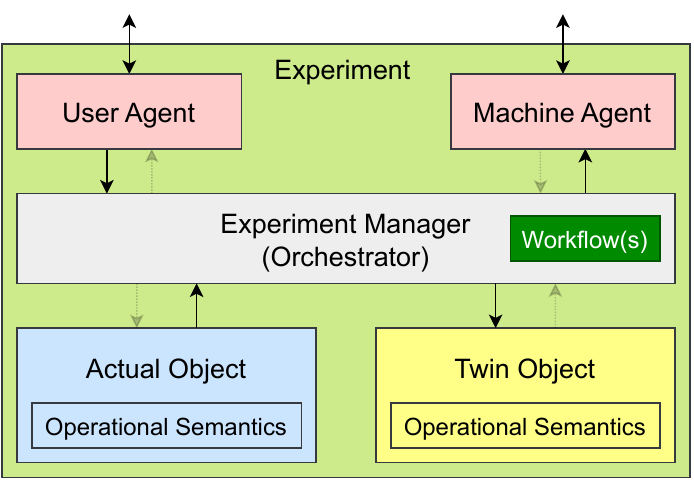}
        \caption{Reference architecture, variant 158.}
        \label{fig:var-157}
    \end{minipage}\hfill
	\begin{minipage}[t]{0.3\textwidth}
		\vspace{0.6cm}
        \centering
        \includegraphics[width=\textwidth]{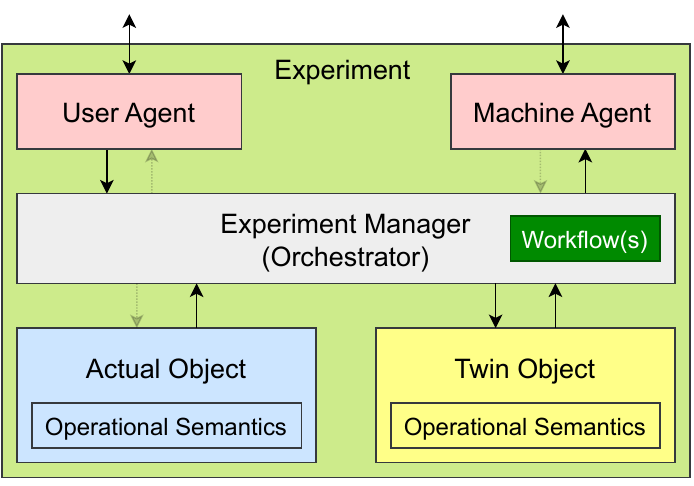}
        \caption{Reference architecture, variant 159.}
        \label{fig:var-158}
    \end{minipage}\\

	\begin{minipage}[t]{0.3\textwidth}
		\vspace{0.6cm}
        \centering
        \includegraphics[width=\textwidth]{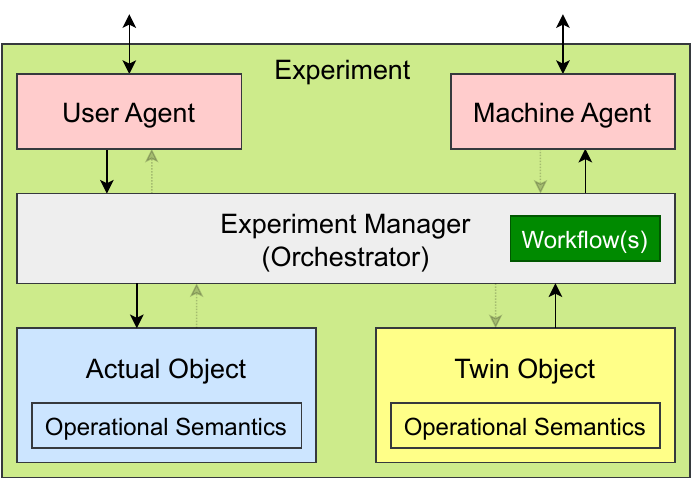}
        \caption{Reference architecture, variant 160.}
        \label{fig:var-159}
    \end{minipage}\hfill
	\begin{minipage}[t]{0.3\textwidth}
		\vspace{0.6cm}
        \centering
        \includegraphics[width=\textwidth]{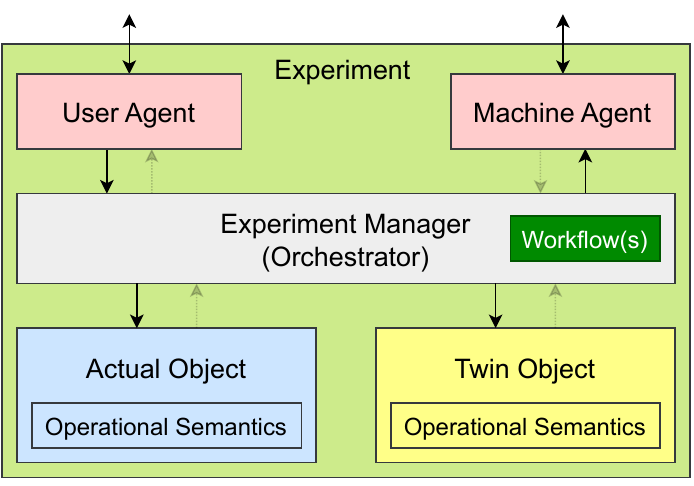}
        \caption{Reference architecture, variant 161.}
        \label{fig:var-160}
    \end{minipage}\hfill
	\begin{minipage}[t]{0.3\textwidth}
		\vspace{0.6cm}
        \centering
        \includegraphics[width=\textwidth]{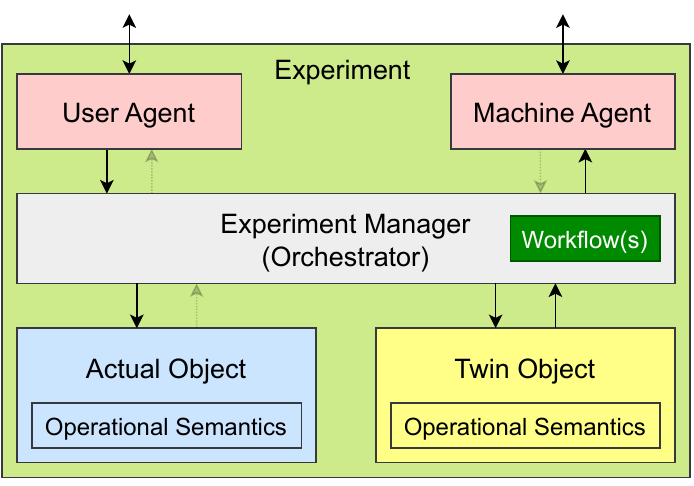}
        \caption{Reference architecture, variant 162.}
        \label{fig:var-161}
    \end{minipage}\\

	\begin{minipage}[t]{0.3\textwidth}
		\vspace{0.6cm}
        \centering
        \includegraphics[width=\textwidth]{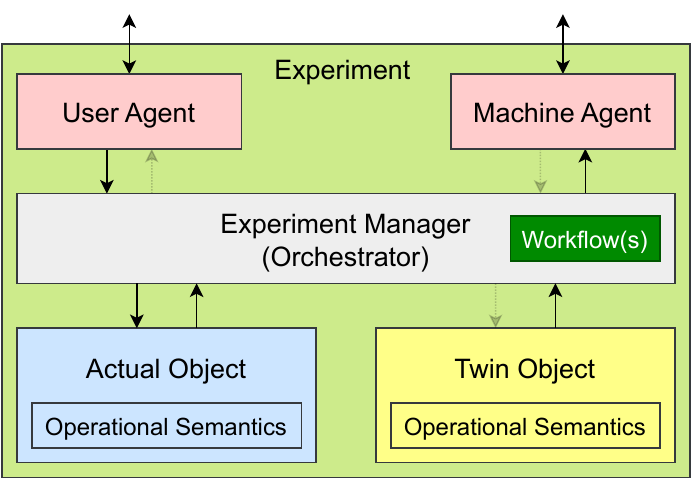}
        \caption{Reference architecture, variant 163.}
        \label{fig:var-162}
    \end{minipage}\hfill
	\begin{minipage}[t]{0.3\textwidth}
		\vspace{0.6cm}
        \centering
        \includegraphics[width=\textwidth]{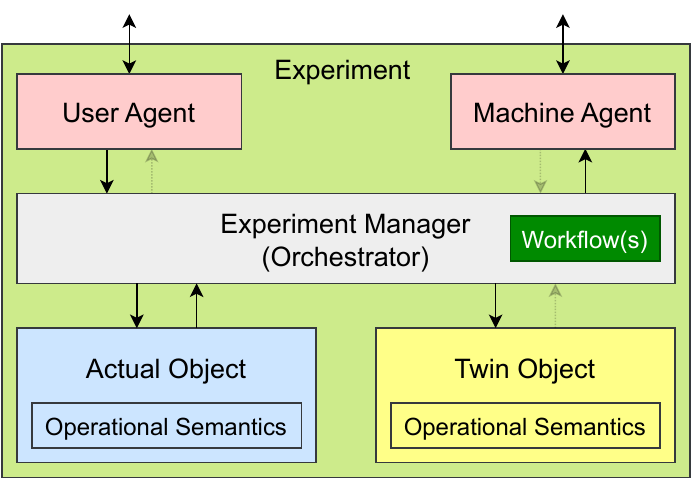}
        \caption{Reference architecture, variant 164.}
        \label{fig:var-163}
    \end{minipage}\hfill
	\begin{minipage}[t]{0.3\textwidth}
		\vspace{0.6cm}
        \centering
        \includegraphics[width=\textwidth]{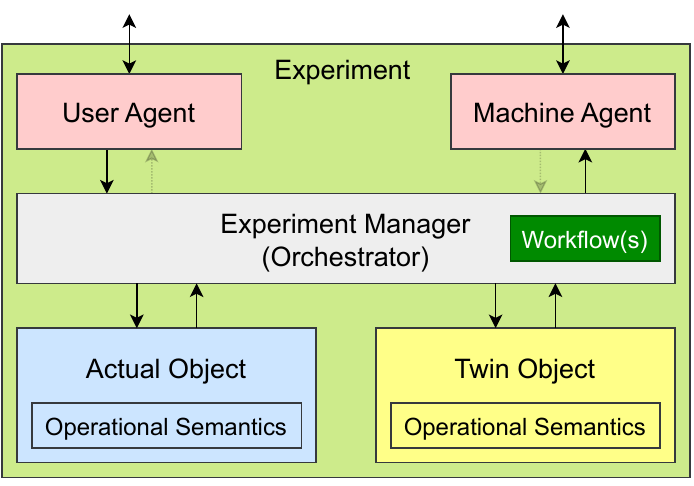}
        \caption{Reference architecture, variant 165.}
        \label{fig:var-164}
    \end{minipage}\\

\end{figure}
\begin{figure}[p]
	\centering
	\begin{minipage}[t]{0.3\textwidth}
		\vspace{0.6cm}
        \centering
        \includegraphics[width=\textwidth]{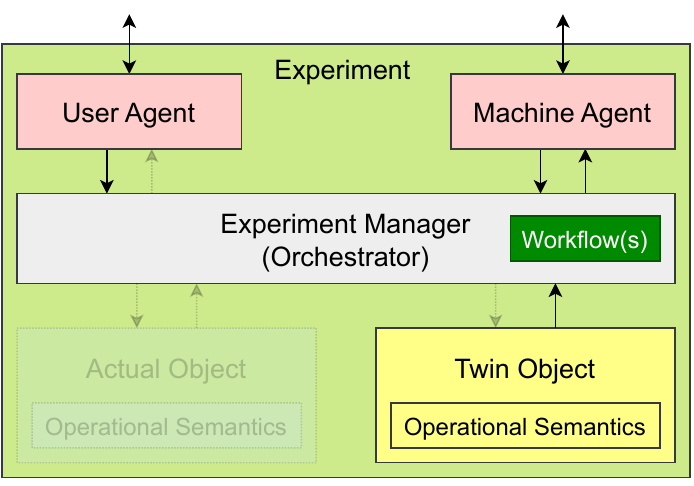}
        \caption{Reference architecture, variant 166, without Actual Object.}
        \label{fig:var-165}
    \end{minipage}\hfill
	\begin{minipage}[t]{0.3\textwidth}
		\vspace{0.6cm}
        \centering
        \includegraphics[width=\textwidth]{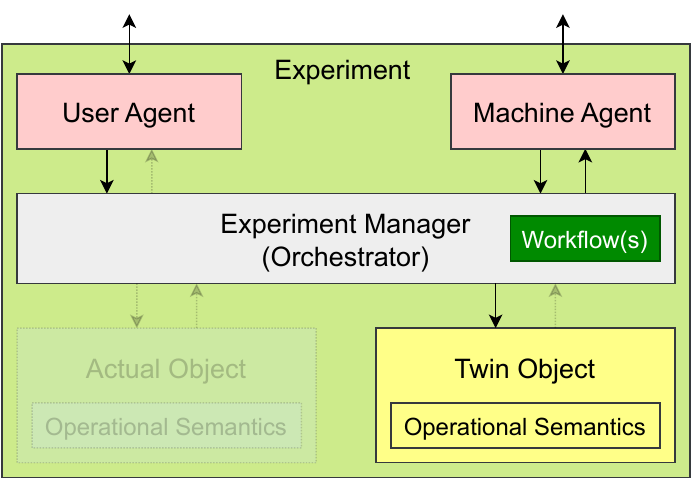}
        \caption{Reference architecture, variant 167, without Actual Object.}
        \label{fig:var-166}
    \end{minipage}\hfill
	\begin{minipage}[t]{0.3\textwidth}
		\vspace{0.6cm}
        \centering
        \includegraphics[width=\textwidth]{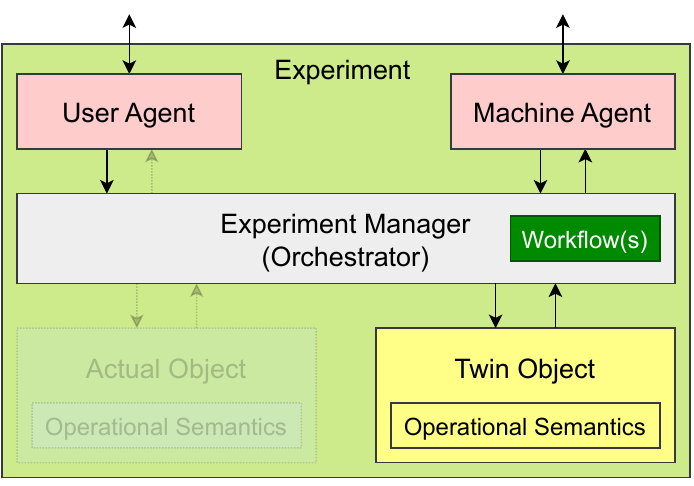}
        \caption{Reference architecture, variant 168, without Actual Object.}
        \label{fig:var-167}
    \end{minipage}\\

	\begin{minipage}[t]{0.3\textwidth}
		\vspace{0.6cm}
        \centering
        \includegraphics[width=\textwidth]{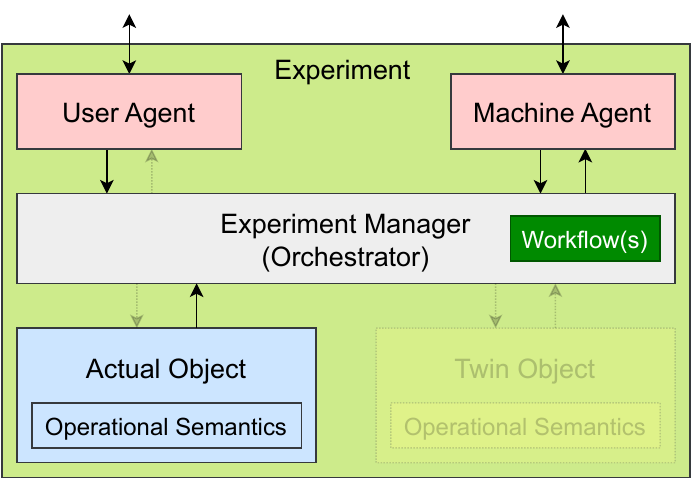}
        \caption{Reference architecture, variant 169, without Twin Object.}
        \label{fig:var-168}
    \end{minipage}\hfill
	\begin{minipage}[t]{0.3\textwidth}
		\vspace{0.6cm}
        \centering
        \includegraphics[width=\textwidth]{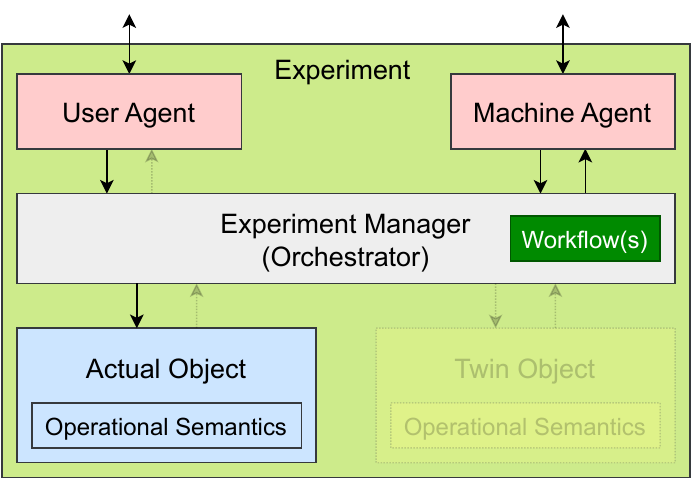}
        \caption{Reference architecture, variant 170, without Twin Object.}
        \label{fig:var-169}
    \end{minipage}\hfill
	\begin{minipage}[t]{0.3\textwidth}
		\vspace{0.6cm}
        \centering
        \includegraphics[width=\textwidth]{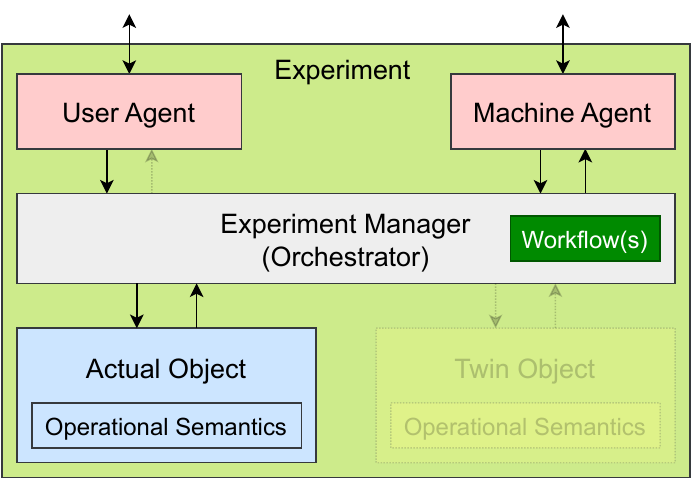}
        \caption{Reference architecture, variant 171, without Twin Object.}
        \label{fig:var-170}
    \end{minipage}\\

	\begin{minipage}[t]{0.3\textwidth}
		\vspace{0.6cm}
        \centering
        \includegraphics[width=\textwidth]{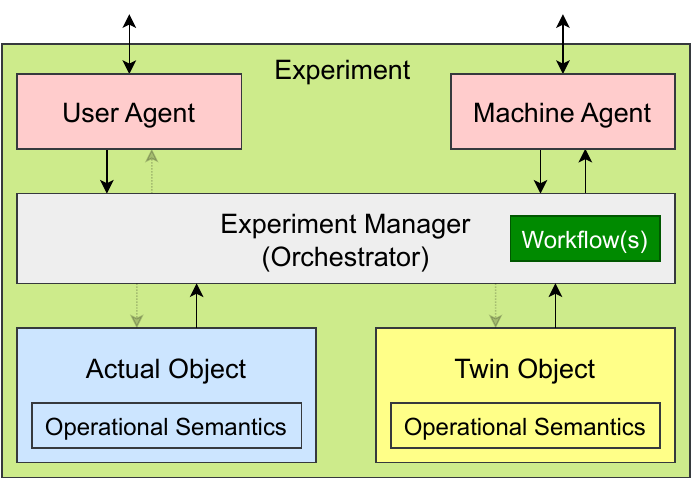}
        \caption{Reference architecture, variant 172.}
        \label{fig:var-171}
    \end{minipage}\hfill
	\begin{minipage}[t]{0.3\textwidth}
		\vspace{0.6cm}
        \centering
        \includegraphics[width=\textwidth]{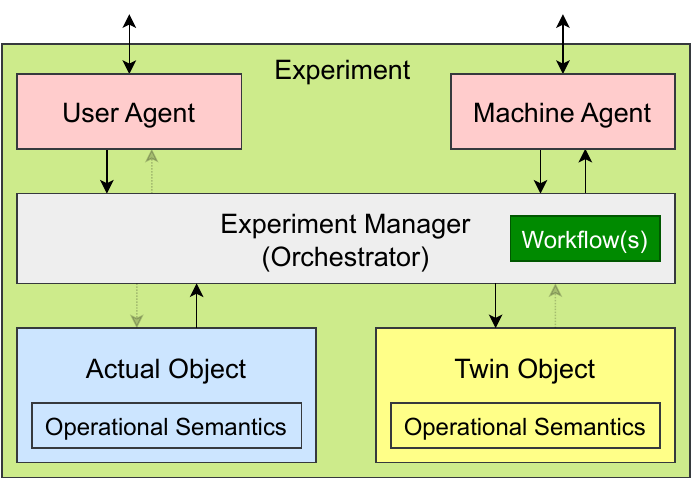}
        \caption{Reference architecture, variant 173.}
        \label{fig:var-172}
    \end{minipage}\hfill
	\begin{minipage}[t]{0.3\textwidth}
		\vspace{0.6cm}
        \centering
        \includegraphics[width=\textwidth]{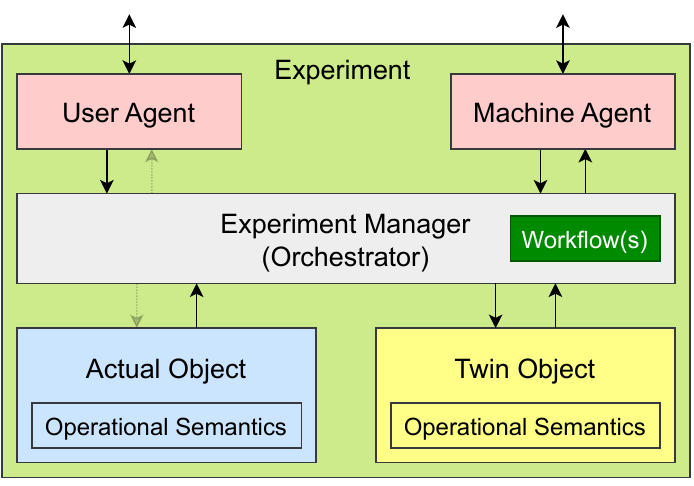}
        \caption{Reference architecture, variant 174.}
        \label{fig:var-173}
    \end{minipage}\\

	\begin{minipage}[t]{0.3\textwidth}
		\vspace{0.6cm}
        \centering
        \includegraphics[width=\textwidth]{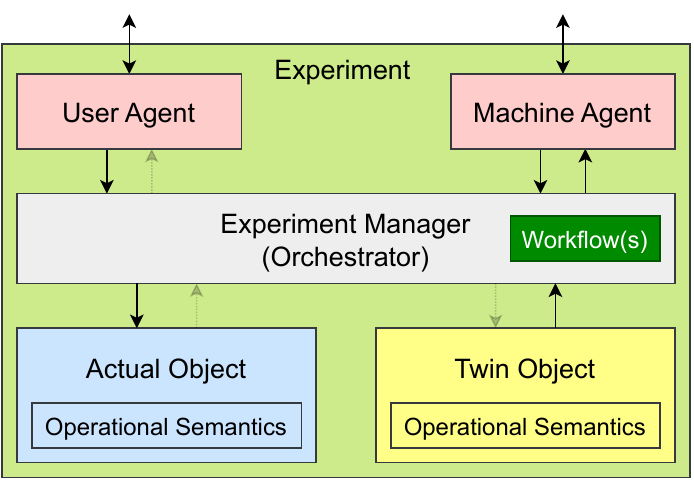}
        \caption{Reference architecture, variant 175.}
        \label{fig:var-174}
    \end{minipage}\hfill
	\begin{minipage}[t]{0.3\textwidth}
		\vspace{0.6cm}
        \centering
        \includegraphics[width=\textwidth]{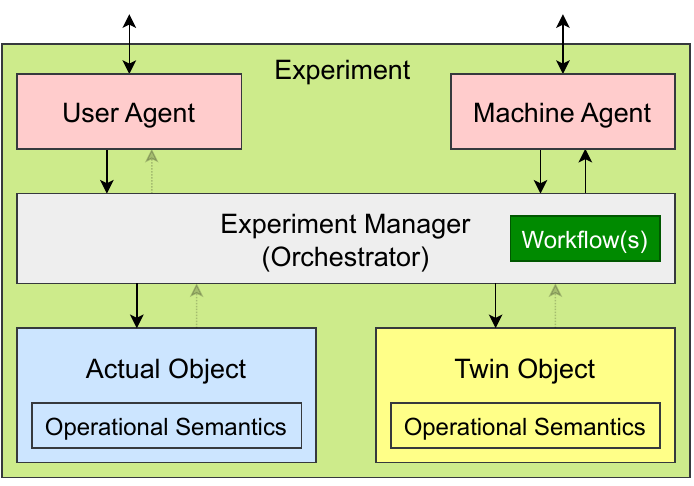}
        \caption{Reference architecture, variant 176.}
        \label{fig:var-175}
    \end{minipage}\hfill
	\begin{minipage}[t]{0.3\textwidth}
		\vspace{0.6cm}
        \centering
        \includegraphics[width=\textwidth]{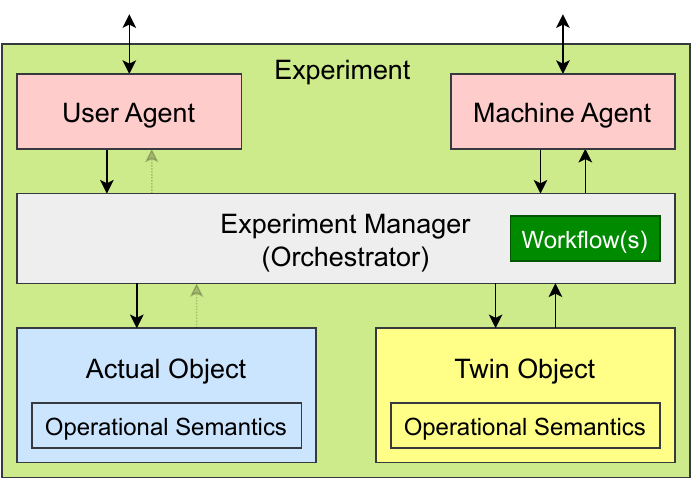}
        \caption{Reference architecture, variant 177.}
        \label{fig:var-176}
    \end{minipage}\\

\end{figure}
\begin{figure}[p]
	\centering
	\begin{minipage}[t]{0.3\textwidth}
		\vspace{0.6cm}
        \centering
        \includegraphics[width=\textwidth]{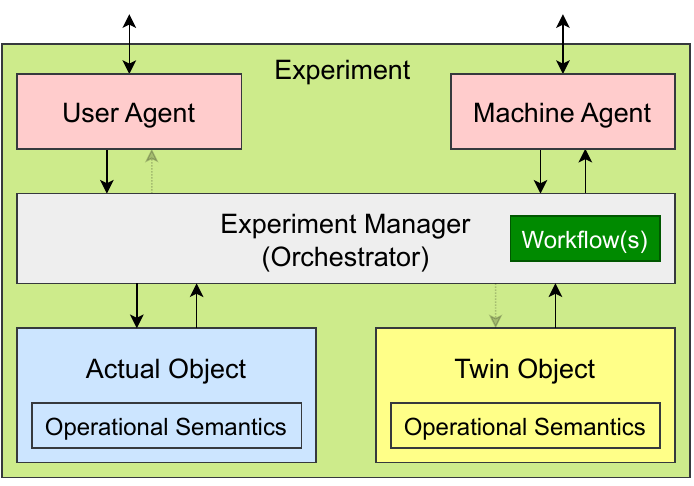}
        \caption{Reference architecture, variant 178.}
        \label{fig:var-177}
    \end{minipage}\hfill
	\begin{minipage}[t]{0.3\textwidth}
		\vspace{0.6cm}
        \centering
        \includegraphics[width=\textwidth]{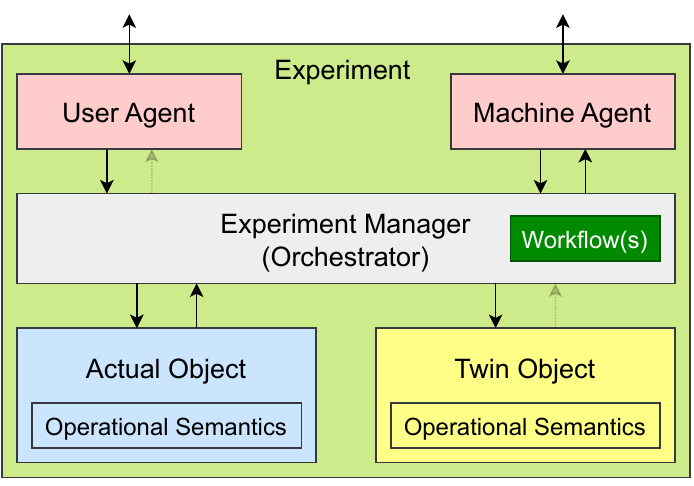}
        \caption{Reference architecture, variant 179.}
        \label{fig:var-178}
    \end{minipage}\hfill
	\begin{minipage}[t]{0.3\textwidth}
		\vspace{0.6cm}
        \centering
        \includegraphics[width=\textwidth]{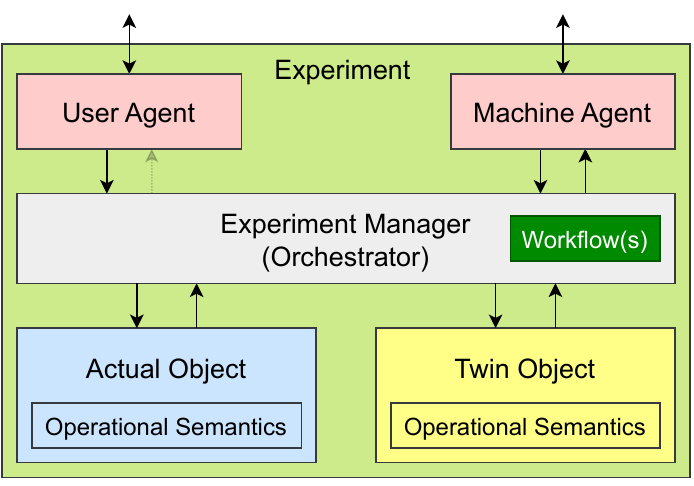}
        \caption{Reference architecture, variant 180.}
        \label{fig:var-179}
    \end{minipage}\\

	\begin{minipage}[t]{0.3\textwidth}
		\vspace{0.6cm}
        \centering
        \includegraphics[width=\textwidth]{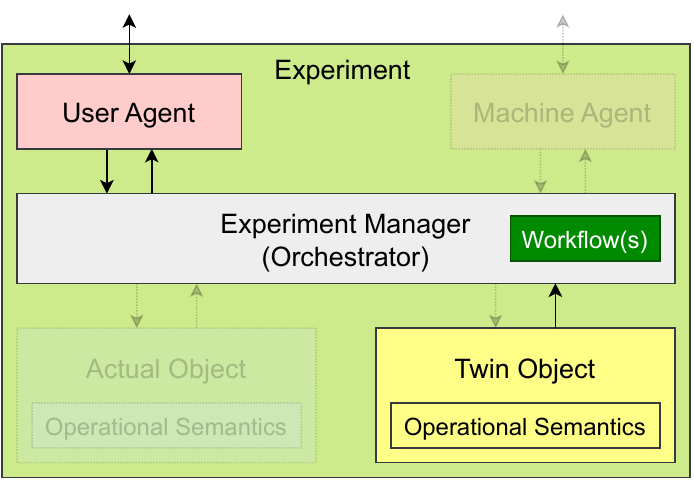}
        \caption{Reference architecture, variant 181, without Actual Object and Machine Agent.}
        \label{fig:var-180}
    \end{minipage}\hfill
	\begin{minipage}[t]{0.3\textwidth}
		\vspace{0.6cm}
        \centering
        \includegraphics[width=\textwidth]{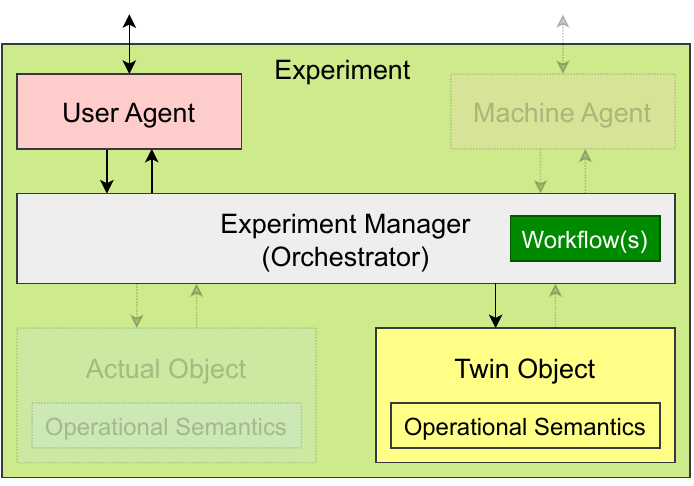}
        \caption{Reference architecture, variant 182, without Actual Object and Machine Agent.}
        \label{fig:var-181}
    \end{minipage}\hfill
	\begin{minipage}[t]{0.3\textwidth}
		\vspace{0.6cm}
        \centering
        \includegraphics[width=\textwidth]{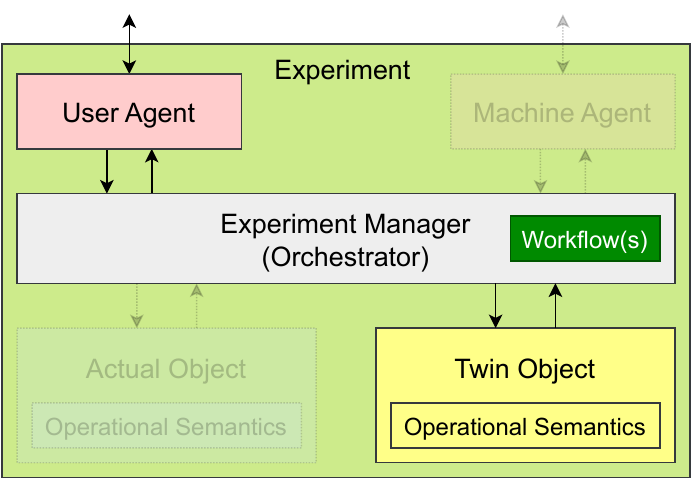}
        \caption{Reference architecture, variant 183, without Actual Object and Machine Agent.}
        \label{fig:var-182}
    \end{minipage}\\

	\begin{minipage}[t]{0.3\textwidth}
		\vspace{0.6cm}
        \centering
        \includegraphics[width=\textwidth]{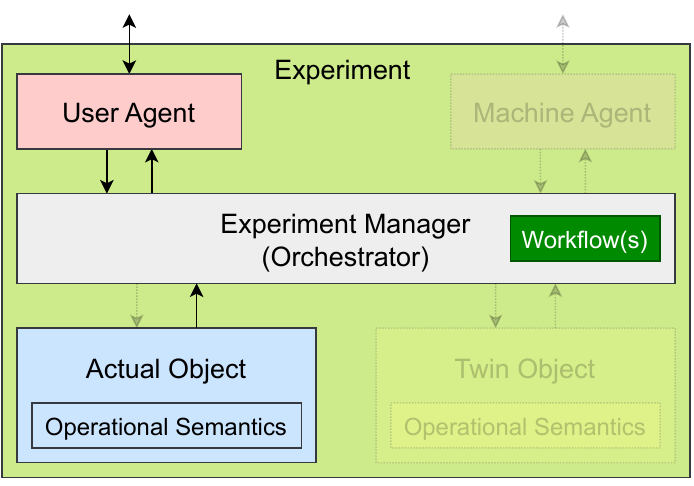}
        \caption{Reference architecture, variant 184, without Twin Object and Machine Agent.}
        \label{fig:var-183}
    \end{minipage}\hfill
	\begin{minipage}[t]{0.3\textwidth}
		\vspace{0.6cm}
        \centering
        \includegraphics[width=\textwidth]{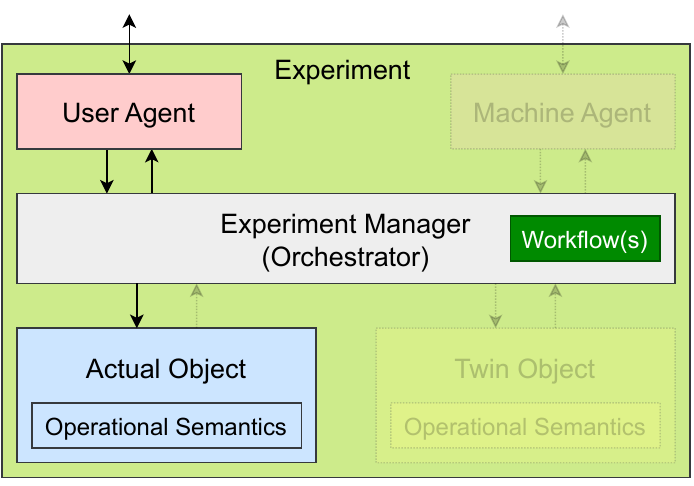}
        \caption{Reference architecture, variant 185, without Twin Object and Machine Agent.}
        \label{fig:var-184}
    \end{minipage}\hfill
	\begin{minipage}[t]{0.3\textwidth}
		\vspace{0.6cm}
        \centering
        \includegraphics[width=\textwidth]{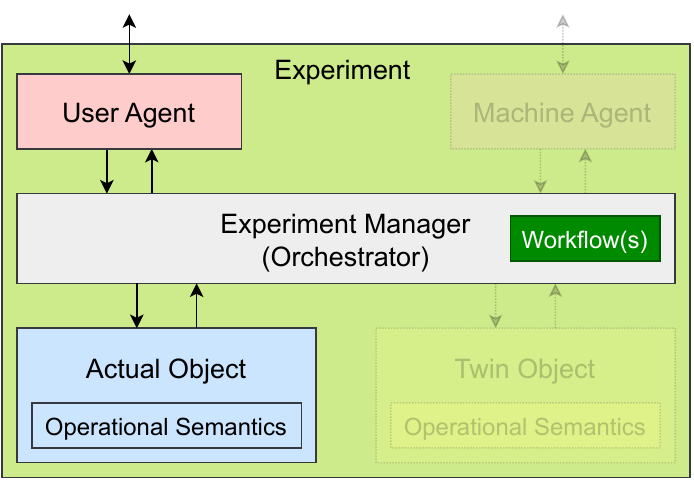}
        \caption{Reference architecture, variant 186, without Twin Object and Machine Agent.}
        \label{fig:var-185}
    \end{minipage}\\

	\begin{minipage}[t]{0.3\textwidth}
		\vspace{0.6cm}
        \centering
        \includegraphics[width=\textwidth]{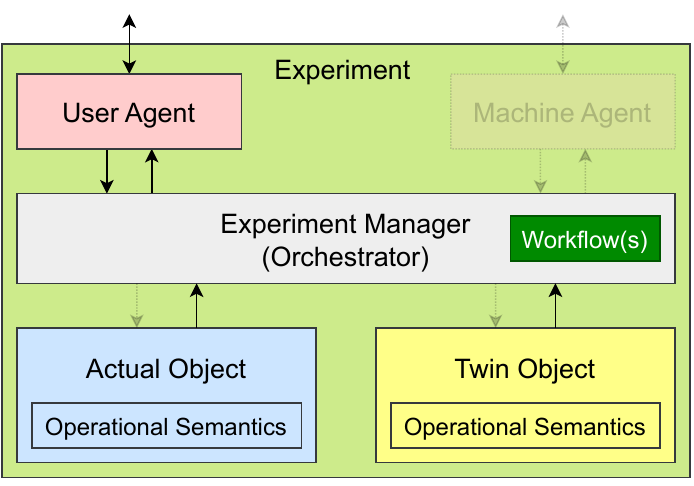}
        \caption{Reference architecture, variant 187, without Machine Agent.}
        \label{fig:var-186}
    \end{minipage}\hfill
	\begin{minipage}[t]{0.3\textwidth}
		\vspace{0.6cm}
        \centering
        \includegraphics[width=\textwidth]{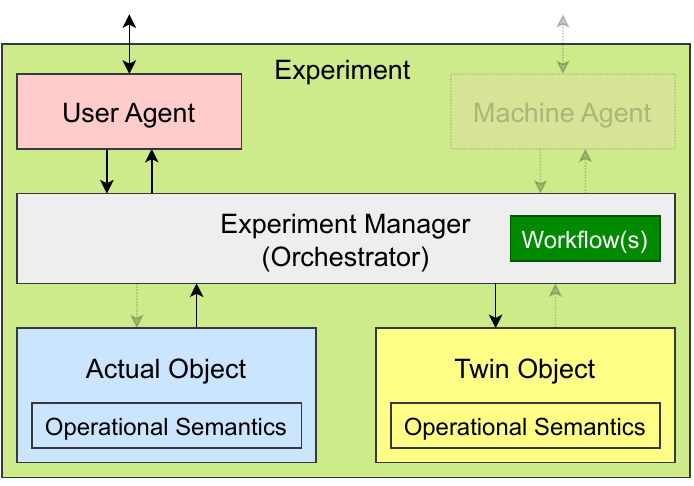}
        \caption{Reference architecture, variant 188, without Machine Agent.}
        \label{fig:var-187}
    \end{minipage}\hfill
	\begin{minipage}[t]{0.3\textwidth}
		\vspace{0.6cm}
        \centering
        \includegraphics[width=\textwidth]{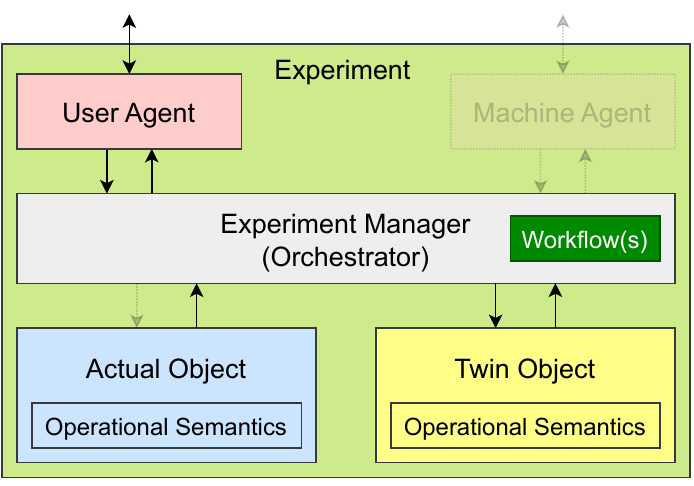}
        \caption{Reference architecture, variant 189, without Machine Agent.}
        \label{fig:var-188}
    \end{minipage}\\

\end{figure}
\begin{figure}[p]
	\centering
	\begin{minipage}[t]{0.3\textwidth}
		\vspace{0.6cm}
        \centering
        \includegraphics[width=\textwidth]{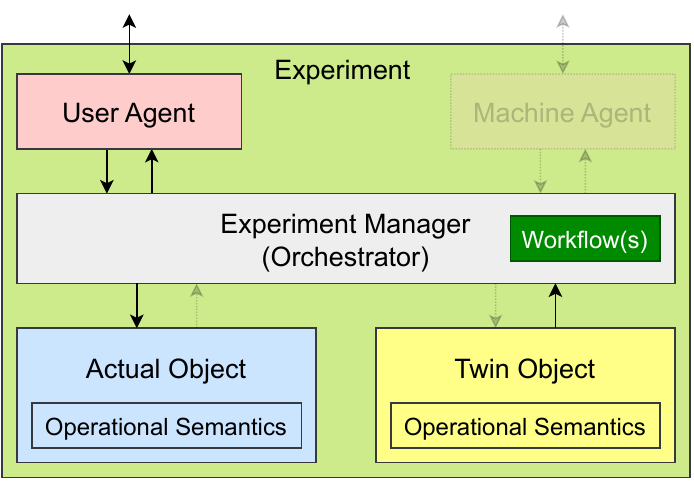}
        \caption{Reference architecture, variant 190, without Machine Agent.}
        \label{fig:var-189}
    \end{minipage}\hfill
	\begin{minipage}[t]{0.3\textwidth}
		\vspace{0.6cm}
        \centering
        \includegraphics[width=\textwidth]{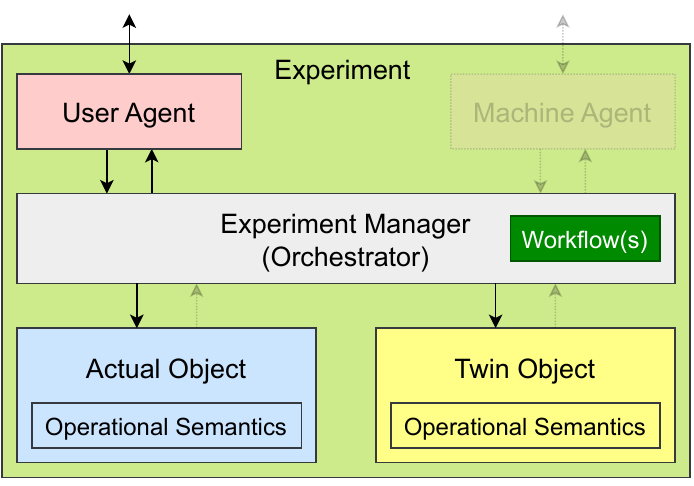}
        \caption{Reference architecture, variant 191, without Machine Agent.}
        \label{fig:var-190}
    \end{minipage}\hfill
	\begin{minipage}[t]{0.3\textwidth}
		\vspace{0.6cm}
        \centering
        \includegraphics[width=\textwidth]{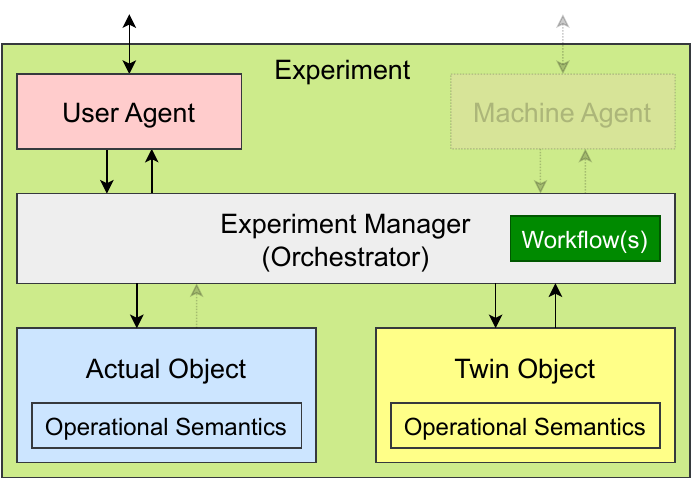}
        \caption{Reference architecture, variant 192, without Machine Agent.}
        \label{fig:var-191}
    \end{minipage}\\

	\begin{minipage}[t]{0.3\textwidth}
		\vspace{0.6cm}
        \centering
        \includegraphics[width=\textwidth]{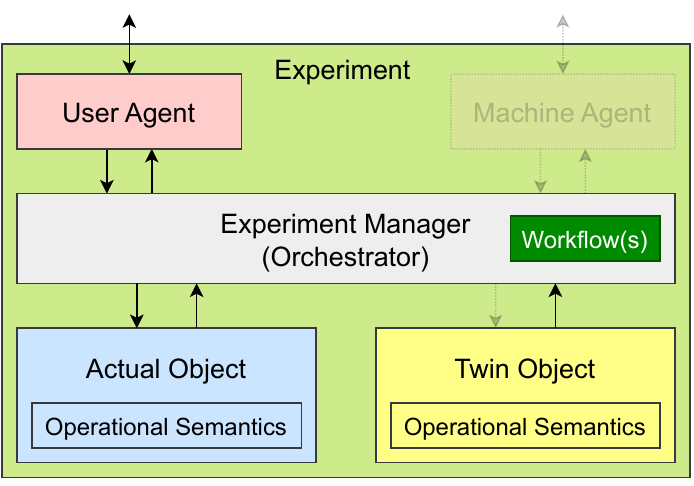}
        \caption{Reference architecture, variant 193, without Machine Agent.}
        \label{fig:var-192}
    \end{minipage}\hfill
	\begin{minipage}[t]{0.3\textwidth}
		\vspace{0.6cm}
        \centering
        \includegraphics[width=\textwidth]{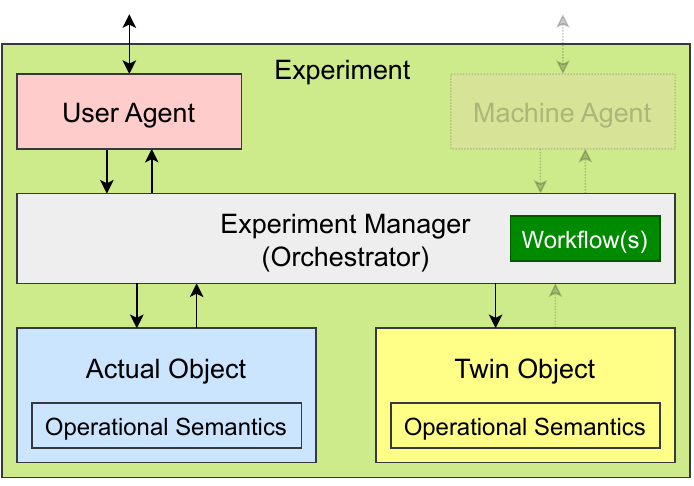}
        \caption{Reference architecture, variant 194, without Machine Agent.}
        \label{fig:var-193}
    \end{minipage}\hfill
	\begin{minipage}[t]{0.3\textwidth}
		\vspace{0.6cm}
        \centering
        \includegraphics[width=\textwidth]{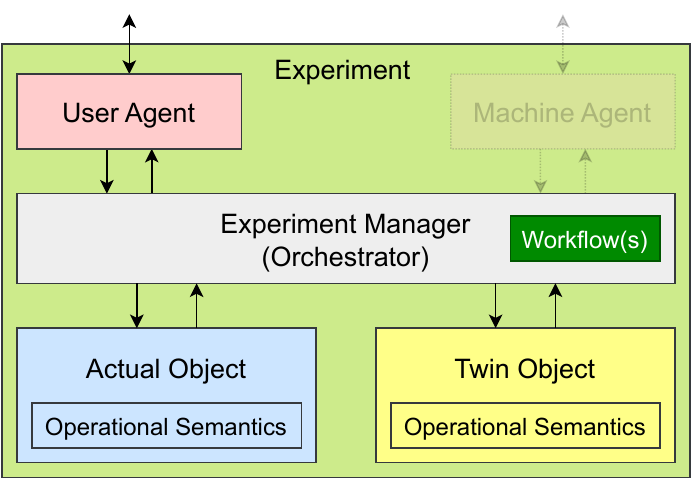}
        \caption{Reference architecture, variant 195, without Machine Agent.}
        \label{fig:var-194}
    \end{minipage}\\

	\begin{minipage}[t]{0.3\textwidth}
		\vspace{0.6cm}
        \centering
        \includegraphics[width=\textwidth]{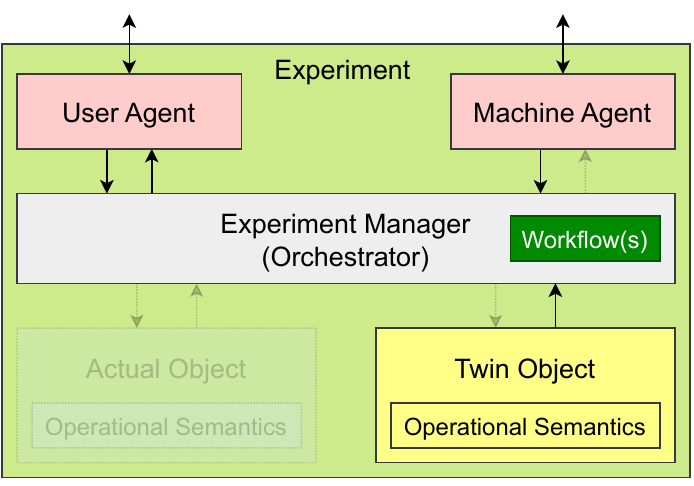}
        \caption{Reference architecture, variant 196, without Actual Object.}
        \label{fig:var-195}
    \end{minipage}\hfill
	\begin{minipage}[t]{0.3\textwidth}
		\vspace{0.6cm}
        \centering
        \includegraphics[width=\textwidth]{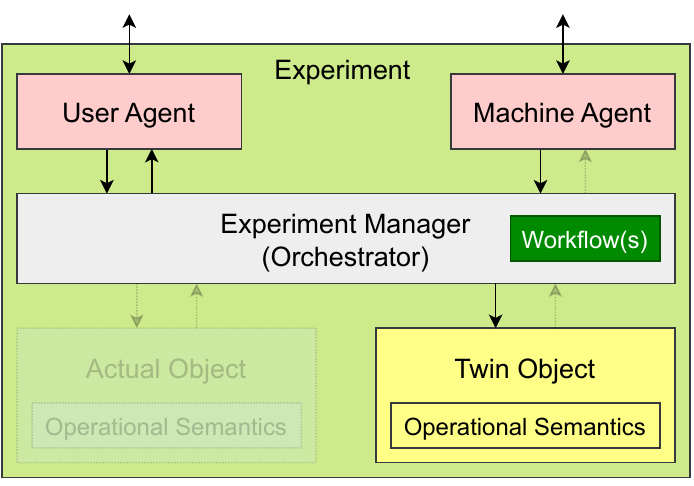}
        \caption{Reference architecture, variant 197, without Actual Object.}
        \label{fig:var-196}
    \end{minipage}\hfill
	\begin{minipage}[t]{0.3\textwidth}
		\vspace{0.6cm}
        \centering
        \includegraphics[width=\textwidth]{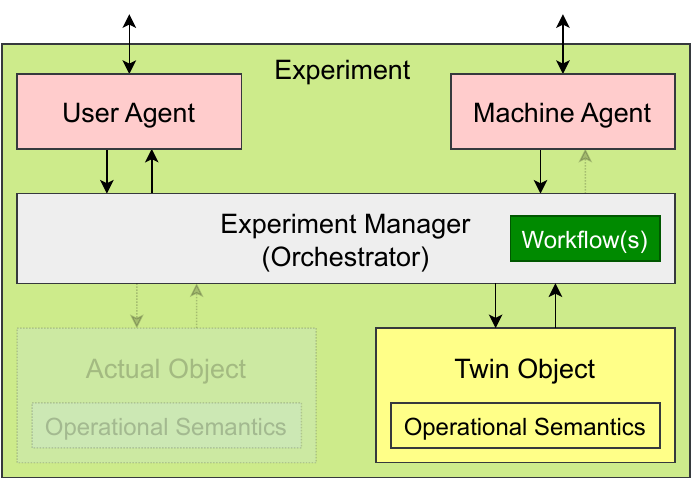}
        \caption{Reference architecture, variant 198, without Actual Object.}
        \label{fig:var-197}
    \end{minipage}\\

	\begin{minipage}[t]{0.3\textwidth}
		\vspace{0.6cm}
        \centering
        \includegraphics[width=\textwidth]{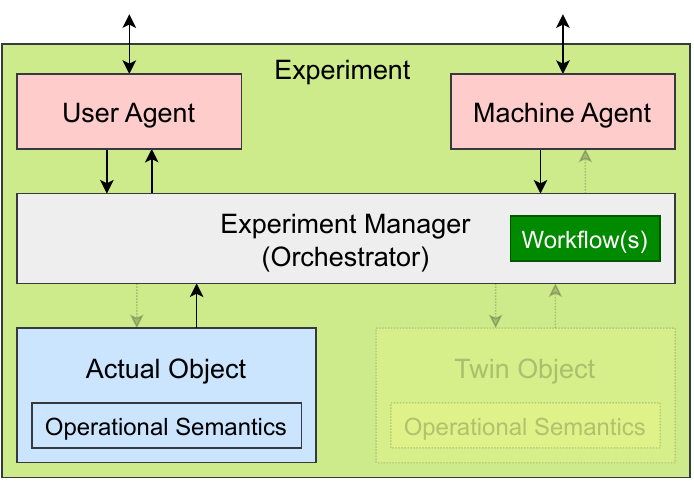}
        \caption{Reference architecture, variant 199, without Twin Object.}
        \label{fig:var-198}
    \end{minipage}\hfill
	\begin{minipage}[t]{0.3\textwidth}
		\vspace{0.6cm}
        \centering
        \includegraphics[width=\textwidth]{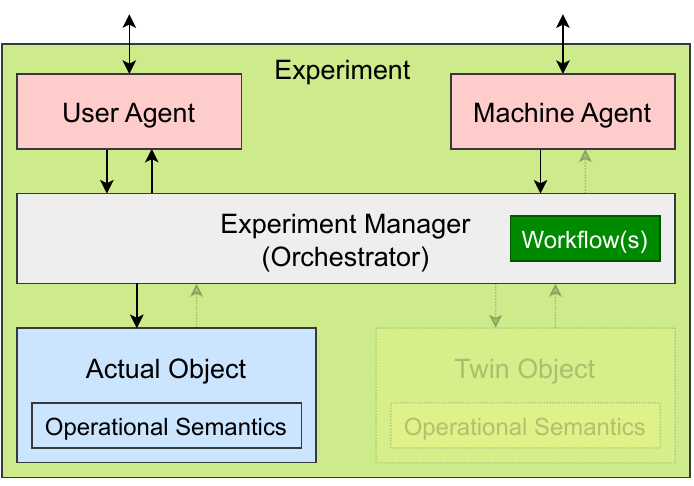}
        \caption{Reference architecture, variant 200, without Twin Object.}
        \label{fig:var-199}
    \end{minipage}\hfill
	\begin{minipage}[t]{0.3\textwidth}
		\vspace{0.6cm}
        \centering
        \includegraphics[width=\textwidth]{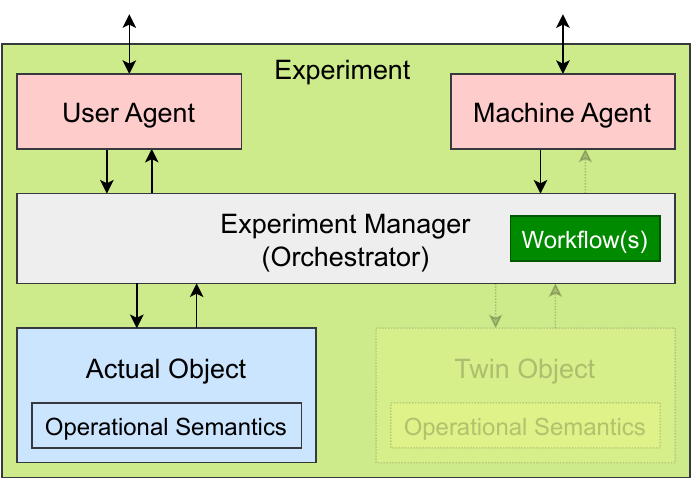}
        \caption{Reference architecture, variant 201, without Twin Object.}
        \label{fig:var-200}
    \end{minipage}\\

\end{figure}
\begin{figure}[p]
	\centering
	\begin{minipage}[t]{0.3\textwidth}
		\vspace{0.6cm}
        \centering
        \includegraphics[width=\textwidth]{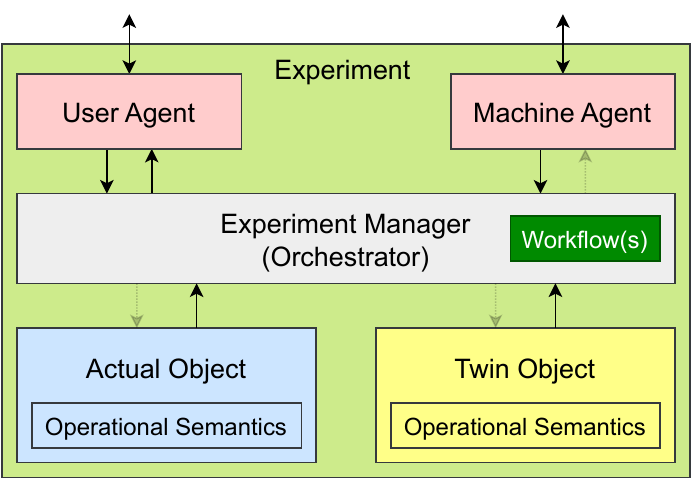}
        \caption{Reference architecture, variant 202.}
        \label{fig:var-201}
    \end{minipage}\hfill
	\begin{minipage}[t]{0.3\textwidth}
		\vspace{0.6cm}
        \centering
        \includegraphics[width=\textwidth]{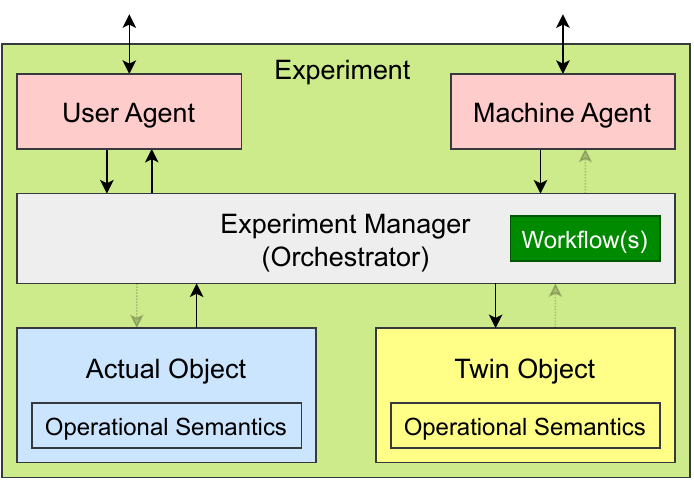}
        \caption{Reference architecture, variant 203.}
        \label{fig:var-202}
    \end{minipage}\hfill
	\begin{minipage}[t]{0.3\textwidth}
		\vspace{0.6cm}
        \centering
        \includegraphics[width=\textwidth]{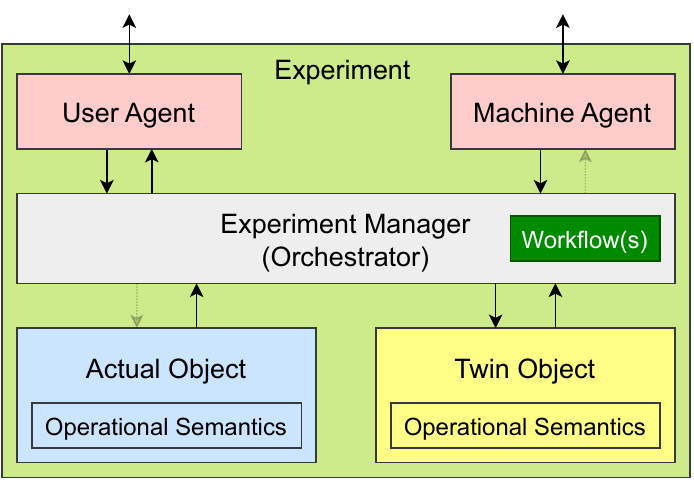}
        \caption{Reference architecture, variant 204.}
        \label{fig:var-203}
    \end{minipage}\\

	\begin{minipage}[t]{0.3\textwidth}
		\vspace{0.6cm}
        \centering
        \includegraphics[width=\textwidth]{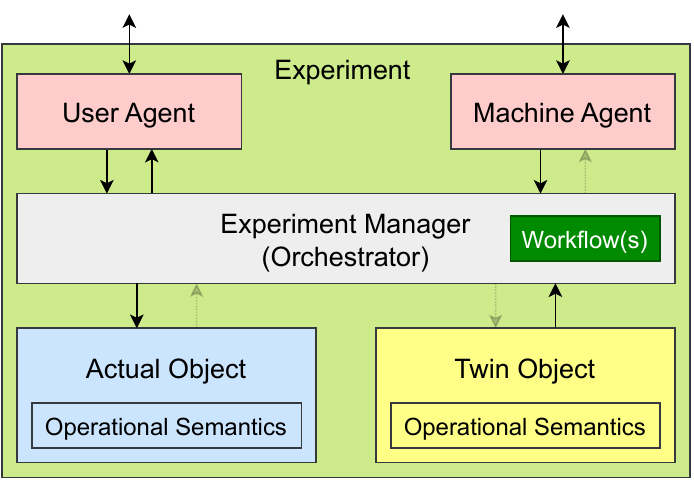}
        \caption{Reference architecture, variant 205.}
        \label{fig:var-204}
    \end{minipage}\hfill
	\begin{minipage}[t]{0.3\textwidth}
		\vspace{0.6cm}
        \centering
        \includegraphics[width=\textwidth]{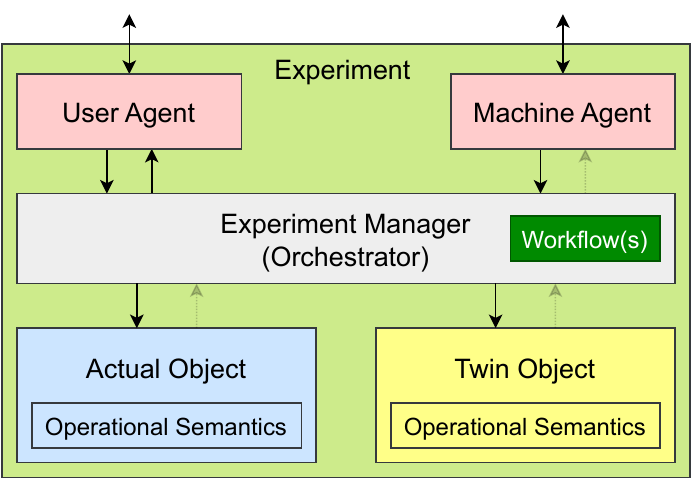}
        \caption{Reference architecture, variant 206.}
        \label{fig:var-205}
    \end{minipage}\hfill
	\begin{minipage}[t]{0.3\textwidth}
		\vspace{0.6cm}
        \centering
        \includegraphics[width=\textwidth]{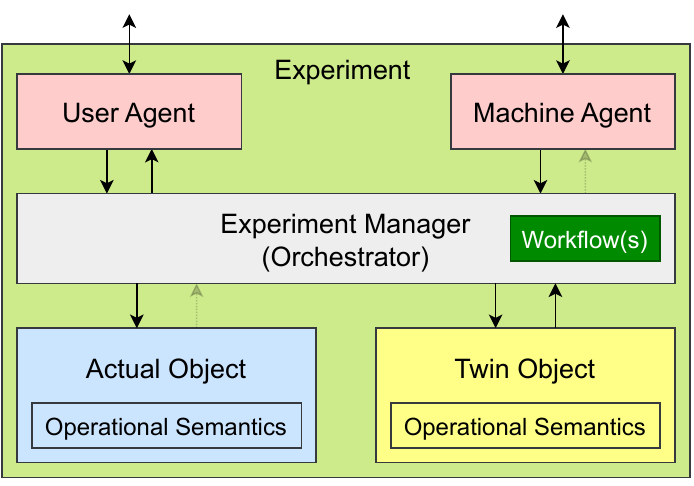}
        \caption{Reference architecture, variant 207.}
        \label{fig:var-206}
    \end{minipage}\\

	\begin{minipage}[t]{0.3\textwidth}
		\vspace{0.6cm}
        \centering
        \includegraphics[width=\textwidth]{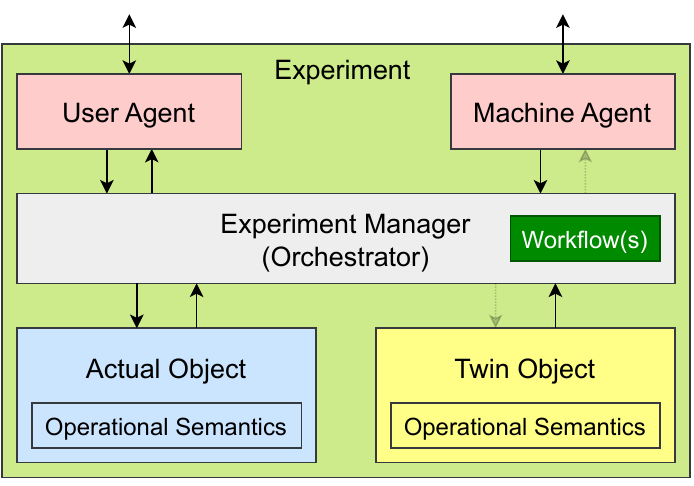}
        \caption{Reference architecture, variant 208.}
        \label{fig:var-207}
    \end{minipage}\hfill
	\begin{minipage}[t]{0.3\textwidth}
		\vspace{0.6cm}
        \centering
        \includegraphics[width=\textwidth]{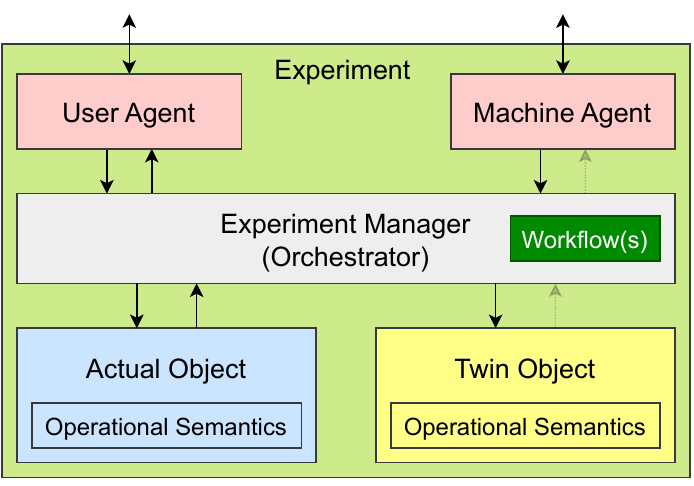}
        \caption{Reference architecture, variant 209.}
        \label{fig:var-208}
    \end{minipage}\hfill
	\begin{minipage}[t]{0.3\textwidth}
		\vspace{0.6cm}
        \centering
        \includegraphics[width=\textwidth]{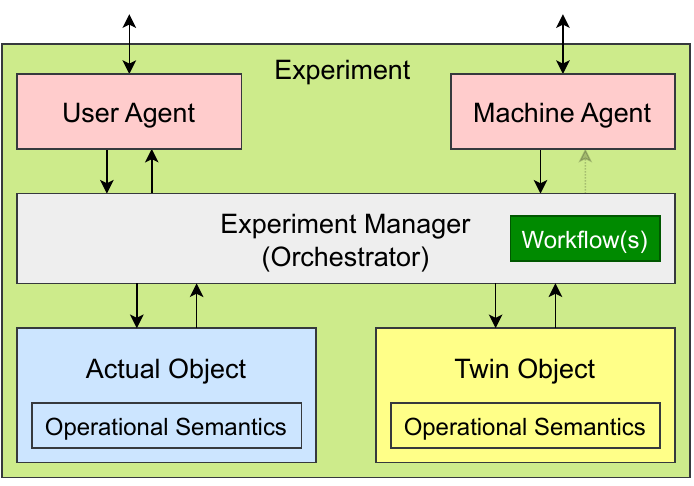}
        \caption{Reference architecture, variant 210.}
        \label{fig:var-209}
    \end{minipage}\\

	\begin{minipage}[t]{0.3\textwidth}
		\vspace{0.6cm}
        \centering
        \includegraphics[width=\textwidth]{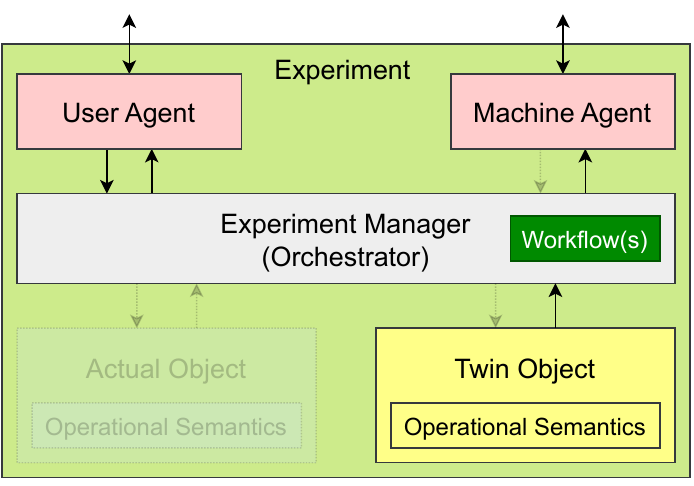}
        \caption{Reference architecture, variant 211, without Actual Object.}
        \label{fig:var-210}
    \end{minipage}\hfill
	\begin{minipage}[t]{0.3\textwidth}
		\vspace{0.6cm}
        \centering
        \includegraphics[width=\textwidth]{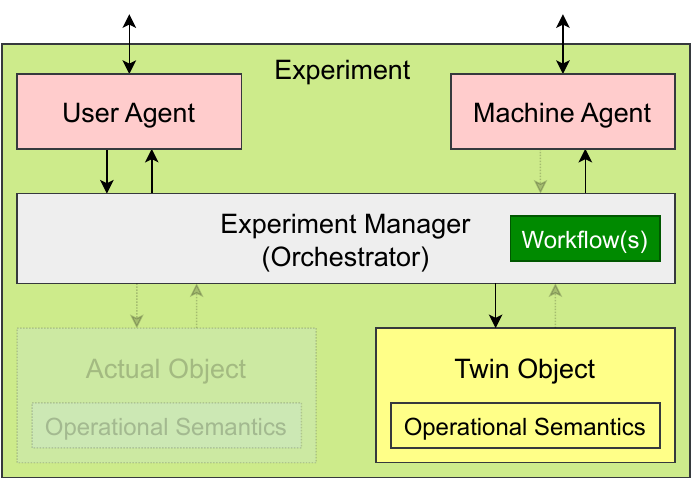}
        \caption{Reference architecture, variant 212, without Actual Object.}
        \label{fig:var-211}
    \end{minipage}\hfill
	\begin{minipage}[t]{0.3\textwidth}
		\vspace{0.6cm}
        \centering
        \includegraphics[width=\textwidth]{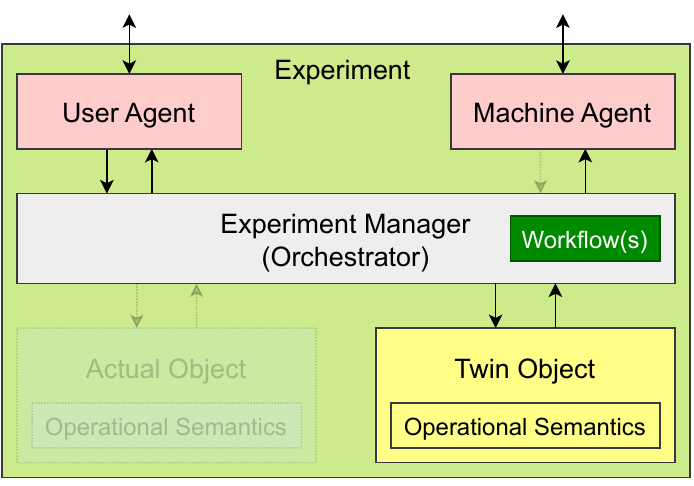}
        \caption{Reference architecture, variant 213, without Actual Object.}
        \label{fig:var-212}
    \end{minipage}\\

\end{figure}
\begin{figure}[p]
	\centering
	\begin{minipage}[t]{0.3\textwidth}
		\vspace{0.6cm}
        \centering
        \includegraphics[width=\textwidth]{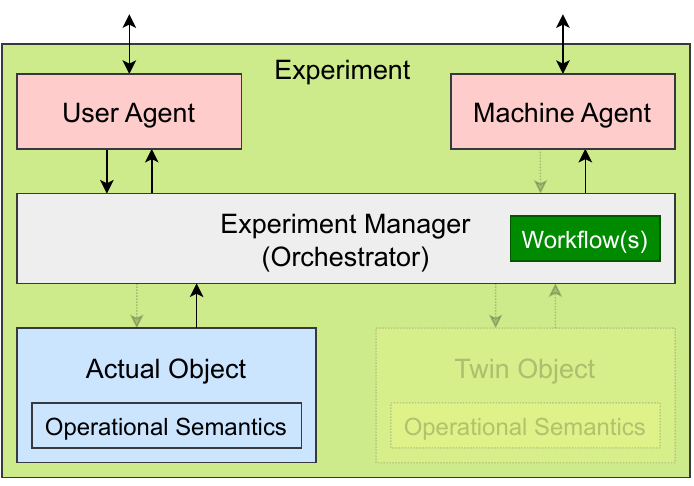}
        \caption{Reference architecture, variant 214, without Twin Object.}
        \label{fig:var-213}
    \end{minipage}\hfill
	\begin{minipage}[t]{0.3\textwidth}
		\vspace{0.6cm}
        \centering
        \includegraphics[width=\textwidth]{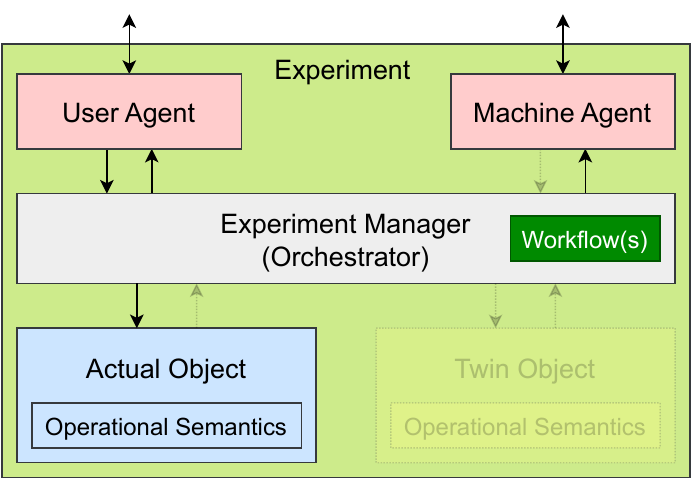}
        \caption{Reference architecture, variant 215, without Twin Object.}
        \label{fig:var-214}
    \end{minipage}\hfill
	\begin{minipage}[t]{0.3\textwidth}
		\vspace{0.6cm}
        \centering
        \includegraphics[width=\textwidth]{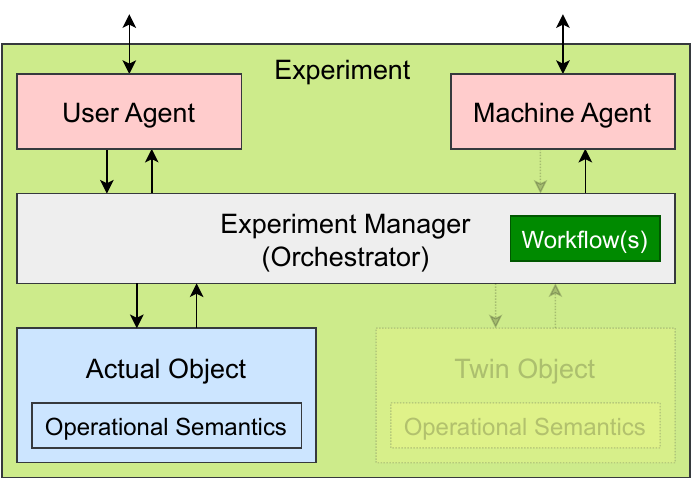}
        \caption{Reference architecture, variant 216, without Twin Object.}
        \label{fig:var-215}
    \end{minipage}\\

	\begin{minipage}[t]{0.3\textwidth}
		\vspace{0.6cm}
        \centering
        \includegraphics[width=\textwidth]{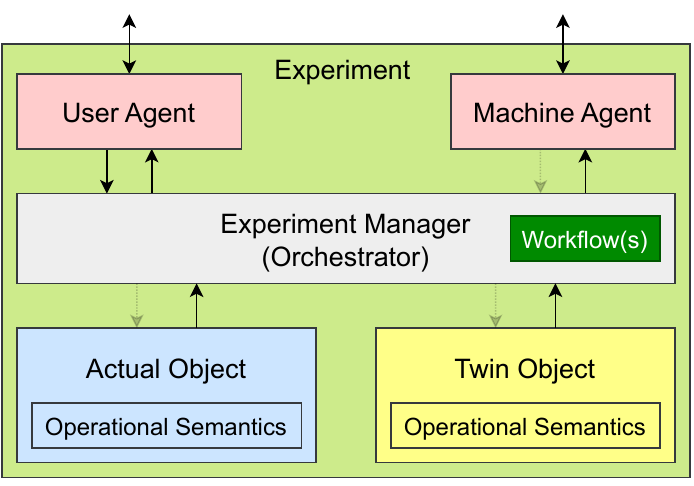}
        \caption{Reference architecture, variant 217.}
        \label{fig:var-216}
    \end{minipage}\hfill
	\begin{minipage}[t]{0.3\textwidth}
		\vspace{0.6cm}
        \centering
        \includegraphics[width=\textwidth]{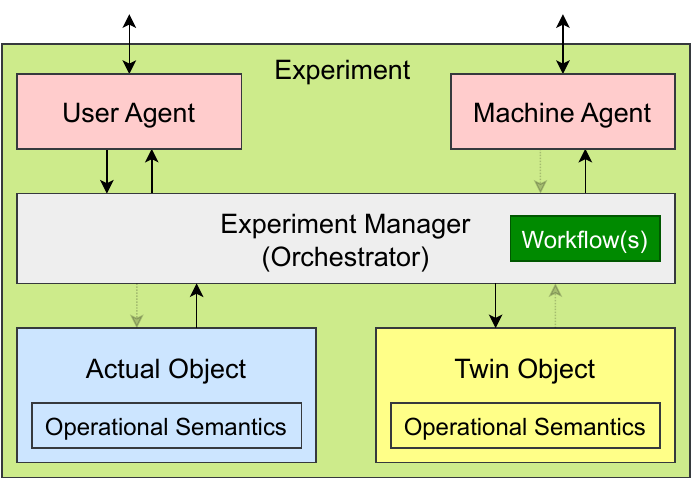}
        \caption{Reference architecture, variant 218.}
        \label{fig:var-217}
    \end{minipage}\hfill
	\begin{minipage}[t]{0.3\textwidth}
		\vspace{0.6cm}
        \centering
        \includegraphics[width=\textwidth]{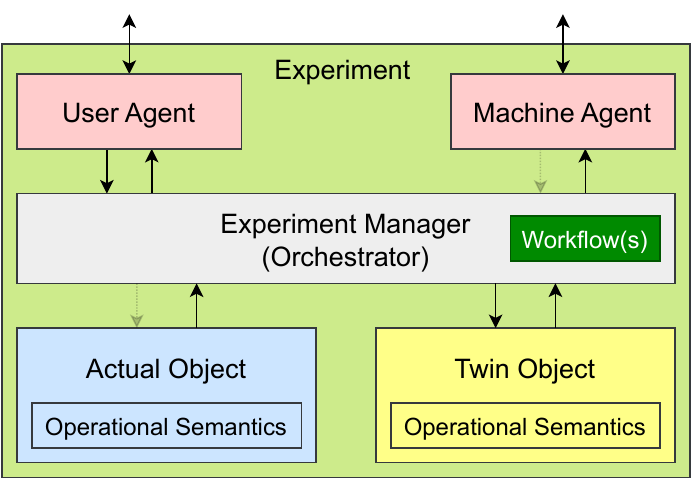}
        \caption{Reference architecture, variant 219.}
        \label{fig:var-218}
    \end{minipage}\\

	\begin{minipage}[t]{0.3\textwidth}
		\vspace{0.6cm}
        \centering
        \includegraphics[width=\textwidth]{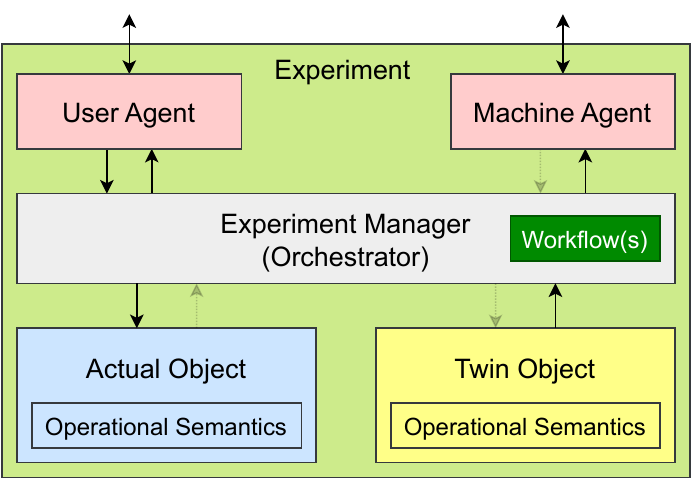}
        \caption{Reference architecture, variant 220.}
        \label{fig:var-219}
    \end{minipage}\hfill
	\begin{minipage}[t]{0.3\textwidth}
		\vspace{0.6cm}
        \centering
        \includegraphics[width=\textwidth]{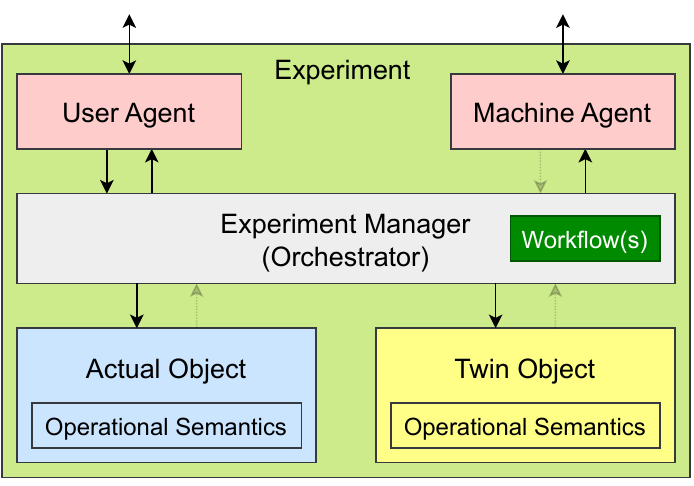}
        \caption{Reference architecture, variant 221.}
        \label{fig:var-220}
    \end{minipage}\hfill
	\begin{minipage}[t]{0.3\textwidth}
		\vspace{0.6cm}
        \centering
        \includegraphics[width=\textwidth]{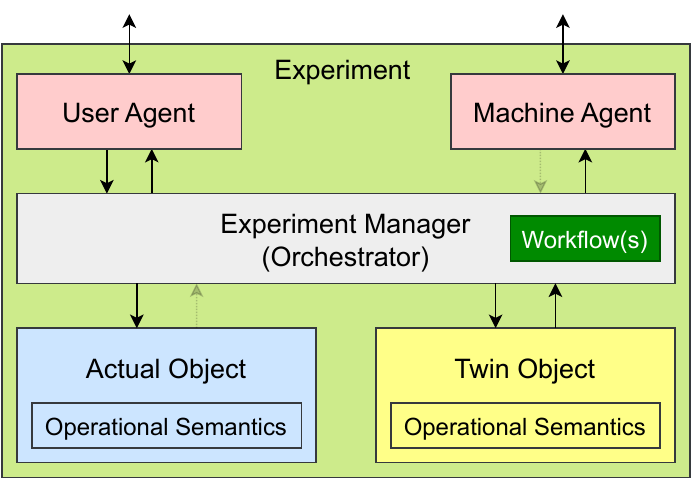}
        \caption{Reference architecture, variant 222.}
        \label{fig:var-221}
    \end{minipage}\\

	\begin{minipage}[t]{0.3\textwidth}
		\vspace{0.6cm}
        \centering
        \includegraphics[width=\textwidth]{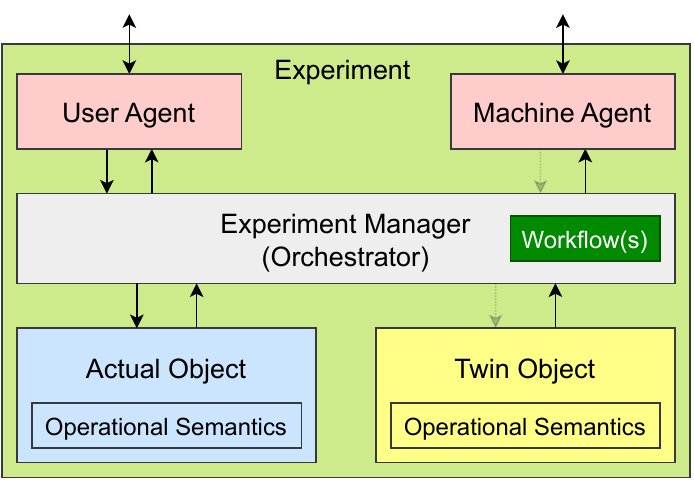}
        \caption{Reference architecture, variant 223.}
        \label{fig:var-222}
    \end{minipage}\hfill
	\begin{minipage}[t]{0.3\textwidth}
		\vspace{0.6cm}
        \centering
        \includegraphics[width=\textwidth]{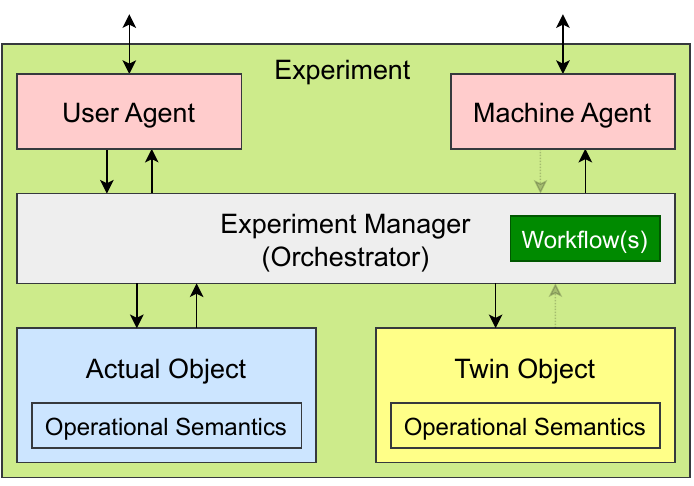}
        \caption{Reference architecture, variant 224.}
        \label{fig:var-223}
    \end{minipage}\hfill
	\begin{minipage}[t]{0.3\textwidth}
		\vspace{0.6cm}
        \centering
        \includegraphics[width=\textwidth]{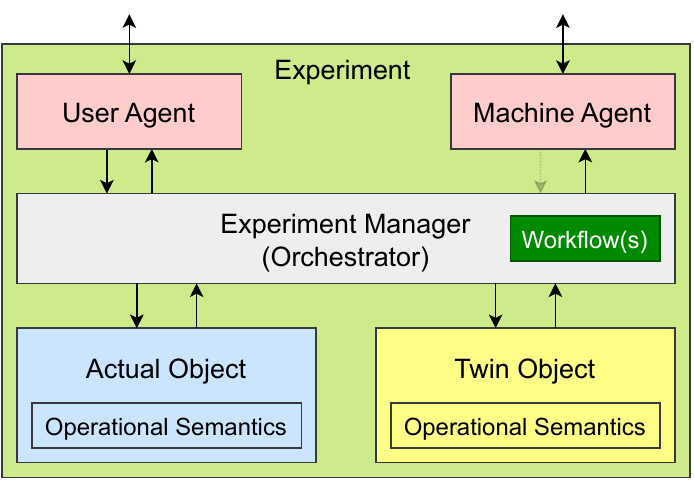}
        \caption{Reference architecture, variant 225.}
        \label{fig:var-224}
    \end{minipage}\\

\end{figure}
\begin{figure}[p]
	\centering
	\begin{minipage}[t]{0.3\textwidth}
		\vspace{0.6cm}
        \centering
        \includegraphics[width=\textwidth]{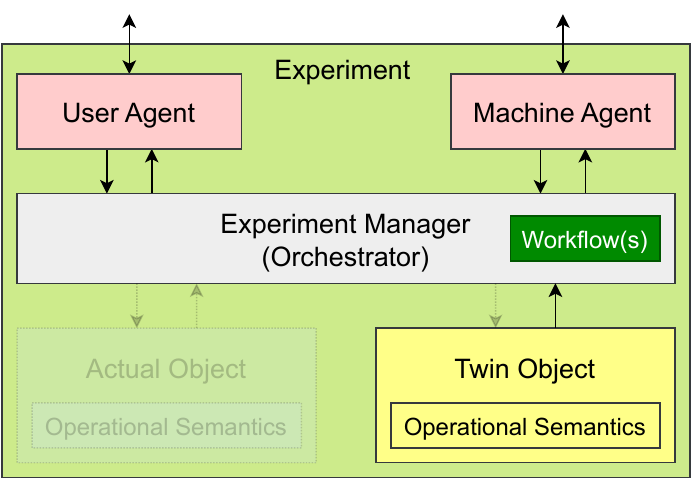}
        \caption{Reference architecture, variant 226, without Actual Object.}
        \label{fig:var-225}
    \end{minipage}\hfill
	\begin{minipage}[t]{0.3\textwidth}
		\vspace{0.6cm}
        \centering
        \includegraphics[width=\textwidth]{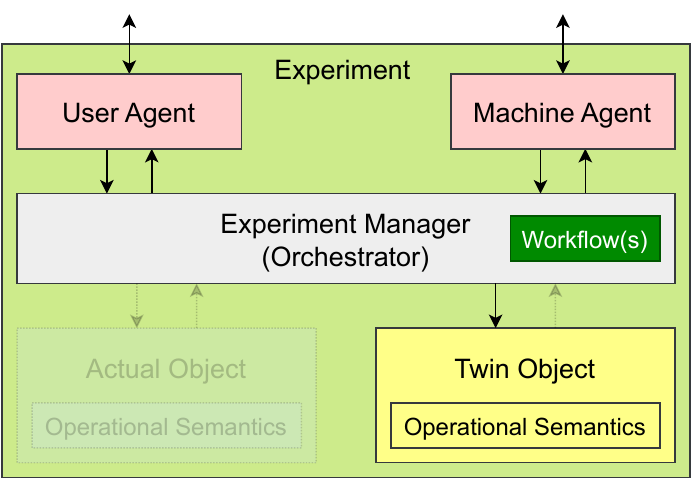}
        \caption{Reference architecture, variant 227, without Actual Object.}
        \label{fig:var-226}
    \end{minipage}\hfill
	\begin{minipage}[t]{0.3\textwidth}
		\vspace{0.6cm}
        \centering
        \includegraphics[width=\textwidth]{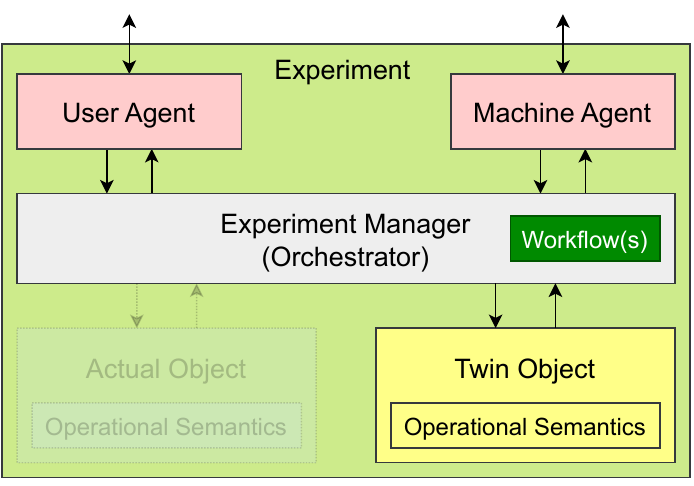}
        \caption{Reference architecture, variant 228, without Actual Object.}
        \label{fig:var-227}
    \end{minipage}\\

	\begin{minipage}[t]{0.3\textwidth}
		\vspace{0.6cm}
        \centering
        \includegraphics[width=\textwidth]{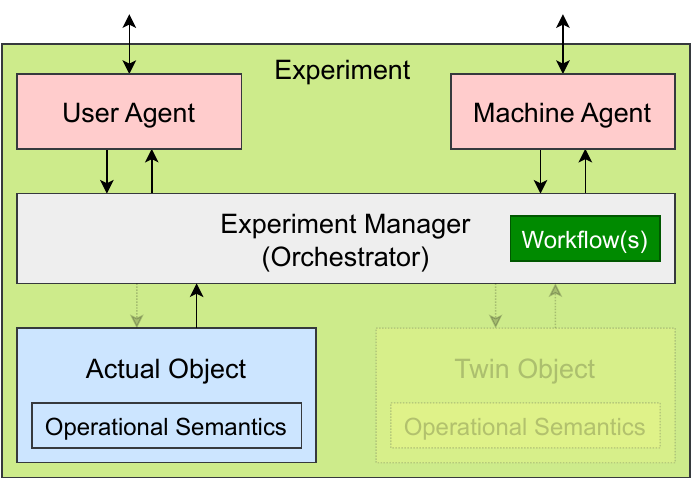}
        \caption{Reference architecture, variant 229, without Twin Object.}
        \label{fig:var-228}
    \end{minipage}\hfill
	\begin{minipage}[t]{0.3\textwidth}
		\vspace{0.6cm}
        \centering
        \includegraphics[width=\textwidth]{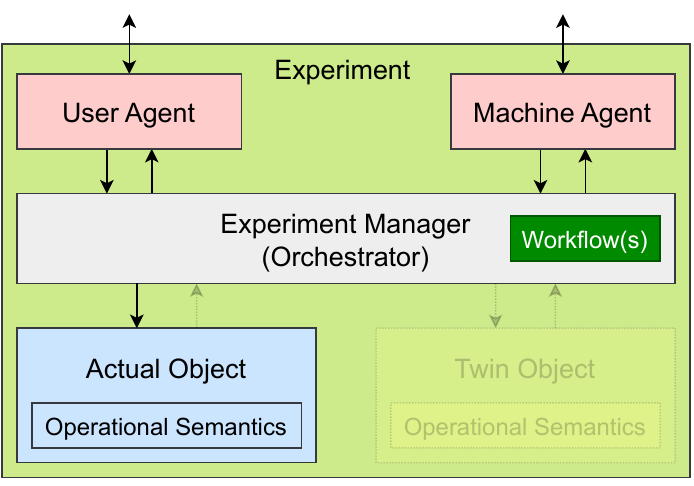}
        \caption{Reference architecture, variant 230, without Twin Object.}
        \label{fig:var-229}
    \end{minipage}\hfill
	\begin{minipage}[t]{0.3\textwidth}
		\vspace{0.6cm}
        \centering
        \includegraphics[width=\textwidth]{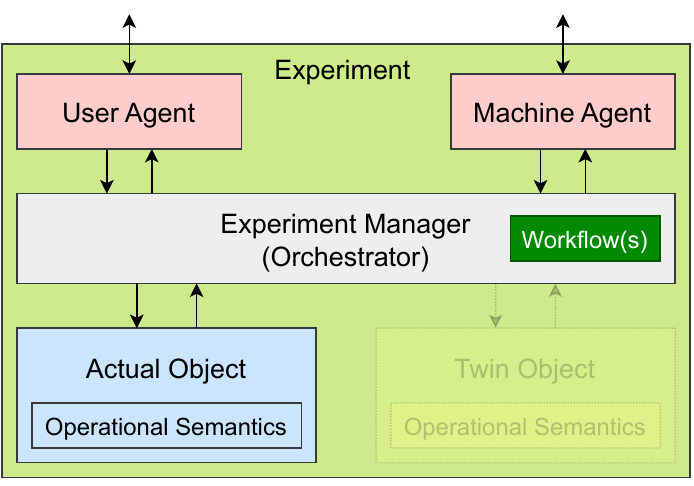}
        \caption{Reference architecture, variant 231, without Twin Object.}
        \label{fig:var-230}
    \end{minipage}\\

	\begin{minipage}[t]{0.3\textwidth}
		\vspace{0.6cm}
        \centering
        \includegraphics[width=\textwidth]{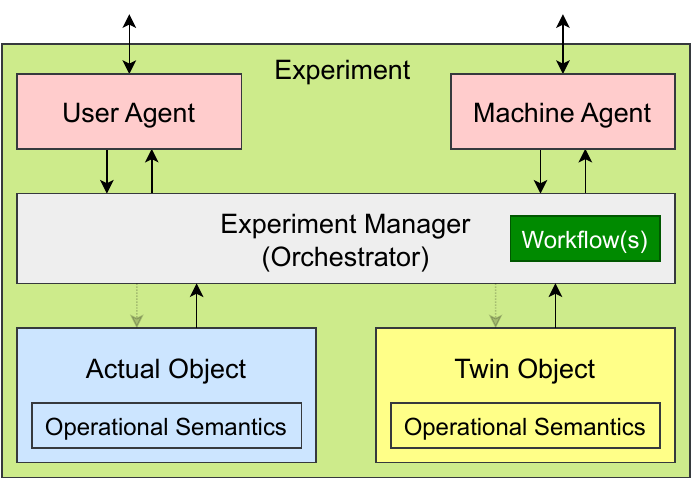}
        \caption{Reference architecture, variant 232.}
        \label{fig:var-231}
    \end{minipage}\hfill
	\begin{minipage}[t]{0.3\textwidth}
		\vspace{0.6cm}
        \centering
        \includegraphics[width=\textwidth]{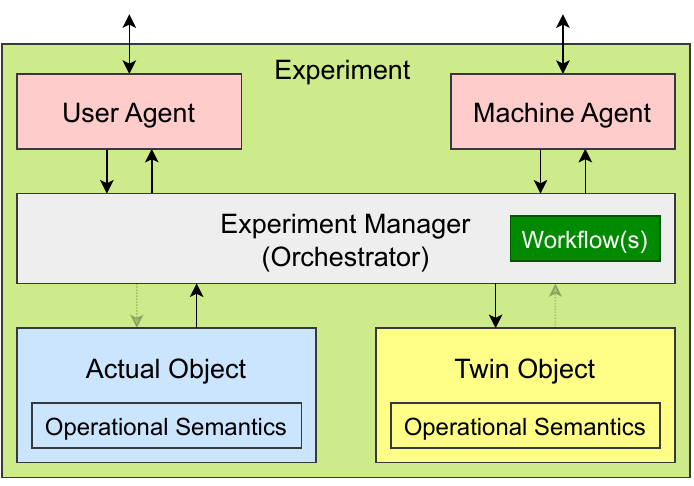}
        \caption{Reference architecture, variant 233.}
        \label{fig:var-232}
    \end{minipage}\hfill
	\begin{minipage}[t]{0.3\textwidth}
		\vspace{0.6cm}
        \centering
        \includegraphics[width=\textwidth]{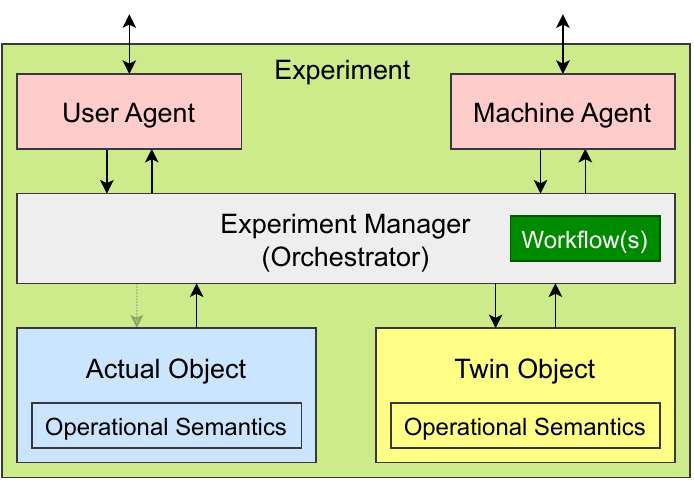}
        \caption{Reference architecture, variant 234.}
        \label{fig:var-233}
    \end{minipage}\\

	\begin{minipage}[t]{0.3\textwidth}
		\vspace{0.6cm}
        \centering
        \includegraphics[width=\textwidth]{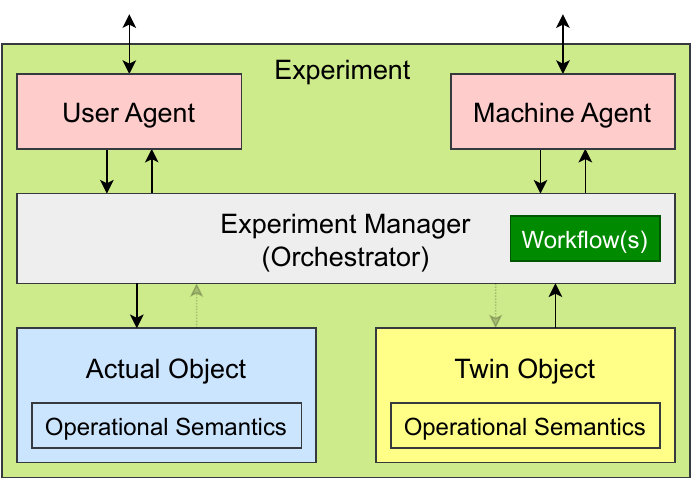}
        \caption{Reference architecture, variant 235.}
        \label{fig:var-234}
    \end{minipage}\hfill
	\begin{minipage}[t]{0.3\textwidth}
		\vspace{0.6cm}
        \centering
        \includegraphics[width=\textwidth]{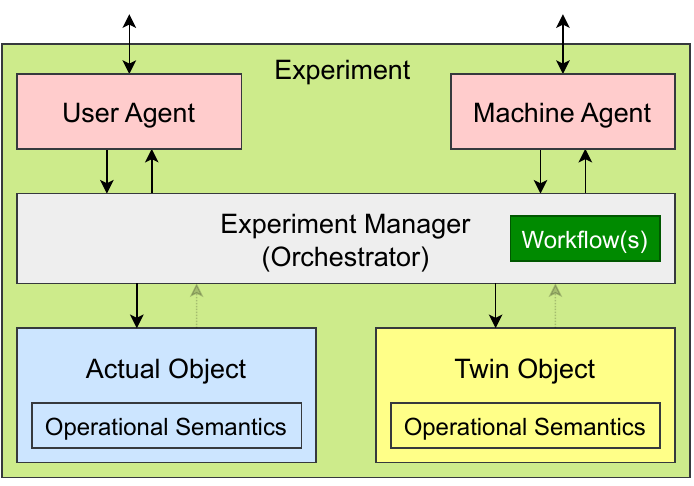}
        \caption{Reference architecture, variant 236.}
        \label{fig:var-235}
    \end{minipage}\hfill
	\begin{minipage}[t]{0.3\textwidth}
		\vspace{0.6cm}
        \centering
        \includegraphics[width=\textwidth]{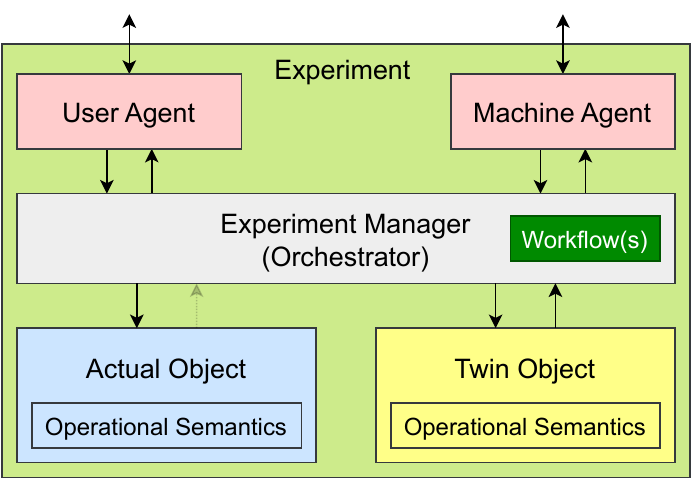}
        \caption{Reference architecture, variant 237.}
        \label{fig:var-236}
    \end{minipage}\\

\end{figure}
\begin{figure}[p]
	\centering
	\begin{minipage}[t]{0.3\textwidth}
		\vspace{0.6cm}
        \centering
        \includegraphics[width=\textwidth]{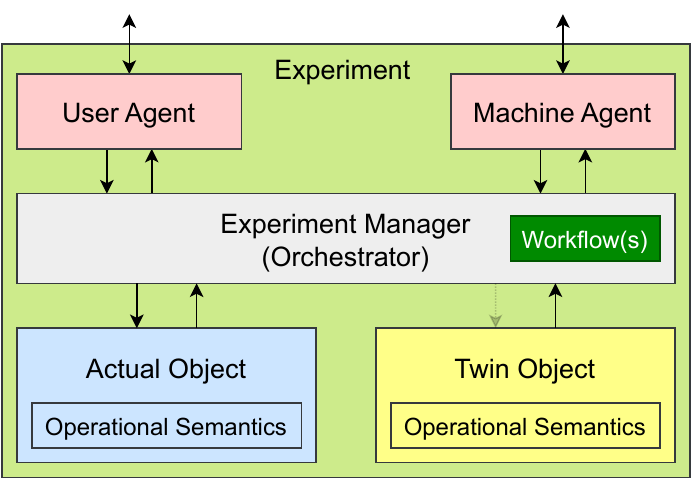}
        \caption{Reference architecture, variant 238.}
        \label{fig:var-237}
    \end{minipage}\hfill
	\begin{minipage}[t]{0.3\textwidth}
		\vspace{0.6cm}
        \centering
        \includegraphics[width=\textwidth]{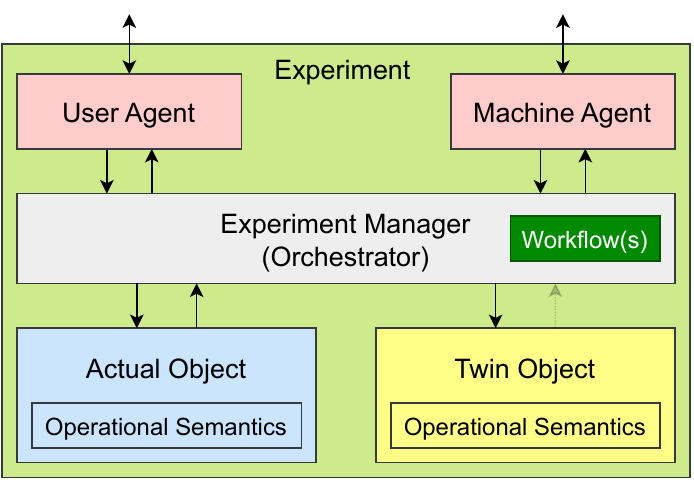}
        \caption{Reference architecture, variant 239.}
        \label{fig:var-238}
    \end{minipage}\hfill
	\begin{minipage}[t]{0.3\textwidth}
		\vspace{0.6cm}
        \centering
        \includegraphics[width=\textwidth]{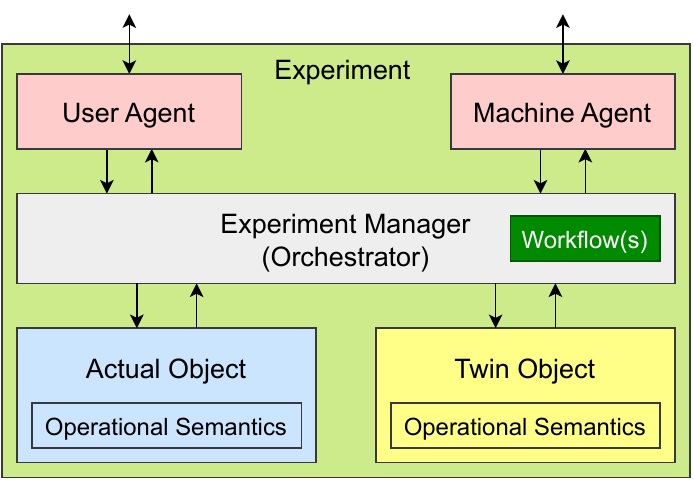}
        \caption{Reference architecture, variant 240.}
        \label{fig:var-239}
    \end{minipage}\\

\end{figure}

\end{document}